\documentclass[pdftex,twocolumn,epjc3]{svjour3}          

\newcommand{\Nb}{\bar{N}}

\newcommand{\muf}{\mu_F}

\newcommand{\g}{{\cal G}}

\def\LogmW1{{{\ln (1-\omega)}}}

\def\zo{\overline{z}_1}
\def\zt{\overline{z}_2}
\def\zo{\overline{z}_1}
\def\zt{\overline{z}_2}
\def\zob{\overline{z}_1}
\def\ztb{\overline{z}_2}
\def\zotb{z_{12}}

\usepackage[table]{xcolor}
\usepackage{colortbl}
\RequirePackage{ifpdf} 
\usepackage{amsmath} 
\usepackage{mathtools}
\usepackage{cases}
\usepackage{bm,amssymb}
\usepackage{amsbsy}
\usepackage{endnotes}
\let\footnote=\endnote

\usepackage{pstricks}
\usepackage[final]{pdfpages} 
\usepackage{ifpdf} 
\usepackage{slashed}

\usepackage{color}
\usepackage{xcolor}
\definecolor{urlblue}{rgb}{0.2,0.4,0.7}
\definecolor{citegreen}{rgb}{0,0.6,0.2}
\definecolor{linkred}{rgb}{0.9,0.2,0.1}

\usepackage{graphics}
\usepackage{etoolbox} 
\usepackage{fixmath}
\usepackage{psfrag}
\usepackage{bm}

\usepackage{notoccite}

\usepackage{amsfonts}
\usepackage{autobreak}
\usepackage{marginnote}
\usepackage{widetext}
\usepackage{tabularx}
\newcolumntype{P}[1]{>{\centering\arraybackslash}p{#1}}
\usepackage{multirow, caption, booktabs}
\usepackage{slashed}
\usepackage{array}
\newcolumntype{P}[1]{>{\centering\arraybackslash}p{#1}}
\allowdisplaybreaks

\usepackage{tikz}
\usetikzlibrary{positioning,arrows}
\usetikzlibrary{decorations.pathmorphing}
\usetikzlibrary{decorations.markings}
\usetikzlibrary{shapes.geometric}
\tikzset{
    vector/.style={decorate, decoration={snake}, draw},
    provector/.style={decorate, decoration={snake,amplitude=2.5pt}, draw},
    antivector/.style={decorate, decoration={snake,amplitude=-2.5pt}, draw},
    fermion/.style={draw=black, postaction={decorate},decoration={markings,mark=at position .55 with {\arrow[draw=black]{>}}}},
    fermionbar/.style={draw=black, postaction={decorate},
                       decoration={markings,mark=at position .55 with {\arrow[draw=black]{<}}}},
    fermionnoarrow/.style={draw=black},
    gluon/.style={decorate, draw=black,decoration={coil,amplitude=4pt, segment length=5pt}},
    scalar/.style={dashed,draw=black, postaction={decorate},decoration={markings,mark=at position .55 with {\arrow[draw=black]{>}}}},
    scalarbar/.style={dashed,draw=black, postaction={decorate},decoration={markings,mark=at position .55 with {\arrow[draw=black]{<}}}},
    scalarnoarrow/.style={dashed,draw=black},
    electron/.style={draw=black, postaction={decorate},decoration={markings,mark=at position .55 with {\arrow[draw=black]{>}}}},
    bigvector/.style={decorate, decoration={snake,amplitude=4pt}, draw},
}

\def\g{\mathcal{\tilde{G}}}
\def\gm{\gamma}

\def\epm1{\frac{1}{\epsilon}}
\def\epm2{\frac{1}{\epsilon^{2}}}
\def\epm3{\frac{1}{\epsilon^{3}}}
\def\epm4{\frac{1}{\epsilon^{4}}}

\def\gm0{\gamma_{0}}
\def\gm1{\gamma_{1}}
\def\gm2{\gamma_{2}}
\def\gm3{\gamma_{3}}
\def\nn{\nonumber}

\def\nn{\nonumber\\}

\RequirePackage[T1]{fontenc}

\smartqed  

\RequirePackage{graphicx}
\RequirePackage{mathptmx}      
\RequirePackage{flushend}
\RequirePackage[numbers,sort&compress]{natbib}
\RequirePackage[colorlinks,citecolor=blue,urlcolor=blue,linkcolor=blue]{hyperref}

\journalname{Eur. Phys. J. C}

\begin{document}

\title{Rapidity distribution at Soft-virtual and beyond for $n$-colorless particles to N$^4$LO in QCD}




\author{Taushif Ahmed\thanksref{e1,addr1,addr3}
        \and
       A. H. Ajjath\thanksref{e2,addr2} 
      \and
         Pooja Mukherjee\thanksref{e3,addr2} 
          \and
         V. Ravindran\thanksref{e4,addr2} 
          \and
         Aparna Sankar\thanksref{e5,addr2} 
}

\thankstext{e1}{e-mail: taushif@mpp.mpg.de }
\thankstext{e2}{e-mail: ajjathah@imsc.res.in}
\thankstext{e3}{e-mail:poojamukherjee@imsc.res.in }
\thankstext{e4}{e-mail:ravindra@imsc.res.in }
\thankstext{e5}{e-mail: aparnas@imsc.res.in}

\institute{Max-Planck-Institut f\"ur Physik, Werner-Heisenberg-Institut, 80805 M\"unchen, Germany\label{addr1}
          \and
          Dipartimento di Fisica and Arnold-Regge Center, Universit\`a di Torino, 
\\ and INFN, Sezione di Torino, Via Pietro Giuria 1, I-10125 Torino, Italy\label{addr3}
          \and
         The Institute of Mathematical Sciences, HBNI, IV Cross Road, Taramani, Chennai 600113, India\label{addr2}
}

\date{Received: date / Accepted: date}

\maketitle

\begin{abstract}
We present a systematic framework to study the threshold contributions of the differential rapidity distribution for the production of any number of colorless particles in the hadronic colliders. This has been achieved based on the universality structure of the soft enhancements associated with the real emissions, along with the factorization property of the differential cross section and the renormalization group invariance. In this formalism, we present a universal soft-collinear operator to compute the soft virtual differential cross section for a generic $2\rightarrow n$ scattering process up to next-to-next-to-next-to-next-to-leading order (N$^4$LO) in perturbative QCD. We also provide a universal operator to perform the threshold resummation to next-to-next-to-next-to-leading logarithmic (N$^3$LL) accuracy. We explicitly present the approximate analytical results of the rapidity distributions at N$^4$LO and N$^3$LL for the Higgs boson production through gluon fusion and bottom quark annihilation, and also for the Drell-Yan production at the hadronic collider. \textcolor{black}{We extend our framework to include the next to threshold contributions for the diagonal partonic channels.}
\end{abstract}
\section{Introduction}
\label{sec:intro}


The upcoming era of high energy physics will confront a huge boom of data driven by the upgraded run of the Large Hadron Collider (LHC). The High-Luminosity LHC will also come into effect in a few years of time. In order to fully exploit the increased quantity of data which will not only help to detect a rare phenomena but also improve the precision, an enhanced amount of effort by the theoretical physicists will be called for. In improving theoretical precision, the higher order quantum chromodynamics (QCD) and electroweak (EW) corrections play an important role. Among different observables, since the differential cross-section allows a wider range of comparisons with the experimental data, over the past few decades several attempts have been made to incorporate the higher order QCD and EW radiative corrections to this observable. The topic of this article is concerning the differential cross-section with respect to rapidity, in particular, we address the question of computing the higher order QCD corrections to this observable for any generic process at a hadron collider with all the final state particles as colorless.


Despite its high importance, unlike the inclusive cross-section, the differential rapidity distribution and its radiative corrections are computed only for a limited number of scattering processes. The rapidity distributions in Drell-Yan and of the scalar Higgs boson were computed to next-to-next-to-leading order (NNLO) QCD in refs.~\cite{Anastasiou:2003yy,Anastasiou:2003ds} and~\cite{Anastasiou:2004xq}, respectively. In case of the scalar Higgs boson produced through gluon fusion, the next-to-NNLO (N$^3$LO) QCD correction was incorporated in ref.~\cite{Dulat:2018bfe}. Shortly before, it was approximated in ref.~\cite{Cieri:2018oms} in the formalism of $q_T$-subtraction. For the Higgs boson production through bottom quark annihilation, it was computed to NNLO in ref.~\cite{Buehler:2012cu}.

Needless to say, achieving a full QCD correction to any order is not easy and with increasing perturbative order, the complexity level increase substantially which often prevents us from achieving it. In absence of the full QCD correction, it is often desirable to find an alternative method averting the full complexity to capture the dominant contribution. In this article, we discuss such a method, called soft-gluon or soft-virtual (SV) approximation. In the soft limit, the momenta of all the real emission diagrams are assumed to be infinitesimally small which leads to an all order exponentiation of this contribution. In ref.~\cite{Ravindran:2006bu,Ravindran:2007sv}, a formalism to incorporate the soft-gluon contribution to the rapidity distribution for the production of a colorless final state in hadron collider is presented which, in this article, is extended to the case of any number of final state colorless particles. The formalism is based on QCD factorization, which dictates that the soft part of the real emission diagrams factorizes from the hard 
contribution, and renormalization group (RG) invariance. The factorized soft part is conjectured to fulfil a Sudakov type differential equation with respect to the final state invariant mass square and as a consequence, it is found to get exponentiated which not only provides us with the fixed order result under soft limit but also enables us to perform a resummation in the soft limit. For the production of arbitrary number of colorless particles in hadronic collision, the soft part essentially remains identical to the case of Sudakov type process since the real emission can only takes place from the initial state partons. Using this idea, in this article, we extend the formalism~\cite{Ravindran:2006bu} to the case of $2\rightarrow n$ scattering, where $n$ denotes the number of final state colorless particles. We show that combining the virtual matrix element that captures the process dependence, the universal soft part and mass-factorization kernel in an elegant way, we can calculate the SV differential rapidity distribution for any generic $2 \rightarrow n$ process. In addition, we also show how naturally it leads to the threshold resummation for the same observable. Needless to say, in some kinematic regions such as when partonic center-of-mass energy becomes very close to threshold, the logarithms often become so large that it endangers the validity of the perturbation theory and we are left with no other choice but to perform an all order resummation in order to get a reliable prediction.

In the literature, several results for the rapidity resummation employing different methods are available. In ref.~\cite{Mukherjee:2006uu}, following the conjecture given in \cite{Laenen:1992ey}, the resummation of rapidity of $W^\pm$ gauge boson and in ref.~\cite{Bolzoni:2006ky} of Drell-Yan are computed in Mellin-Fourier (M-F) space. A detailed theoretical underpinnings and phenomenological implications of threshold resummation of rapidity are examined in ref.~\cite{Bonvini:2010tp} emphasizing the role of
prescriptions that take care of diverging series at a given
logarithmic accuracy. Our method belongs to a category, so called direct QCD approach~\cite{Catani:1989ne}, which is based on ref.~\cite{Ravindran:2006bu,Ravindran:2007sv,Banerjee:2017cfc} resums the soft gluons in two dimensional Mellin space (M-M). In \cite{Lustermans:2019cau}, the merits of different approaches are discussed in details.

Over the past decade, the formalism~\cite{Ravindran:2006bu} for the $2 \rightarrow 1$ scattering has been tested extensively through the computations of the rapidity distribution of several Sudakov type process. In ref.~\cite{Ravindran:2006bu}, the observable was computed partially at N$^3$LO under the SV approximation for the Higgs boson production through gluon fusion and for di-lepton in Drell-Yan. Later, some of us completed the full SV correction at N$^3$LO in ref.~\cite{Ahmed:2014uya} by incorporating the missing part from real emission diagrams. One of the salient features of our formalism is that the soft part that enters into the rapidity distribution is shown to be connected to the respective part of inclusive cross-section through a very simple relation involving gamma function of the dimensional regulator. This relation was used to extract the required soft part from the respective quantity of the SV cross-section~\cite{Ahmed:2014cla,Anastasiou:2014vaa,Catani:2014uta}. The SV rapidity distribution at N$^3$LO is also computed for the Higgs boson production through bottom quark annihilation~\cite{Ahmed:2014era}. The threshold resummed results at next-to-next-to-leading log (NNLL) for the Higgs boson~\cite{Banerjee:2017cfc} and Drell-Yan~\cite{Banerjee:2018vvb} are also obtained. In this article, we extend the formalism for computing the SV differential rapidity distribution and its resummation to any $2 \rightarrow n$ scattering process at the hadron collider and we provide the results to next-to-N$^3$LO (N$^4$LO) and next-to-NNLL (N$^3$LL) in terms of generic quantities in Online Resource. We also provide the approximate results of the rapidity distributions at N$^4$LO and N$^3$LL explicitly for the Higgs boson production through gluon fusion and bottom quark annihilation, and for the Drell-Yan production. The approximation comes from the unavailability of the full four loop virtual matrix elements and the soft-collinear distributions. \textcolor{black}{In recent times,
there has been a surge of interest to understand the next-to SV (NSV) contributions to inclusive as well as differential distributions, see \cite{Laenen:2008ux,Laenen:2010uz,Bonocore:2014wua,Bonocore:2015esa,Bonocore:2016awd,DelDuca:2017twk,Bahjat-Abbas:2019fqa,Soar:2009yh,Moch:2009hr,deFlorian:2014vta,Beneke:2018gvs,Beneke:2019mua,Beneke:2019oqx} for more details.  In \cite{Ajjath:2020ulr,Ajjath:2020lwb,Ajjath:2020sjk}, some of us have demonstrated how these contributions can be systematically included for single colorless particle productions.  This was done for the diagonal partonic channels. 
In the context of rapidity distributions, it was shown that like soft virtual terms,
NSV terms can also be resummed to all orders both in z and double
Mellin $\{N_1,N_2\}$ spaces \cite{Ajjath:2020lwb}. The \textit{goal} of this article is to present the formal methodology of computing threshold rapidity corrections  including NSV terms for any generic process of $2 \rightarrow n$ kind at hadron collider.}   

The paper is organized as follows: in section~\ref{sec:SVrapidity}, we introduce the notion of soft-virtual correction in the context of differential rapidity distribution and then describe our formalism in details in section~\ref{sec:phi}. The universality of soft part leads us to define a quantity called the differential soft-collinear operator that essentially captures the process independent part is also introduced in section~\ref{sec:phi}. In the next section~\ref{sec:resum}, we extend our formalism to incorporate the threshold resummation of the rapidity distribution.\textcolor{black}{ In section ~\ref{sec:NSV}, we discuss how next-to soft virtual corrections can be included in the rapidity distribution in particular for diagonal channels.}
All the results to N$^4$LO and N$^3$LL in terms of generic quantities are presented in Online Resource in Mathematica format which can be used for any generic $2\rightarrow n$ process once the corresponding virtual matrix element becomes available provided the soft distribution function for Sudakov type process is known.

\section{Soft-virtual rapidity distribution}
\label{sec:SVrapidity}
We begin by introducing the regime of soft gluon contribution to differential rapidity distribution for the  production of $n$-number of 
colorless particles in hadron collisions within the framework of perturbative QCD. Our prescription that we will develop subsequently to capture this contribution is within the scope of QCD improved parton model where the collinear divergences factorize to all orders in strong coupling constant $\alpha_s=g_s^2/4 \pi$. We consider a generic hadronic collision between two hadrons $H_{1,(2)}$ having momentum $P_{1(2)}$ that produces a final state consisting of $n$-number of colorless particles, denoted as $F_i(q_i)$
\begin{align}
\label{eq:had-collision}
    H_1(P_1) + H_2(P_2) \rightarrow \sum_{i=1}^n F_i(q_i)+X\,.
\end{align}
Through the quantity $X$, we represent an inclusive hadronic state. $q_i$ stands for the momentum of corresponding colorless particle $F_i$. We denote the invariant mass square of the final state by $q^2$ which is related to the momenta $\{q_i\}$ through $q^2=\left(\sum_i q_i \right)^2$. Without loss of generality, the rapidity of the final state invariant mass system is defined as
\begin{align}
\label{eq:rapidity}
    y \equiv \frac{1}{2} \log \left( \frac{P_2 \cdot q}{P_1 \cdot q}\right)\,.
\end{align}
The differential rapidity distribution at the hadronic level can be written as
\begin{align}
\label{eq:rapidity-dist}
    &\frac{d^2}{dydq^2}\sigma\left(\tau,q^2,y\right) = \sigma_B(\tau,q^2) W(\tau,q^2,y) \quad \text{with} \nonumber\\
    &W = \sum_{a,b=q,{\bar q},g} \int_{0}^1 dx_1 \int_{0}^1 dx_2 ~\hat{f}_a\left(x_1\right) \hat{f}_b\left(x_2\right) 
    \int_0^1 dz \delta(\tau-z x_1 x_2) \nonumber\\
    &\qquad\qquad~~\times \int  \left[ dPS_{m+n}\right] \left| \overline{\rm M}_{ab} \right|^2 \delta\left( y-\frac{1}{2} \log \left( \frac{P_2 \cdot q}{P_1 \cdot q}\right)\right) 
    \nonumber\\  &\qquad\qquad~~
    \times  \delta\big(q^2-\big(\sum_i q_i \big)^2\big) \,,  
\end{align}
where $\sigma_B$ is the leading order inclusive cross section.
The dimensionless variables, $\tau$ and $z$, are respectively defined as the ratios of invariant mass square of the final state to the square of hadronic ($S$) and partonic ($\hat s$) center-of-mass energies i.e.
\begin{align}
\label{eq:tau-z}
    \tau \equiv \frac{q^2}{S} \quad \text{and} \quad z \equiv \frac{q^2}{\hat s}\,.
\end{align}
We denote the fraction of the initial state hadronic momentum carried by the partons ($a,b$) that take part in the scattering at the partonic level as $x_{1(2)}$, and these are constrained through the relation $\tau=zx_1x_2$ as reflected by the presence of the respective $\delta$-function in the definition of $W$. \textcolor{black}{
The remaining delta functions reflects the momentum conservation and also the rapidity of final state invariant mass system as defined by \eqref{eq:rapidity}.
$\hat f_{a(b)}$ are the bare parton distribution functions (PDF). 
The spin and color averaged square of the scattering matrix element is denoted by $|{\overline{\rm M}}_{ab}|^2$. The corresponding $m+n$-particle phase space is given by $[dPS_{m+n}]$ where the integer $n$ indicates the number of colorless particles at the final state. Note that the numerical value of the integer $m$ depends on the number of radiated partons which is solely controlled by the perturbative order we are interested in.}

The scattering matrix element or equivalently the partonic differential distribution can be perturbatively expanded in powers of strong coupling constant $a_s(\mu_R^2) \equiv \alpha_s(\mu_R^2)/4\pi$ with $\mu_R$ being the renormalization scale. In this article, we confine ourselves to the regime where the leading order (LO) processes can only be initiated through color neutral quark or gluon channels,
\begin{align}
\label{eq:partonic-process}
    q(p_1)+{\bar q}(p_2) \rightarrow \sum_{i=1}^n F_i(q_i) ~~ \text{and} ~~ g(p_1)+g(p_2) \rightarrow \sum_{i=1}^n F_i(q_i)
\end{align}
with the corresponding momenta $p_{1(2)}$. Moreover, we are interested in computing the differential rapidity distribution only in the soft limit which constrained all the partonic radiation to be only soft i.e. those can only have nearly vanishing momenta. The \textit{goal} of this article is to present a prescription to compute the leading contribution of the differential rapidity distribution in the soft limit which is commonly referred as soft-virtual (SV) contribution. \textcolor{black}{In addition, we also discuss how the current formalism can be extended to study the next-to-soft virtual (NSV) contribution.} In order to define the soft limit for the rapidity distribution, we choose to work with a set of symmetric scaling variables $x_{1(2)}^0$ instead of $y$ and $\tau$ which are related through
\begin{align}
\label{eq:tau-y-x0}
    y \equiv \frac{1}{2} \log \left( \frac{x_1^0}{x_2^0}\right) \quad \text{and} \quad \tau \equiv x_1^0 x_2^0\,.
\end{align}
Note that unlike the inclusive cross-section, the choice of variables which one needs to take in order to define the soft limit is not unique and as it turns out, our choice of these new set of variables is crucial for our prescription. In terms of these variables, the partonic contributions arising from the subprocesses are found to depend on the ratios 
\begin{align}
\label{eq:scaling-x}
    z_i \equiv \frac{x_i^0}{x_i}\,,
\end{align}
which play the role of scaling variables at the partonic level. \textcolor{black}{After performing the mass-factorization and also evaluating the $\delta$-function integration over $z$, the $W(\tau,q^2,y)$ in \eqref{eq:rapidity-dist} can be rewritten as }
\begin{align}
\label{eq:W-intermsof-x0}
    W(x_1^0,x_2^0,q^2) =& \sum_{a,b} \int_{x_1^0}^1 \frac{dz_1}{z_1} \int_{x_2^0}^1 \frac{dz_2}{z_2} f_a\left(\frac{x_1^0}{z_1},\mu_F^2\right) f_b\left(\frac{x_2^0}{z_2},\mu_F^2\right) 
    \nonumber\\  
    &\times \int \left[ dPS_{n}\right] \Delta_{d,ab}\left( z_1,z_2,q^2,\{p_j\cdot q_k\},\mu_F^2\right)
    \nonumber\\
     &\times  \delta\left( y-\frac{1}{2} \log \left( \frac{P_2 \cdot q}{P_1 \cdot q}\right)\right)    \delta\big(q^2-\big(\sum_i q_i \big)^2\big)\,,
\end{align}
\textcolor{black}{where $\mu_F$ is the factorization scale. In \eqref{eq:W-intermsof-x0} $, \Delta_{d,ab}$ is the partonic coefficient function which can be expressed as }

\begin{align}
\label{eq:Deltadab}
\Delta_{d,ab} =& \left| \overline{\cal M}_{ab} \right|^2 (\Gamma^T)^{-1} \otimes \Bigg[
\frac{\int  \left[ dPS_{m+n}\right] \left| \overline{\rm M}_{ab} \right|^2 \delta_1 \delta_2}{\int  \left[ dPS_{n}\right] \left| \overline{\cal M}_{ab} \right|^2 \delta_1 \delta_2}  \Bigg] \otimes (\Gamma)^{-1}\,,
\end{align}
\textcolor{black}{where $\delta_1$ and $\delta_2$ are the shorthand notations for the two $\delta$-functions appearing in the last line of \eqref{eq:W-intermsof-x0}. In the above equation, ${\cal M}_{ab}$ is the UV renormalized pure virtual corrections for the $2\rightarrow n$ scattering while the quantity inside the square brackets encapsulates all the real emission contributions as it is already normalised by the pure virtual contributions after the phase space integration. The $\Gamma$'s are the mass-factorization kernels which are introduced to cancel the initial state collinear singularity present in both virtual and real emission contributions in \eqref{eq:Deltadab}. In the next section, we will study the IR factorisation property of ${\cal M}_{ab}$.}
%
%
Being a scattering process containing $n$-number of final state colorless particles, the partonic coefficient function does, in fact, depend on the Mandelstam variables constructed out of all the independent external momenta which is concisely denoted through $\{p_j \cdot q_k\}$.  In order to find the definition of soft limit in terms of the new partonic scaling variables, we take the double Mellin moment of $W$ with respect to the variables $N_{1(2)}$ which turns out to be 
\begin{align}
\label{eq:W-Mellin}
    W(N_1,N_2) &\equiv \int dx_1^0 (x_1^0)^{N_1-1} \int dx_2^0 (x_2^0)^{N_2-1} W(x_1^0,x_2^0)
    \nonumber\\
    &=\sum_{ab} f_a(N_1)f_b(N_2) \Delta_{d,ab}(N_1,N_2)\,.
\end{align}
All the quantities with functional dependence of $N_{1(2)}$ are in Mellin space where the soft limit is defined by the simultaneous limit of $N_{1(2)} \to \infty$. In terms of partonic scaling variables this condition gets translated to $z_{1(2)} \to 1$. Note that we normalize the coefficient function $\Delta_{d,ab}$ in such a way that at the leading order $W$ satisfies
\begin{align}
\label{eq:norm}
    W(x_1^0,x_2^0,q^2,\mu_R^2) = W_B \equiv \delta(1-x_1^0) \delta(1-x_2^0)\,.
\end{align}
In the following section, we will present the prescription to calculate the infrared safe SV differential rapidity distribution to N$^4$LO in QCD for any $2\rightarrow n$ scattering process which can be computed order by order in perturbation theory:
%
\begin{align}
\label{eq:Deltad-expand}
    \Delta_{d,ab}^{\rm sv} = a_s^{\lambda}(\mu_R^2)\sum_{k=0}^{\infty} a_s^k(\mu_R^2) \Delta_{d,ab}^{(k),{\rm sv}}\,.
\end{align}
$\lambda$ is the order of strong coupling constant at the leading order partonic process. Here the arguments in $  \Delta_{d,ab}^{\rm sv}$ are suppressed.  

\subsection{Universal soft-collinear operator for SV rapidity distribution}
\label{sec:phi}

In this section, we setup a framework to compute the soft-virtual corrections to the rapidity distribution to all orders in strong coupling constant. The infrared safe SV rapidity distribution can be obtained by combining the ultraviolet (UV) renormalized virtual matrix elements with the soft gluon contribution and performing appropriate mass factorization to get rid of initial state collinear singularities.It is well-known that the combined soft and collinear divergences, conveniently denoted as infrared (IR), in virtual matrix elements factorize from the corresponding UV renormalized part to all orders in perturbation theory and thereby in dimensional regularization we can write
\begin{align}
\label{eq:M-Mfin}
        {\cal M}_{ab,{\rm fin}}\left(\{p_j\},\{ q_k\},\mu_R^2\right)  &= \lim_{\epsilon \rightarrow 0} Z_{ab,\rm IR}^{-1}(q^2,\mu_R^2,\epsilon) \nonumber\\& \times 
        {\cal M}_{ab}\left(\{p_j\}, \{q_k\},\epsilon\right) 
\end{align}
with the space-time dimensions $D=4+\epsilon$. Without loss of generality, we choose the renormalization scale to be equal to the scale of aforementioned factorization which, of course, in general can be different. Upon multiplying the  renormalization constant $Z_{ab,{\rm IR}}$, the IR divergent part of the UV renormalized matrix element ${\cal M}_{ab}$ gets compensated and we end up with the finite part of the matrix element ${\cal M}_{ab,{\rm fin}}$. The renormalization constant is a universal quantity as it is independent of the details of the process, it only depends on the nature of external color particles. It is fully independent of the number and nature of external colorless particles. {\color{black}The exponentiation of the Sudakov form factor in dimensional regularization was first demonstrated in refs.~\cite{Mueller:1979ih,Sen:1981sd,Magnea:1990zb}. For the multiparton amplitudes, t}he universal structure of the IR divergences in dimensional regularization, up to the single pole in $\epsilon$ at two-loop, was first depicted in ref.~\cite{Catani:1998bh} and subsequently {\color{black}analyzed} in ref.~\cite{Sterman:2002qn}.  For scattering involving only two external colored particles, the universality of the single pole at two-loop was established in terms of soft and collinear anomalous dimensions in ref.~\cite{Ravindran:2004mb} which was verified at three-loop in ref.~\cite{Moch:2005tm}. An all order conjecture on the form of IR divergences in QCD for any generic scattering process is depicted in terms of soft anomalous dimension matrix in ref.~\cite{Becher:2009cu,Gardi:2009qi}. {\color{black}An elaborated discussion on the universality of subleading infrared poles in form factors is given in ref.~\cite{Dixon:2008gr}}. We provide the results of the renormalization constant to four loops in Online Resource.

Before moving further, we wish to iterate a well-known fact. For processes involving only conserved operator, such as Drell-Yan, the coupling constant renormalization is sufficient to get rid of all the UV divergences. However, for other processes, such as the Higgs boson production in heavy quark effective theory, an additional renormalization which often called the operator renormalization is needed. This is a property inherent to the operator itself. 

In order to get the infrared safe and finite differential rapidity distribution, we need to combine the UV renormalized virtual matrix element to the real emission contributions in the soft limit and perform mass factorization which ensures the removal of collinear singularities arising from the initial state colored particles. Therefore, the universal nature of IR divergences in virtual matrix element implies that the combined contribution from the real emission diagrams and mass-factorization kernels must exhibit the same universality. By employing the criteria of universal IR structure and imposing the finiteness property of the rapidity distribution, we develop the prescription to compute the rapidity distribution under SV approximation for any generic $2 \rightarrow n$ scattering process and present the result in terms of universal quantities to N$^4$LO QCD. Once the pure virtual matrix element for any process of the kind under consideration becomes available, our expression can immediately be employed to calculate the SV rapidity distribution at that order in QCD.

In this article, we extend the prescription, which was introduced for the Sudakov type process in ref.~\cite{Ravindran:2006bu}, to the case of $2 \rightarrow n$ scattering. We propose that the coefficient function for the rapidity distribution in \eqref{eq:Deltad-expand} can be written as a Mellin convolution of the pure virtual contribution ${\cal F}$, soft-collinear distribution $\Phi_{d}$ and mass-factorization kernel $\Gamma$, which read as
%

\begin{align}
\label{eq:SV-master-formula}
    \Delta_{d,a\bar a}^{\rm sv} &=  | {\cal M}^{(0)}_{a\overline a} |^2   |{\cal F}_{a\overline a}(\{p_j\cdot q_k\},q^2,\epsilon)|^2 \delta(\zo) \nonumber\\& \times \delta(\zt) 
    \otimes {\cal C} \exp\bigg(2\Phi_{d,a\overline a} \left(z_1,z_2,q^2,\epsilon\right) \nonumber\\& 
    - {\cal C} \log \Gamma_{a\overline a}(z_1,\mu_F^2,\epsilon) \delta(\zt)
    \nonumber\\&
    - {\cal C} \log \Gamma_{a\overline a}(z_2,\mu_F^2,\epsilon)\delta(\zo)\bigg)\,,
\end{align}
where $\delta(\overline z_l) = \delta(1-z_l)$ for $l=1,2.$

The pure virtual contribution is captured through the form factor ${\cal F}_{a\bar a}$ that is defined as
\begin{align}
\label{eq:FF-defn}
    {\cal F}_{a\overline a} = 1+\sum_{k=1}^{\infty} a_s^k {\cal F}_{a\overline a}^{(k)} \equiv 1+\sum_{k=1}^{\infty} a_s^k \frac{\langle {\cal M}_{a\overline a}^{(0)}|{\cal M}_{a\overline a}^{(k)}\rangle}{\langle {\cal M}_{a\overline a}^{(0)}|{\cal M}_{a\overline a}^{(0)} \rangle}\,,
\end{align}
where ${\cal M}_{a\overline a}^{(k)}$ represents the $k$-th order UV renormalized matrix element of the underlying partonic level process $a(p_1)+\overline a(p_2) \rightarrow \sum_{i=1}^n F_i(q_i)$. The symbol ``${\cal C}$'' stands for the convolution whose actions on a distribution $g(z_1,z_2)$ is defined as
\begin{align}
\label{eq:Cordered-expoen}
    {\cal C}e^{g(z_1,z_2)} &= \delta(\zo) \delta(\zt)+ \frac{1}{1!} g(z_1,z_2)+ \frac{1}{2!} \left(g \otimes g\right)(z_1,z_2) \nonumber\\& +\cdots\,,
\end{align}
where $\otimes$ denotes Mellin convolution. In this article in the context of SV corrections, we encounter only $\delta(1-z_i)$ and ${\cal D}_j(z_i)$, where
\begin{align}
\label{eq:plus-distr}
    {\cal D}_j(z_i) \equiv \left[ \frac{\log^j(1-z_i)}{(1-z_i)} \right]_+\,.
\end{align}
The contribution from the real emission diagrams is contained in $\Phi_{d,a\overline a}$ which we call as soft-collinear distribution. The soft divergences arising from the real emission and virtual diagrams, which are respectively encapsulated in $\Phi_d$ and ${\cal F}$, get cancelled.  The final state collinear singularity is guaranteed to go away, as dictated by Kinoshita-Lee-Nauenberg (KLN) theorem, once the sum over all the final states is performed. The mass factorization kernel takes care of the initial state collinear singularities. As a result, the coefficient function $\Delta^{\rm sv}_{d,a\bar a}$ in \eqref{eq:SV-master-formula} becomes finite. By demanding the finiteness of this quantity we can put a constraint on the soft-collinear distribution which turns out to be a Sudakov type renormalization group (RG) equation. This has profound implications which not only reveals a significant amount of insights about the IR world but also it enables us to perform threshold resummation as we will see in the next section. To be more precise, the solution of the RG equation results an all order exponentiation of the soft-collinear distribution. So, the whole job of computing the SV correction depends on our ability to determine and explore the unknown distribution $\Phi_{d,a\overline a}$ to which we now turn to\footnote{It is interesting to compare the equation \eqref{eq:SV-master-formula} to the corresponding one for computing the SV cross-section which is proposed in another publication~\cite{Ahmed:2020nci}.}.

As we have discussed, the soft-collinear distribution essentially captures the contribution arising from real emission diagrams which only can occur from colored partons. Naturally, $\Phi_{d,a\overline{a}}$ for Sudakov form factor i.e. $2\rightarrow 1$ and $2 \rightarrow n$ scattering should essentially be identical. The presence of more Mandelstam variables in the latter process just makes it more involved in its kinematic dependence when it is expressed in terms of $\{p_j\cdot q_k\}$. However, in terms of the total invariant mass square of the final state colorless particles i.e. $q^2$, it has to be exactly same as that of Sudakov process. In ref.~\cite{Ravindran:2006bu}, it was conjectured to satisfy a integro-differential RG equation to all orders in QCD coupling constant.
The underlying reason behind this all order conjecture is inspired by the akin integro-differential Sudakov equation~\cite{Mueller:1979ih,Collins:1980ih,Sen:1981sd} fulfilled by the form factor whose solution is present explicitly to five loops order in massless QCD in refs.~\cite{Moch:2005tm,Ravindran:2005vv,Ravindran:2006cg,Ahmed:2017gyt}. 
By integrating the differential rapidity distribution, we get the inclusive cross-section. Upon taking the Mellin moment with respect to the same Mellin variable $N$ of this relation we get 
\begin{align}
\label{eq:mellin-rapidity-Xsec}
    \int_0^1 dx_1^0 \int_0^1 dx_2^0 (x_1^0x_2^0)^{N-1} \frac{d\sigma}{dy} = \int_0^1 d\tau \tau^{N-1} \sigma\,.
\end{align}
By taking the limit $N\rightarrow \infty$ on both sides of this relation, we can relate the soft-collinear distributions in rapidity and that of inclusive cross-section. This is remarkable in a sense that given the soft-collinear distribution for inclusive cross-section, we can automatically calculate it for the rapidity. Since this is the only quantity that is unknown in comparison to the ingredients for the computation of SV cross-section, we can immediately calculate the SV rapidity distribution. The $\Phi_{d,a\overline a}$ for the Sudakov form factor is determined to NNLO in refs.~\cite{Ravindran:2006bu} and in~\cite{Ahmed:2014uya} at N$^3$LO in QCD.
In the current article, for the first time we present the general analytical form of $\Phi_{d,a\overline a}$ in terms of universal quantities at N$^4$LO for any generic $2\rightarrow n$ scattering. One of the most notable features of this quantity is it satisfies the maximally non-Abelian property i.e.
\begin{align}
\label{eq:Max-Non-Abelian}
    \Phi_{d,gg} = \frac{C_A}{C_F} \Phi_{d,q\bar q}\,,
\end{align}
where the quadratic Casimirs in Adjoint and fundamental representations of SU($n_c$) are denoted by $C_A=n_c$ and $C_f=(n_c^2-1)/2n_c$, respectively. The property says that the soft-collinear distribution for quark and gluon initiated processes which are respectively denoted by $q\bar q$ and $gg$ are related by simple scaling of quadratic Casimirs. This essentially signifies the universality of the real emission in the soft limit i.e. it is independent of the details of the process, it solely depends on the nature of the external partons. Needless to say, it is also quark flavour blind. The relation in \eqref{eq:Max-Non-Abelian} was  explicitly verified to NNLO in refs.~\cite{Ravindran:2006bu} and at N$^3$LO in~\cite{Ahmed:2014uya}. The flavour dependence of the $\Phi_{d,a\bar a}$ was exploited in ref.~\cite{Ahmed:2014era} in order to calculate the SV  rapidity distribution at N$^3$LO for the Higgs boson production in bottom quark annihilation.  We expect the Casimir scaling to hold true to all orders in perturbation theory since it originates entirely from the soft-collinear part of the differential cross-section, and therefore it would indeed be interesting to see whether truly it holds beyond N$^3$LO with generalised Casimir scaling~\cite{Moch:2018wjh}.

We decompose all the quantities into its singular (sing) and finite (fin) parts as
\begin{align}
\label{eq:phid-sing-fin}
    &\Phi_{d,a\overline a}=\Phi_{d,a\overline a,{\rm sing}}+\Phi_{d,a\overline a,{\rm fin}}\,,\nonumber\\ 
    &\log \Gamma_{a\overline a}=\log \Gamma_{a\overline a,{\rm sing}}+\log \Gamma_{a\overline a,{\rm fin}}\,,
\end{align}
\eqref{eq:SV-master-formula} can be recast into
\begin{align}
\label{eq:SV-master-formula-recast}
    \Delta_{d,a\bar a}^{\rm sv} &=  | {\cal M}^{(0)}_{a\overline a} |^2 |{\cal F}_{a{\overline a},{\rm fin}}(\{p_j\cdot q_k\},q^2,\mu_R^2)|^2 \delta(\zo) \delta(\zt) 
 \nonumber\\  & 
    \otimes {\cal C} \exp\bigg(2\Phi_{d,{a\overline a},{\rm fin}} \left(z_1,z_2,q^2,\mu_R^2\right)
   \nonumber\\&  - {\cal C} \log \Gamma_{a{\overline a},{\rm fin}}(z_1,\mu_F^2,\mu_R^2) \delta(\zt)
    \nonumber\\ 
    & - {\cal C} \log \Gamma_{a{\overline a},{\rm fin}}(z_2,\mu_F^2,\mu_R^2) \delta(\zo)\bigg)
    \otimes\textbf{I}_{d,a{\overline{a}}}\,,
\end{align}
where
\begin{align}
\label{eq:Isv}
    \textbf{I}_{d,a\overline a} &= |Z_{a{\bar a},{\rm IR}}(q^2,\mu_R^2, \epsilon)|^2 \delta(\zo) \delta(\zt)  \nonumber\\&
    \otimes \; {\cal C} \exp\bigg(2\Phi_{d,{a\overline a},{\rm sing}} \left(z_1,z_2,q^2,\mu_R^2,\epsilon\right)
    \nonumber\\
    &- {\cal C} \log \Gamma_{d,a\overline a,{\rm sing}}(z_1,\mu_R^2,\epsilon) \delta(\zt) \nonumber\\&
    - {\cal C} \log \Gamma_{d,a\overline a,{\rm sing}}(z_2,\mu_R^2,\epsilon) \delta(\zo)
    \bigg)\,.
\end{align}
Through the decomposition of the quantities into singular and finite parts in \eqref{eq:phid-sing-fin}, we put together all the singular components of the rapidity distribution  into $\textbf{I}_{d,a\bar a}$ which must be unit distribution $ \delta(\zo) \delta(\zt)$ in order to get a finite $\Delta^{\rm sv}_{d,a\bar a}$. In \eqref{eq:SV-master-formula-recast}, the form factor and the leading order matrix element are the only process dependent quantity. The remaining part which comprises of the finite segments of the soft-collinear distribution and mass factorization kernel is a process independent universal quantity which we call as \textit{differential soft-collinear operator}
\begin{align}
\label{eq:soft-collinear-operator}
{\textbf{S}}_{d,a\overline a} (z_1,z_2,q^2,\mu_R^2,\mu_F^2) \equiv &{\cal C} \exp\bigg(2\Phi_{d,{a\overline a},{\rm fin}} \left(z_1,z_2,q^2,\mu_R^2\right) \nonumber\\&
    - {\cal C} \log \Gamma_{a{\overline a},{\rm fin}}(z_1,\mu_R^2,\mu_F^2) \delta(\zt)
    \nonumber\\ 
    & - {\cal C} \log \Gamma_{a{\overline a},{\rm fin}}(z_2,\mu_R^2,\mu_F^2) \delta(\zo)\bigg)\,.
\end{align}
\textcolor{black}{The expression of $\textbf{S}_{d,a\overline{a}}$ being process independent can be used for any generic $2\rightarrow n$ scattering process.} Since we are confining our discussion to only those scattering processes with initial state quark-antiquark pair of same flavours or a pair of gluon, we conveniently rewrite the \eqref{eq:SV-master-formula-recast} as
\begin{align}
\label{eq:SV-master-formula-recast-I}
    \Delta^{\rm sv}_{d,I}=|{\cal M}^{(0)}_I|^2 |{\cal F}_{I,{\rm fin}}|^2 \delta(\zo) \delta(\zt) \otimes \textbf{S}_{d,I}
\end{align}
where $I=q{\bar q},gg$. We can calculate the SV coefficient function for the rapidity distribution order by order in perturbation theory by expanding it in powers of $a_s$ according to \eqref{eq:Deltad-expand}.
We provide the results of $\textbf{S}_{d,I}$ and $\Delta_{d,I}^{\rm sv}$ for any generic process, which can be expressed in terms of universal light-like cusp ($A^I$), eikonal or soft ($f^I$) and virtual or collinear ($B^I$) anomalous dimensions along with the process dependent form factors, in \ref{secA:SCdist} and \ref{secA:SVDelta} up to N$^2$LO QCD, respectively. The results beyond  N$^2$LO can be found in the Online Resource provided with this article. 
Though the scale dependence can be restored employing the renormalization group evolution, nonetheless for users' convenience we provide the results of $\Delta_{d,I}^{\rm sv}$ by keeping the explicit scale dependence in Online Resource. Once the virtual matrix element becomes available, one can directly use our generic results to calculate the SV rapidity distribution for any generic process of the kind under consideration. The anomalous dimensions are expanded in powers of $a_s(\mu_R^2)$ as
\begin{align}
\label{eq:AD-expand}
    X^I(\mu_R^2) = \sum_{j=1}^{\infty} a_s^j(\mu_R^2) X^I_j\,, 
\end{align}
where $X=A,B,f$. Thanks to recent calculations, the light-like  cusp anomalous dimensions are available to four loops~\cite{Korchemsky:1987wg,Moch:2004pa,Vogt:2004mw,Henn:2019swt,vonManteuffel:2020vjv} in QCD. The soft and collinear anomalous dimensions can be extracted~\cite{Ravindran:2004mb,Moch:2005tm} from the quark and gluon collinear anomalous dimensions~\cite{Becher:2006mr,Becher:2009qa} through the conjecture~\cite{Ravindran:2004mb}
\begin{align}
\label{eq:conjecture}
    \gamma^I=2 B^I + f^I
\end{align}
to three loops. At four loop, only partial results are available in refs.~\cite{Davies:2016jie,vonManteuffel:2020vjv,Das:2019btv,Das:2020adl}. 

We present the new results of $\Delta_{d,I}^{\rm sv}$ for $2\rightarrow 1$ processes such as the Drell-Yan and the Higgs boson productions through gluon fusion as well as bottom quark annihilation at N$^4$LO in \ref{secD:N4LO}. This has been achieved by using the explicit results of the recently computed four loop cusp anomalous dimension~\cite{Korchemsky:1987wg,Moch:2004pa,Vogt:2004mw,Henn:2019swt,vonManteuffel:2020vjv} along with the form factors for Drell-Yan and the Higgs boson productions which are approximately available to fourth order in QCD~\cite{Henn:2016men,Lee:2016ixa,vonManteuffel:2016xki,Lee:2017mip,vonManteuffel:2019wbj,vonManteuffel:2020vjv}. The previously missing coefficients of $ \delta(\zo) \delta(\zt)$ in $\Delta_{d,I}^{\rm sv}$ at N$^4$LO were due to the missing  ${\cal O}(\epsilon^0)$ results of form factors and the soft gluon contributions at four loop. The full explicit results of the quark and gluon form factors corresponding to Drell-Yan and the Higgs boson production in gluon fusion are available at three loop to ${\cal O}(\epsilon^2)$ in ref.~\cite{Gehrmann:2010ue,Gehrmann:2010tu} which are also required. The corresponding partial four loop form form factors are computed in several articles over the past few years~\cite{Henn:2016men,Lee:2016ixa,vonManteuffel:2016xki,Lee:2017mip,vonManteuffel:2019wbj,vonManteuffel:2020vjv}. For the case of Higgs production through bottom quark annihilation, the three loop~\cite{Gehrmann:2014vha,Lee:2017mip} and partial four loop results are available in ref.~\cite{Lee:2017mip}. Moreover, the one- and two-loop results are needed to expand to ${\cal O}(\epsilon^5)$  and ${\cal O}(\epsilon^3)$, respectively. Similarly for the coefficient which contributes to the ${\cal O}(\epsilon^0)$ part of four loop form factor, the one-, two-, and three-loop results are needed to order ${\cal O}(\epsilon^6)$, ${\cal O}(\epsilon^4)$ and ${\cal O}(\epsilon^2)$, respectively. The four loop explicit results which we present in this article are still incomplete due to the unavailability of the full explicit results for form factors as well as soft contributions resulting from the real emission processes at four loop, so far these are the state-of-the-art available results in the literature.

In Online Resource, the explicit expressions of all the anomalous dimensions including the QCD $\beta$-functions to three loops can be found. In the following section, we describe how our formulation of SV rapidity distribution naturally leads us to the soft gluon resummation for any generic $2 \rightarrow n$ scattering process.

\section{Threshold resummation and its universal soft-collinear operator}
\label{sec:resum}
Here in this section, we develop the  resummation formalism for the differential distribution with respect to the rapidity variable $y$, for the production of $n$-colorless particles  which also paves the way for a wider range of comparisons with the experiments. Earlier we have seen that there exists an operator, $\textbf{S}_{d,I}$, referred as the   \textit{differential soft-collinear operator}, which embeds the universality of all the  soft enhancements associated with the soft gluon emissions in the production of $n$-colorless particles in the hadronic collision. The universality lies in the fact that the operator, $\textbf{S}_{d,I}$ in \eqref{eq:soft-collinear-operator}, for the SV differential cross-section depends only on the initial state partons and is completely independent of the hard process under study. Besides being the process independent operator, interestingly it also exhibits an exponential behaviour. Recall that the threshold resummation~\cite{Ahmed:2020nci} relies on the fact that the soft contribution exponentiates to all orders in perturbation theory, owing to the Sudakov differential equation and the renormalization group invariance. Following the same argument we proceed towards the resummation formalism for differential cross-section as well. 

The relevance of resummation of differential cross-section arises from the fact that, in the limit $z_{1(2)}\rightarrow 1$, the logarithms of type 
$\Big(a_s^n \log^{m_1} (1-z_1)\log^{m_2} (1-z_2)\Big)/\big((1-z_1)(1-z_2)\big)$ for $m_1+m_2 \leq 2(n-1)$, give rise to large contributions which could potentially spoil the reliability of the perturbative series. Hence a systematic way of exponentiating these large logarithms and resumming them to all orders in perturbation theory becomes indispensable. In ref.~\cite{Catani:1989ne} it was shown, in the context of differential distribution with respect to the Feynman variable $x_F$, that  the potential logarithms which give dominant contributions in certain kinematic regions  can be resummed to all orders in perturbation theory in Mellin-Mellin (M-M) space approach. This approach was also extended to rapidity distributions in the earlier works by one of the authors of this paper (See~\cite{Banerjee:2018vvb,Banerjee:2017cfc} for details).
Note that this 
approach is different from the Mellin-Fourier (M-F) approach \cite{Laenen:1992ey} proposed by Laenen \& Sterman which was earlier discussed in the introduction. In M-F formalism partonic cross-section is expressed in terms of scaling variable $z$ and rapidity variable $y$ and then the threshold limit is taken only for $z\rightarrow 1$ which resums delta ($\delta(1-z)$) and distributions ($[\frac{\log(1-z)}{1-z}]_{+}$) in $z$, but for rapidity variable $y$ only delta ($\delta(y)$) piece is resummed.
In ref.~\cite{Banerjee:2018vvb}, one of the authors of this paper has made a detailed numerical
comparison of M-M approach against the M-F approach and found that both the approaches converges
to a few percent correction to the fixed order prediction at NNLL level.
In the following we further extend the M-M approach and derive the resummation formalism for the production of $n$-colorless particles in a partonic collision.

Within the framework of M-M approach, both the partonic scaling variables $z_{1(2)}$ are simultaneously taken to the threshold limit 1 and the corresponding delta, $\delta(1-z_i)$, and plus distributions, $\Big[\frac{\log^{m_i}(1-z_i)}{1-z_i}\Big]_+$, are resummed to all orders in perturbation theory. Due to the involvement of convolutions in the $z_{1(2)}$ space, the resummation is performed in two dimensional Mellin space where
the differential cross-section is expressed in terms of simple normal products along with the aforementioned exponential structure in form. In the following we derive the generic formalism in terms of the Mellin variables $N_1$ and $N_2$ corresponding to the $z_1$ and $z_2$ variables, respectively. Hence the threshold limit $z_{1(2)}\rightarrow 1$ translate to $N_{1(2)}\rightarrow \infty$ in Mellin space and the large logarithms proportional to $\log N_{1(2)}$ are resummed to all orders in perturbation theory.

To derive the all order behaviour of the SV differential cross-section, $\Delta^{\mathrm{sv}}_{d,I} (z_1,z_2)$, in the two dimensional Mellin space with $\Nb_i = N_i e^{\gamma_E}$, we begin with the Mellin moment of the same, which takes the following form:
\begin{align}
\label{SVmellin}
 &\tilde\Delta^{\mathrm{sv}}_{d,I}(\bar N_1,\bar N_2) &= \bigg[\prod_{i=1,2}    \int_0^1  dz_i   z_i^{N_i-1}   \bigg]\Delta^{\mathrm{sv}}_{d,I} (\{p_j\cdot q_k\},z_1,z_2) \,.
 \end{align}
$\gamma_E$ is the Euler-Mascheroni constant.
In the previous section in \eqref{eq:SV-master-formula}, $\Delta^{\mathrm{sv}}_{d,I}$ is decomposed into constituents corresponding to the virtual as well as the soft-collinear real emission contributions. Now in this section, we further decompose those contributions into a process dependent and a process independent quantities. We denote the process dependent coefficient $C_{d,0}^I$ in the context of $2\rightarrow n$ scattering process as,
\begin{align}
\label{eq:resc0}
C_{d,0}^I\left(\{p_j\cdot q_k\},q^2,\muf^2\right)  
 &= | {\cal M}^{(0)}_{I} |^2 |{\cal F}_{I,{\rm fin}}(\{p_j\cdot q_k\},q^2,\mu_R^2)|^2 \nonumber\\& \times S_{res,\delta}^{d,I}(q^2,\mu_R^2,\mu_F^2) .
\end{align}
Here $C_{d,0}^I$ accounts for all the finite contributions coming from the virtual corrections 
and the coefficients proportional to $\delta(\zo)\delta(\zt)$ of the real emission contributions. Besides, it also contains the finite part of the mass factorized kernel $\Gamma_{I,\rm fin}$ in terms of $\log(\mu_F^2/\mu_R^2)$ which results from the coupling constant renormalization. The quantity $S^{d,I}_{res,\delta}$ which we name as the \textit{differential soft-collinear operator for threshold resummation}, embeds the $\delta(1-z_i)$ contributions from the soft distribution function $\Phi_{d,I}$ and from $\Gamma_{I,\rm fin}$ in the following way: 
\begin{align}
\label{eq:sdelta}
S_{res,\delta}^{d,I}(q^2,\mu_R^2,\mu_F^2) &= \exp\left({2\Phi_{d,I,\delta}(q^2,\mu_R^2,\mu_F^2)   -2  \log \Gamma_{{I,\delta}}}(\mu_F^2)\right)\,.
\end{align}
The subscript $\delta$ indicates $\delta(\zo)\delta(\zt)$ coefficients of the aforementioned quantities. In a similar way, we denote the process independent contributions to $\Delta_{d,I}^{\mathrm{sv}}$ as $\Phi_{d,I}^{res}$ which comprises of the terms proportional to plus distributions
from $\Phi_{d,I}$ and $\Gamma_{I,\rm fin}$. Mathematically it can be written as,
\begin{align}
\label{eq:resphi}
\Phi^{res}_{d,I}(z_1,z_2,q^2,\mu_F^2)  =~& 2 \Phi_{d,I,{\cal D}} (z_1,z_2,q^2)  
\nonumber\\&
-  {\cal C} \log \Gamma_{I,{\cal D}}(z_1,\mu_F^2)\delta(\zt)
\nonumber\\&
- {\cal C} \log \Gamma_{ I,{\cal D}}(z_2,\mu_F^2)\delta(\zo)\,,
\end{align}
where the subscript  ${\cal D}$ indicates the terms proportional to plus distribution which includes, ${\cal D}_{i}(z_1) \delta(\zt)$ , ${\cal D}_{i}(z_2) \delta(\zo)$
and ${\cal D}_{i}(z_1) {\cal D}_{j}(z_2)$. Now following the approach given in ref.~\cite{Ravindran:2006bu}, the above equation can be written in an integral form which is given as,

\begin{align}
\label{phiint}
\Phi^{res}_{d,I}(z_1,z_2,q^2,\mu_F^2) =&
\Bigg[\delta(\ztb)~\Bigg({ 1 \over \zob} \Bigg\{ \int_{\mu_F^2}^{q^2 \zob}
{d \lambda^2 \over \lambda^2} A^I\left(a_s(\lambda^2)\right) 
\nonumber\\&
	+ \textbf{D}^I_d\left(a_s(q^2\zob)\right) \Bigg\} \Bigg)_+  
	+ {1 \over 2} \Bigg( {1 \over \zob \ztb } \nonumber \\& 
	\Bigg\{A^I(a_s( \zotb)) 
	+ {d \textbf{D}^{I}_d(a_s(\zotb))\over d\log \zotb} \Bigg\}\Bigg)_+ \nonumber \\& 
+ (z_1 \leftrightarrow z_2) \Bigg]\,,
\end{align}

here the subscript + indicates the standard plus distribution and the other constants are defined as $\bar z_i = (1-z_i)$ and $\zotb=q^2 \zob \ztb$. The finite functions, $ \textbf{D}_d^I = \sum_{i=1}^\infty a_s^i \textbf{D}_{d,i}^I$, are related to the threshold exponent $\textbf{D}^I_i$ of inclusive cross section owing to the relation given in \eqref{eq:mellin-rapidity-Xsec}
( See \cite{Ravindran:2006bu,Banerjee:2017cfc} for more details). For completeness, we provide the coefficients $A^I_i$ and $ \textbf{D}_{d,i}^I$ in the Online Resource.

Consequently the SV differential cross-section decomposes into a process dependent and a process independent way and can be re-written in the following form:
%
 \begin{align}
\label{eq:SVresum}
        \Delta^{\rm sv}_{d,I} \left(\{p_j\cdot q_k\},z_1,z_2,q^2,\muf^2\right) &=

         C_{d_,0}^I\left(\{p_j\cdot q_k\},q^2,\muf^2\right) \nonumber\\& \times \delta(\zo)\delta(\zt) 
         \nonumber\\&
         \otimes {\cal C} \exp( \Phi^{res}_{d,I} (z_1,z_2,q^2,\mu_F^2))  \nonumber\\&
    \otimes  \textbf{I}_{d,I}\,.
\end{align}
Substituting \eqref{eq:SVresum} in \eqref{SVmellin} and after doing the two dimensional Mellin transformation systematically, we obtain
\begin{align} \label{SVresumfin}
     \tilde \Delta_{d,I}^{\mathrm{sv}}(\bar N_1,\bar N_2) &=  
     C_{d,0}^I\left(\{p_j\cdot q_k\},q^2,\muf^2\right) \nonumber\\& \times
     \exp\Big(
      G_{d,0}^I(q^2,\mu_F^2) + 
 G_{d,\Nb}^I(q^2,\mu_F^2,\omega)\Big)\,,
\end{align}

with $\omega = \beta_0 a_s(\mu_R^2) \log(\Nb_1 \Nb_2) $. The first coefficient of QCD $\beta$-function is denoted by $\beta_0 \equiv (11 C_A-2 n_f)/3$, $n_f$ is the number of active light quark flavours. Here, the decomposition  in the exponent is done in such a way that the coefficient $ G_{d,\Nb}^I$ contains  $N_{1(2)}$ dependent terms, and the remaining ones are embedded in $ G_{d,0}^I$. Besides this, $ G_{d,\Nb}^I(q^2,\mu_F^2,\omega)$ also vanishes in the limit $\omega\rightarrow 1$. Needless to say that both of these coefficients has a universal structure in terms of the anomalous dimensions $A^I$ and process independent coefficients $ \textbf{D}_d^I$ and thus are dependent only on the incoming partons.  
Further we combine the $N_{1(2)}$ independent coefficients  $ G_{d,0}^I$ with $C_{d,0}^I$ from \eqref{SVresumfin} and define,
\begin{align}
\label{eq:g0bexpanded}
g_{d,0}^I (\{p_j\cdot q_k\},q^2,\muf^2) &= C_{d,0}^I\left(\{p_j\cdot q_k\},q^2,\muf^2\right) \nonumber\\& \times
                    \exp\left( {G}_{d,0}^I(q^2,\muf^2)\right)\,
                
\end{align}
which can be expanded in terms of $a_s(\mu_R^2)$ as,


\begin{eqnarray}
\label{eq:g0bexp}
~~~~  g_{d,0}^I = \sum_{i=0}^\infty a_s^i(\mu_R^2)  g_{d,0}^{I,i} \,.
\end{eqnarray}

From \eqref{eq:g0bexpanded} it can be seen, that the coefficient $g_{d,0}^I$ contains finite contribution from virtual corrections, \textit{differential soft-collinear operator for threshold resummation} and $N$ independent terms coming from Mellin transformation of plus distribution.
Consequently, \eqref{SVresumfin} gets modified as,
\begin{align}
\label{eq:res2}
\tilde\Delta^{\mathrm{sv}}_{d,I}(\Nb_1,\Nb_2)
&=  {g}_{d,0}^I\left(\{p_j\cdot q_k\},q^2,\muf^2\right) \nonumber\\& \times
\exp\Big(  G_{d,\Nb}^I\left( q^2, \muf^2,\omega \right)\Big)\,.
\end{align}
where the exponent ${ G}_{d,\Nb}^I$ can be organized as a resummed perturbation series in Mellin space as,
\begin{align}
\label{eq:GNbexp}
 G^I_{d,\Nb}(q^2,\mu_F^2,\omega)&={ g}^I_{d,1}(\omega) \log(\Nb_1 \Nb_2) 
\nonumber\\& + \sum_{i=0}^\infty a_s^i(\mu_R^2){ g}^I_{d,i+2}(\omega,q^2,\mu_F^2, \mu_R^2)\,.  
\end{align} 
The explicit form in \eqref{eq:res2} when expanded till $k$-th order in powers of $a_s(\mu_R^2)$, gives the logarithmically enhanced contributions to the fixed order results $\tilde\Delta^{\mathrm{sv}}_{d,I}(\Nb_1,\Nb_2)$ up to the same order. 
The successive terms in the above series given in \eqref{eq:GNbexp} along with the corresponding terms in \eqref{eq:g0bexp} define the resummed accuracy as LL, NLL, NNLL, N$^3$LL and so on. In general for N$^k$LL accuracy, terms up to $ g_{d,{k+1}}^I$ must be included  along with  $ g_{d,0}^I$  up to order $a_s^k(\mu_R^2)$. The general expression for the coefficients $ g_{d,0}^{I,i}$ and $ g_{d,i}^I$ up to $\mathrm{N^3LL}$ are provided in the Online Resource.

The coefficients $ G^I_{d,\Nb}$ remains unaltered even for $2\rightarrow n$ scattering process owing to its universality. However, the process dependent coefficient function $ g_{d,0}^I$ changes for the production of $n$-colorless particles due to the inclusion of process specific form factor via
\eqref{eq:resc0} and \eqref{eq:g0bexpanded}.
The results of these coefficients appear as a product of $N_1$ and $N_2$ in the Mellin space, and all those terms which are only function of  $N_1$ or $N_2$ cancel internally.

We have also observed that the coefficients $ g_{d,0}^I$ and $ G_{d,\Nb}^I$ coincides with their inclusive counterparts $ g_{0}^I$ and $ G_{\Nb}^I$ respectively in the limit $N_1 \rightarrow N_2 \rightarrow N$, provided the coefficients $\textbf{D}_{d}^I$ in \eqref{phiint} is expressed in terms of $\textbf{D}^I$ of inclusive soft distribution function using the relation \eqref{eq:mellin-rapidity-Xsec}. Hence we infer that all the above observations which hold true for $2\rightarrow 1$ scattering processes are further extended and verified for any generic system of $n$-colorless particles in the final state.

\section{Beyond soft virtual rapidity distribution}
\label{sec:NSV}
Having obtained the results for the SV part of coefficient functions, namely $\Delta^{\rm {sv}}_{d,I}$, both
in $z_i$ and $N_i$ spaces, we would like to extend our approach to include NSV terms.  
These are logarithms of the form $\log^k(1-z_i), k\ge 0 $  present in coefficient functions
that are often comparable or even larger than SV contributions. Recently,
some of us have developed a formalism to study the NSV structure in inclusive reactions, namely, the production of leptons pairs in Drell-Yan process, Higgs boson production in gluon fusion or in bottom quark annihilation
\cite{Ajjath:2020ulr} and
deep inelastic scattering and semi-inclusive $e^+e^-$ annihilation processes \cite{Ajjath:2020sjk}. 
Later the formalism has been extended to rapidity distributions in 
\cite{Ajjath:2020lwb}. 
We considered the observables such as invariant mass (total cross section) and/or rapidity distributions of a lepton pair (a Higgs boson) in Drell-Yan (gluon fusion or bottom quark annihilation). We restricted ourselves to only diagonal channels where either quark and anti quark
annihilate (light quarks in Drell-Yan, bottom quarks for Higgs boson) or gluon fusion in Higgs boson production.
We used the principles, namely mass factorization, RG invariance that we had used 
for studying SV logarithms. Remarkably, the IR structure of the NSV terms in the coefficient functions
can be understood in terms of the corresponding contributions in real emission processes and mass factorization
kernels which are the building blocks.
In the case of production of a single colorless state, we found that a part of  the NSV terms is controlled by certain universal anomalous dimensions similar to the case of SV terms and the remaining terms in NSV  depend on the process under 
study beyond  second order in $a_s$ \cite{Ajjath:2020ulr,Ajjath:2020lwb}. In particular, the \textit{ differential soft-collinear operator} for the NSV terms
turns out to be process dependent. To the end, we have shown that both SV and NSV logarithms can be systematically
summed up to all orders both in $z_i$ and $N_i$ space and this allowed us to predict
certain SV and NSV logarithms for all $i>n$ at every order in $a_s^i$, using results known up to order $a_s^n$. 

In the following, we apply the same formalism to study the NSV logarithms present in the rapidity of a state of $n$-colorless particles. We restrict ourselves to the diagonal coefficient functions (dCFs). For the diagonal channels, it is straightforward to show that the NSV terms in
dCFs arise only from diagonal partonic sub processes and diagonal part of mass-factorization kernels. The pure virtual
part of the partonic sub-processes does not contain any NSV terms and hence can  be factored out from them.  Hence, in the mass factorization formula, we need to keep both SV as well as NSV terms in 
$\Phi_{d,a \overline a}$ and in $\Gamma_{a \overline a}$. The latter is process independent and is known to third order \cite{Moch:2004pa,Vogt:2004mw}.
Recall that the challenging task was to determine $\Phi_{d,a \overline a}$. It contains singular and finite terms and the singularities are from IR region .The mass factorization demands that the soft part of singularities must cancel with those from pure virtual part of the partonic channels and the initial state collinear singularities should cancel against those from mass factorization kernels. The cancellation of singularities and the knowledge of dCFs allowed us to
parametrise the SV part of $\Phi_{d,a \overline a}$ in terms of $\delta(1-z_l)$, plus distributions containing logarithms
and we can use the same method to determine the NSV logarithms.  We find that the logarithmic structure of finite part in $\Phi_{d,a\overline a}$ does not depend on the process under consideration,
while their numerical coefficients will depend on the process.  As it was done for the SV case, these numerical coefficients can be determined using the inclusive results by working in the Mellin $N$ space. To end, we obtain
\begin{align}
\label{eq:SVNSV-master-formula-recast}
    \Delta_{d,a\bar a}^{\rm {sv+ nsv}} =&  | {\cal M}^{(0)}_{a\overline a} |^2 |{\cal F}_{a{\overline a},{\rm fin}}(\{p_j\cdot q_k\},q^2,\mu_R^2)|^2 \delta(\bar{z}_1) \delta(\bar{z}_2) 
\nonumber\\
&\otimes
\tilde{ \textbf{S}}_{d,a \overline a }
    \otimes\tilde{ \textbf{I}}_{d,a{\overline{a}}}\,,
\end{align}
with
\begin{align}
\label{eq:nsoft-collinear-operator}
{\tilde{\textbf{S}}}_{d,a\overline a}  \equiv & \tilde{{\cal C}} \exp\bigg(2 \tilde{\Phi}_{d,{a\overline a},{\rm fin}} \left(z_1,z_2,q^2,\mu_R^2\right)
  \nonumber\\ &
    -  \tilde{{\cal C}} \log \tilde{\Gamma}_{a{\overline a},{\rm fin}}(z_1,\mu_R^2,\mu_F^2) \delta(\bar{z}_2)
    \nonumber\\ &
    -  \tilde{{\cal C}} \log \tilde{\Gamma}_{a{\overline a},{\rm fin}}(z_2,\mu_R^2,\mu_F^2)\delta(\bar{z}_1)\bigg)\,,
\end{align}
and 
\begin{align}
\label{eq:nIsv}
    \tilde{\textbf{I}}_{d,a\overline a} =& |Z_{a{\bar a},{\rm IR}}(q^2,\mu_R^2, \epsilon)|^2 \delta(\bar{z}_1)  \delta(\bar{z}_2)
      \nonumber\\&
    \otimes \; \tilde{{\cal C}} \exp\bigg(2 \tilde{\Phi}_{d,{a\overline a},{\rm sing}} \left(z_1,z_2,q^2,\mu_R^2,\epsilon\right)
    \nonumber\\&
    -  \tilde{{\cal C}} \log \tilde{\Gamma}_{d,a\overline a,{\rm sing}}(z_1,\mu_R^2,\epsilon)\delta(\bar{z}_2)
    \nonumber\\&
    -  \tilde{{\cal C}} \log \tilde{\Gamma}_{d,a\overline a,{\rm sing}}(z_2,\mu_R^2,\epsilon)\delta(\bar{z}_1)
    \bigg)\,,
\end{align}
which again reduces to identity in $z_1,z_2$ space, namely $\delta(1-z_1) \delta(1-z_2)$.  
\textcolor{black}{We provide the results of NSV part of the dCFs in \eqref{eq:SVNSV-master-formula-recast} denoted by $\Delta_{d,I}^{\rm nsv}$ and $\tilde{\textbf{S}}_{d,I}$ for any generic process with $I=({gg, q\bar{q}})$ in \ref{secF:NSVDelta} and  \ref{secG:SCdistNSV} up to N$^2$LO QCD, respectively. The results beyond  N$^2$LO can be found in the Online Resource provided with this article.}
In summary,
the coefficient functions for the diagonal terms taking into account the NSV contributions reduce to \eqref{eq:SV-master-formula-recast-I}
with the replacement of $\textbf{S}_{d,a \overline a }$ by  $\tilde{\textbf{S}}_{d,a \overline a }$.  
Note that the modified \textit{differential soft-collinear operator} $\tilde{\textbf{S}}_{d,a \overline a }$  contain additional NSV terms in 
$\tilde{\Phi}_{d,a \overline a}$ and $\tilde{\Gamma}_{a \overline a}$ and we need to keep both SV and NSV terms when
$ \tilde{{\cal C}}$ operator acts on the functions like exponential and logarithms in \eqref{eq:Cordered-expoen}, \eqref{eq:nsoft-collinear-operator} and \eqref{eq:nIsv}. We find that the NSV terms being
proportional to ${\cal D}_k(z_i)\log^l(1-z_j), k,l\ge 0 ,i,j=1,2$ and $\log^k(1-z_i) \delta(1-z_j), k \ge 0,i,j=1,2 $, do not alter the $\delta(1-z_1) \delta(1-z_2)$, ${\cal D}_k(z_1) {\cal D}_l(z_2)$ 
and ${\cal D}_k(z_i)\delta(1-z_j),i,j=1,2,k\ge 0$ terms  in the coefficient functions. This results in
\begin{align}
\label{eq:nSVresum}
        \Delta^{\rm {sv+nsv}}_{d,I}
=&~
         C_{d_,0}^I\left(\{p_j\cdot q_k\},q^2,\muf^2\right) \delta(\bar{z}_1)\delta(\bar{z}_2) 
         \nonumber\\&
         \otimes \tilde{{\cal C}} \exp( \tilde{\Phi}^{res}_{d,I} (z_1,z_2,q^2,\mu_F^2))
    \otimes  \tilde{\textbf{I}}_{d,I}\,,
\end{align}
with
\begin{align}
\label{eq:npsiint}
\tilde{\Phi}^{res}_{d,I} =& {\delta(\overline z_1) \over 2} \Bigg(\!\!\displaystyle {\int_{\mu_F^2}^{q^2 \overline z_2}
{d \lambda^2 \over \lambda^2}}\! {\cal P}^I\left(a_s(\lambda^2),\zt\right) +
{\cal Q}^I_d\left(a_s(q_2^2),\zt\right)
\Bigg)_+ 
\nonumber\\&
+ {1 \over 4} \Bigg( {1 \over \overline z_1 }
\Bigg\{{\cal P}^I(a_s(q_{12}^2),\zt )
+ {\color{black} 2 }L^I(a_s(q_{12}^2) ,\zt)
\nonumber\\&
+ q^2{d \over dq^2} 
\left({\cal Q}^{I}_d(a_s(q_{12}^2 ),\zt) +  {\color{black} 2 }\varphi_{d,I}^f(a_s(q_{12}^2 ),\zt)\right)
\Bigg\}\Bigg)_+
\nonumber\\&
+ \overline z_1 \leftrightarrow \overline z_2,
\nonumber\\
\end{align}
where $q_l^2 = q^2~(1-z_l) $ and $q^2_{12}=q^2 \overline z_1 \overline z_2$. The function ${\cal P}^I$ in \eqref{eq:npsiint} is given as
\begin{align}
{\cal P}^I (a_s, \overline z_l)= 2 \bigg({A^I (a_s) \over (\overline z_l)_+} +  L^I(a_s, \overline z_l)\bigg) \,,
\end{align}
with $A^I$ is the cusp  anomalous dimension and  $L^I(a_s, \overline z_l) \equiv C^I(a_s) \log(\overline z_l)  + D^I(a_s)$.
The subscript $+$ denotes plus distribution.
The function ${\cal Q}^I_d$ in \eqref{eq:npsiint} is given as
\begin{eqnarray}
{\cal Q}^I_d(a_s,\overline z_l) = {2 \over \overline z_l}  \textbf{D}_d^I(a_s) + 2 \varphi^f_{d,I} (a_s,\overline z_l)\,.
\end{eqnarray}
The constants $C^I$ and $D^I$ can be obtained from the the splitting functions given in \cite{Vogt:2004mw,Moch:2004pa} and are  known to three loops in QCD.
The finite function $\varphi_{d,I}^{f}$ 
depends on $a_s$ and $z_l, l=1,2$ and we use explicit fixed order results to 
parametrize in the following way,
\begin{align}
\label{eq:Phidf}
\varphi_{d,I}^f(a_s(\lambda^2),\overline z_l) 
&= \sum_{i=1}^\infty \sum_{k=0}^i a_s^i(\lambda^2) \varphi^{I,(k)}_{d,i} \log^k \overline z_l\,.
\end{align} 
We determine the upper limit on the sum over $k$ by studying the  dimensionally regularised Feynman integrals that contribute
partonic cross sections in fixed order perturbation theory.   
We know that $z$ space result in the case of SV part \cite{Ravindran:2006bu,Ahmed:2014uya,Ahmed:2014era} can predict all the
distributions to dCFs to all orders starting from $a_s^n$ provided the lower order results up to $a_s^{n-1}$ are known. 
In the present case, the inclusion of NSV part can predict terms of the form $\delta(\overline z_l) \log^k\overline z_j, n+1 \le k \le 2n-1 $ and ${\cal D}_i(z_l) \log^k\overline z_j$ for $i,k=0,1,\cdots,n; i+k < 2n-1$ at every order $a_s^n$ provided the form factors, differential soft-collinear and mass factorization kernels are known to order $a_s^{n-1}$. Using the second order inclusive results, some of the authors obtained the third order NSV contributions to dCFs, for the first time in \cite{Ajjath:2020ulr}, that are contributing to Drell-Yan process and Higgs boson productions.
Further the rapidity NSV coefficients $\varphi_{d,i}^{I,(k)}$ and dCFs to third order are obtained in \cite{Ajjath:2020lwb} using third order inclusive results in \cite{Ajjath:2020ulr}
and also with the use of \eqref{eq:mellin-rapidity-Xsec}. The complete third order results for the Higgs production in gluon fusion are already known in \cite{Dulat:2018bfe,Lustermans:2019cau}.
For the DY, the third order prediction of \cite{Ajjath:2020lwb} using the same  formalism for $2 \rightarrow 1$,
 is in complete agreement with the \cite{Lustermans:2019cau} for terms of the type ${\cal D}_i(z_l) \log^j(\overline z_m)$, $i,j\ge 0 , l,m=1,2$. The remaining $\delta(\overline z_l) \log^j(\overline z_m)$ terms in DY and the complete NSV predictions for Higgs production in bottom quark  annihilation channel at third order are new and can be found in \cite{Ajjath:2020lwb}.  
\textcolor{black}{The generic expression for the NSV rapidity coefficients $\varphi^{I,(k)}_{d,i}$ in terms of NSV inclusive counterparts  $\varphi^{(k)}_{I,i}$ and anomalous dimensions like cusp $A^I$, soft $f^I$, SV coefficients $\tilde{\cal G}^{I,(k)}_{j}$ and $C^I$ and $D^I$ from NSV part of splitting function, is given as,}

\begin{align}
\label{eq:Phirel}
\begin{autobreak}
\varphi^{I,(0)}_{d,1} =  
             C^{I}_1 \zeta_2  + \varphi^{(0)}_{I,1} - \frac{1}{2} A^I_1 \,,
             \quad \varphi^{I,(1)}_{d,1} = \varphi^{(1)}_{I,1} \,,
\end{autobreak}
\nonumber\\
\begin{autobreak}
\varphi^{I,(0)}_{d,2} =
            \varphi^{(0)}_{I,2}
            - \frac{1}{2} A^I_2
     - \frac{1}{2} \beta_0 f^I_1
     + \zeta_2 C^{I}_2
     - \zeta_2 \beta_0 D^{I}_1 
     + 3 C^{I}_1 \beta_0 \zeta_3
     - \varphi^{(1)}_{I,1} \zeta_2 \beta_0\,,
\end{autobreak}
\nonumber\\
\begin{autobreak}
\varphi^{I,(1)}_{d,2} = 
    \varphi^{(1)}_{I,2}
     - C^{I}_1 \zeta_2 \beta_0 \,,
       \quad \varphi^{I,(2)}_{d,2} = \varphi^{(2)}_{I,2} \,,
\end{autobreak}
\nonumber\\
\begin{autobreak}
\varphi^{I,(0)}_{d,3} =
    \varphi^{(0)}_{I,3}
     - \frac{1}{2} A^I_3
     - \frac{1}{2} \beta_1 f^I_1
     - \beta_0 f^I_2
     + 6 \beta_0 C^{I}_2 \zeta_3
     - 2 \beta_0^2 \tilde{\cal G}^{I,(1)}_{1}
     - 4 \beta_0^2 D^{I}_1 \zeta_3
     + 4 \varphi^{(2)}_{I,2} \beta_0 \zeta_3
     + \zeta_2 C^{I}_3
     - \zeta_2 D^{I}_1 \beta_1
     - 2 \zeta_2 \beta_0 D^{I}_2 
     - 2 \zeta_2 \varphi^{(1)}_{I,2} \beta_0
     + 3 C^{I}_1 \zeta_3 \beta_1
     + \frac{38}{5} C^{I}_1 \zeta_2^2 \beta_0^2
     + 2 \varphi^{(0)}_{I,1} \zeta_2 \beta_0^2
     - A^I_1 \zeta_2 \beta_0^2
     - 6 \varphi^{(1)}_{I,1} \beta_0^2 \zeta_3
     - \varphi^{(1)}_{I,1} \zeta_2 \beta_1\,,
     \end{autobreak}
     \nonumber\\
\begin{autobreak}
\varphi^{I,(1)}_{d,3} =
        \varphi^{(1)}_{I,3}
     - 2 \zeta_2 \beta_0 C^{I}_2
     - 4 \zeta_2 \varphi^{(2)}_{I,2} \beta_0
     - 4 C^{I}_1 \beta_0^2 \zeta_3
     - C^{I}_1 \zeta_2 \beta_1
     + 2 \varphi^{(1)}_{I,1} \zeta_2 \beta_0^2\,,
\end{autobreak}
\nonumber\\
\begin{autobreak}     
\varphi^{I,(2)}_{d,3} = 
         \varphi^{(2)}_{I,3} \,,
         \quad \varphi^{I,(3)}_{d,3} = \varphi^{(3)}_{I,3} \,,
\end{autobreak}
\nonumber\\
\begin{autobreak}
\varphi^{I,(0)}_{d,4} =
        \varphi^{(0)}_{I,4}
     - \frac{1}{2} A^I_4
     - \frac{1}{2} \beta_2 f^I_1
     - \beta_1 f^I_2
     + 6 C^{I}_2 \zeta_3 \beta_1
     - \frac{3}{2} \beta_0 f^I_3
     - 5 \beta_0 \beta_1 \tilde{\cal G}^{I,(1)}_{1}
     + 9 \beta_0 \zeta_3 C^{I}_3
     + 6 \beta_0 \zeta_3 \varphi^{(2)}_{I,3}
     - 10 \beta_0 D^{I}_1 \zeta_3 \beta_1 
     - 3 \beta_0^2 \tilde{\cal G}^{I,(1)}_{2}
     - 12 \beta_0^2 \zeta_3 D^{I}_2
     - 6 \beta_0^3 \tilde{\cal G}^{I,(2)}_{1}
     - 18 \varphi^{(1)}_{I,2} \beta_0^2 \zeta_3
     + 4 \varphi^{(2)}_{I,2} \zeta_3 \beta_1
     + \zeta_2 C^{I}_4
     - 2 \zeta_2 \beta_1 D^{I}_2
     - \zeta_2 D^{I}_1 \beta_2
     - 3 \zeta_2 \beta_0 D^{I}_3 
     - 3 \zeta_2 \beta_0 \varphi^{(1)}_{I,3}
     + 6 \zeta_2 \beta_0^2 \varphi^{(0)}_{I,2}
     - 3 \zeta_2 \beta_0^2 A^I_2
     - 3 \zeta_2 \beta_0^3 f^I_1
     - 2 \zeta_2 \varphi^{(1)}_{I,2} \beta_1
     - \frac{36}{5} \zeta_2^2 \beta_0 \varphi^{(3)}_{I,3}
     + \frac{114}{5} \zeta_2^2 \beta_0^2 C^{I}_2
     - \frac{57}{5} \zeta_2^2 \beta_0^3 D^{I}_1 
     + \frac{138}{5} \zeta_2^2 \varphi^{(2)}_{I,2} \beta_0^2
     + 3 C^{I}_1 \zeta_3 \beta_2
     + 90 C^{I}_1 \beta_0^3 \zeta_5
     + 30 C^{I}_1 \zeta_2 \beta_0^3 \zeta_3
     + 19 C^{I}_1 \zeta_2^2 \beta_0 \beta_1
     + 12 \varphi^{(0)}_{I,1} \beta_0^3 \zeta_3
     + 5 \varphi^{(0)}_{I,1} \zeta_2 \beta_0 \beta_1
     - 6 A^I_1 \beta_0^3 \zeta_3 
     - \frac{5}{2} A^I_1 \zeta_2 \beta_0 \beta_1
     - 15 \varphi^{(1)}_{I,1} \beta_0 \zeta_3 \beta_1
     - \varphi^{(1)}_{I,1} \zeta_2 \beta_2
     - \frac{114}{5} \varphi^{(1)}_{I,1} \zeta_2^2 \beta_0^3\,,
\end{autobreak}
\nonumber\\
\begin{autobreak}
\varphi^{I,(1)}_{d,4} =
        \varphi^{(1)}_{I,4}
     + 18 \beta_0 \zeta_3 \varphi^{(3)}_{I,3}
     - 12 \beta_0^2 C^{I}_2 \zeta_3
     - 36 \varphi^{(2)}_{I,2} \beta_0^2 \zeta_3
     - 2 \zeta_2 C^{I}_2 \beta_1
     - 3 \zeta_2 \beta_0 C^{I}_3
     - 6 \zeta_2 \beta_0 \varphi^{(2)}_{I,3}
     + 6 \zeta_2 \varphi^{(1)}_{I,2} \beta_0^2
     - 4 \zeta_2 \varphi^{(2)}_{I,2} \beta_1 
     - 10 C^{I}_1 \beta_0 \zeta_3 \beta_1
     - C^{I}_1 \zeta_2 \beta_2
     - \frac{57}{5} C^{I}_1 \zeta_2^2 \beta_0^3
     + 12 \varphi^{(1)}_{I,1} \beta_0^3 \zeta_3
     + 5 \varphi^{(1)}_{I,1} \zeta_2 \beta_0 \beta_1\,,
\end{autobreak}
\nonumber\\
\begin{autobreak}
\varphi^{I,(2)}_{d,4} =
     \varphi^{(2)}_{I,4}
     - 9 \zeta_2 \beta_0 \varphi^{(3)}_{I,3}
     + 6 \zeta_2 \varphi^{(2)}_{I,2} \beta_0^2\,,
\end{autobreak}
\nonumber\\
\begin{autobreak}     
\varphi^{I,(3)}_{d,4} = 
         \varphi^{(3)}_{I,4} \,,
         \quad \varphi^{I,(4)}_{d,4} = \varphi^{(4)}_{I,4} \,,
\end{autobreak}
\end{align}
where $\beta_i$ are the coefficients of QCD-$\beta$ function which are known to five loops \cite{Chetyrkin:2017bjc,Luthe:2017ttg,Herzog:2017ohr,Baikov:2016tgj}. The SV coefficients $\tilde{\cal G}^{I,(k)}_{j}$ from SV part of soft-collinear distribution as well as the NSV anomalous dimensions $C^I$ and $D^I$ for $I=g,q$ up to third order are provided in the Online Resource supplied with this article. The NSV inclusive coefficients $\varphi^{(k)}_{I,j}$ for $I=g,q$ up to third order can be found in \cite{Ajjath:2020ulr}. For completeness, we also provide the explicit results of $\varphi^{I,(k)}_{d,j}$ for Higgs production and Drell-Yan process up to third order along with the partial fourth order results in the Online Resource.   

The $z$ space result for coefficient functions expressed in the integral representation in \eqref{eq:npsiint} is suitable for studying large $N$ behaviour of the 
rapidity distribution and hence we use the modified \textit{differential soft-collinear operator}
to obtain the resummed result in $N$ space taking into account the NSV terms.  
The Mellin moment of dCFs  is found to be
\begin{align}
\label{eq:res2}
\tilde\Delta^{\mathrm{sv+nsv}}_{d,I}(\Nb_1,\Nb_2)
=&  {g}_{d,0}^I\left(\{p_j\cdot q_k\},q^2,\muf^2\right) \times
\nonumber\\&
\exp\Big( \tilde G_{d,\Nb}^{I,\rm{sv+nsv}}\left( q^2, \muf^2,\omega \right)\Big)\,,
\end{align}
where the exponent ${\tilde G}_{d,\Nb}^{I,\rm{sv+nsv}}$ can be organized as a resummed perturbation series in Mellin space as,

\begin{align}
\label{eq:GdNb}
{ \tilde G}_{d,\Nb}^{I,\rm{sv+nsv}}= &~~
  \bigg(g_{d,1}^I(\omega)
+\frac{1}{N_1} \overline{g}_{d,1}^I(\omega)\bigg) \log \overline{N}_1
\nonumber\\&
+ \sum\limits_{i=0}^\infty a_s^i \bigg( \frac{1}{2}  g_{d,i+2}^I(\omega)
+ \frac{1}{N_1} \overline{g}_{d,i+2}^I(\omega) \bigg)
\nonumber \\&
 +\frac{1}{N_1}
\sum\limits_{i=0}^{\infty} a_s^i h^I_{d,i}(\omega, \overline{N}_1) + (\overline{N}_1 \leftrightarrow \overline{N}_2) \,,
\end{align}
with
\begin{align}
\label{hg}
        h^I_{d,0}(\omega, \overline{N}_l) = h^I_{d,00}(\omega) + h^I_{d,01}(\omega) \log \overline{N}_l, 
        \nonumber\\
         h^I_{d,i}(\omega, \overline{N}_l) = \sum_{k=0}^{i} h^I_{d,ik}(\omega)~ \log^k \overline{N}_l \,.
\end{align}
The NSV resummation coefficients are $\overline{g}_{d,i}^I$ and $h^I_{d,i}$.  The coefficient $\overline g_{d,1}^I$ is found to be zero. The coefficients $\overline g_{d,i+2}^I$ depend on universal cusp anomalous dimension $A^I$ and $\textbf{D}_d^I$, while $h_{d,i}^I$s are determined by the NSV coefficients $\varphi_{d,I}^f$ as well as by $C^I, D^I$ from $\mathcal P^I(a_s,\overline z_l)$ as given in \eqref{nsvex}.  The resummation coefficients $g^{I,i}_{d,0},g^I_{d,i}(\omega),\overline {g}^I_{d,i}(\omega)$ and $h^I_{d,i}(\omega)$  contain leading, next-to-leading, $\cdots$, 
SV and NSV logarithms in the dCFs.  

{Rescaling the constants by $\beta_0$ as $\overline A_i^I = A_i^I /\beta_0^i $, $\overline{ \textbf{D}}_{d,i}^I = \textbf{D}_{d,i}^I/ \beta_0^i$, $\overline{ {C}}_{d,i}^I = 
{C}_{d,i}^I/ \beta_0^i$, $\overline{ {D}}_{d,i}^I = 
{D}_{d,i}^I/ \beta_0^i$ and $\overline \beta_i = \beta_i /\beta_0^{i+1}$, we present below the results of the NSV exponents $\overline{g}^I_{d,i}(\omega)$ and $h^I_{d,ij}(\omega)$ after setting $\mu_R^2=\mu_F^2=q^2$. The full list of these exponents with explicit dependence on $\mu_R^2$ and $\mu_F^2$ are provided in the Online Resource supplied with this article.}
\begin{align}
\label{nsvex}
\begin{autobreak}
{\overline g}^I_{d,2}(\omega) = 
               \frac{1}{2}\overline{A}^I_1  L_\omega \,,
\end{autobreak} 
\nonumber\\
\begin{autobreak}
{\overline g}^I_{d,3}(\omega) =  
               \frac{\beta_0}{(1-\omega)} \bigg[                               \frac{1}{2}\overline{\textbf{D}}^{I}_{d,1} 
               + \overline{A}^I_2   \bigg( - \frac{1}{2} \omega \bigg) 
               + \overline{A}^I_1   \bigg(- \frac{1}{2} 
               + \frac{1}{2} \omega \overline{\beta}_1
               + \frac{1}{2} L_\omega \overline{\beta}_1  \bigg) \bigg] \,,
\end{autobreak} 
\nonumber\\
\begin{autobreak}
{\overline g}^I_{d,4}(\omega) =     
               \frac{\beta_0^2}{(1-\omega)^2} \bigg[ \frac{1}{2}\overline{\textbf{D}}^{I}_{d,2} 
               + \overline{\textbf{D}}^{I}_{d,1}   \bigg( \frac{1}{2}
               - \frac{1}{2} L_\omega \overline{\beta}_1 \bigg)
               + \overline{A}^I_3   \bigg( - \frac{1}{2} \omega
               + \frac{1}{4} \omega^2 \bigg) 
               +  \overline{A}^I_2   \bigg(
               - \frac{1}{2}
               + \frac{1}{2} \omega \overline{\beta}_1
               - \frac{1}{4} \omega^2 \overline{\beta}_1
               + \frac{1}{2} L_\omega \overline{\beta}_1  \bigg) \nonumber
               +  \overline{A}^I_1   \bigg(
               - \frac{1}{2} \zeta_2
               - \frac{1}{4} \omega^2 \overline{\beta}_2
               + \frac{1}{4} \omega^2 \overline{\beta}_1^2 
               + \frac{1}{2} L_\omega \overline{\beta}_1
               - \frac{1}{4} L_\omega^2 \overline{\beta}_1^2 \bigg) \bigg]\,,
\end{autobreak}
\nonumber\\
\begin{autobreak}
    h^I_{d,00}(\omega) =

        \overline{D}^{I}_1  L_\omega
          \,,
\end{autobreak}
\nonumber\\
\begin{autobreak}
        h^I_{d,01}(\omega) =

        - \overline{C}^{I}_1  
           L_\omega  \,,
\end{autobreak}
\nonumber\\
\begin{autobreak}
    h^I_{d,10}(\omega) =

       \frac{1}{(1-\omega)} \Bigg[ \overline{D}^{I}_2   \bigg(
          - \omega \beta_0
          \bigg)

       + \overline{D}^{I}_1   \bigg(
           \omega \beta_0 \overline{\beta}_1
          + L_\omega \beta_0 \overline{\beta}_1
          \bigg)

       + \overline{C}^{I}_1   \bigg(
           \beta_0 \zeta_2
          \bigg)

       + \varphi^{I,(0)}_{d,1} \Bigg]
         \,,
\end{autobreak}
\nonumber\\
\begin{autobreak}
   h^I_{d,11}(\omega) =

       \frac{1}{(1-\omega)} \Bigg[ \overline{C}^{I}_2   \bigg( \omega \beta_0 \bigg)

       + \overline{C}^{I}_1   \bigg( - \omega \beta_0 \overline{\beta}_1
          - L_\omega \beta_0 \overline{\beta}_1  \bigg) - \varphi^{I,(1)}_{d,1} \Bigg]
         \,,
\end{autobreak}
\nonumber\\
\begin{autobreak}
    h^I_{d,21}(\omega) =

        \frac{1}{(1-\omega)^2} \Bigg[ \overline{C}^{I}_3   \bigg(
           \omega \beta_0^2
          - \frac{1}{2} \omega^2 \beta_0^2
          \bigg)

       + \overline{C}^{I}_2   \bigg( - \omega \beta_0^2 \overline{\beta}_1
          + \frac{1}{2} \omega^2 \beta_0^2 \overline{\beta}_1
          - L_\omega \beta_0^2 \overline{\beta}_1 \bigg)

       + \overline{C}^{I}_1   \bigg(\beta_0^2 \zeta_2
          + \frac{1}{2} \omega^2 \beta_0^2 \overline{\beta}_2
          - \frac{1}{2} \omega^2 \beta_0^2 \overline{\beta}_1^2
          + \frac{1}{2} L_\omega^2 \beta_0^2 \overline{\beta}_1^2
          \bigg)

       - \varphi^{I,(1)}_{d,2}
          + L_\omega \beta_0 \overline{\beta}_1 \varphi^{I,(1)}_{d,1} \Bigg] \,,
        
\end{autobreak}
\nonumber\\
\begin{autobreak}
    h^I_{d,22}(\omega) =
       \frac{1}{(1-\omega)^2} \Bigg[ \varphi^{I,(2)}_{d,2} \Bigg] \,,
         
\end{autobreak}
\nonumber\\
\begin{autobreak}
    h^I_{d,32}(\omega) =
        \frac{1}{(1-\omega)^3} \Bigg[ \varphi^{I,(2)}_{d,3}
          - 2 L_\omega \beta_0 \overline{\beta}_1 \varphi^{I,(2)}_{d,2} \Bigg] \,,
         
\end{autobreak}
\nonumber\\
\begin{autobreak}
    h^I_{d,33}(\omega) =
       \frac{1}{(1-\omega)^3} \Bigg[ - \varphi^{I,(3)}_{d,3} \Bigg] \,,
         
\end{autobreak}
\nonumber\\
\begin{autobreak}
    h^I_{d,42}(\omega) =
          \frac{1}{(1-\omega)^4} \Bigg[ \varphi^{I,(2)}_{d,4}
          + 6 \zeta_2 \varphi^{I,(4)}_{d,4}
          - 9 \beta_0 \zeta_2 \varphi^{I,(3)}_{d,3}
          + 6 \beta_0^2 \zeta_2 \varphi^{I,(2)}_{d,2}
          + 2 \omega \beta_0^2 \overline{\beta}_2 \varphi^{I,(2)}_{d,2}
          - 2 \omega \beta_0^2 \overline{\beta}_1^2 \varphi^{I,(2)}_{d,2}
          - 3 L_\omega \beta_0 \overline{\beta}_1 \varphi^{I,(2)}_{d,3}
          - 2 L_\omega \beta_0^2 \overline{\beta}_1^2 \varphi^{I,(2)}_{d,2}
          + 3 L_\omega^2 \beta_0^2 \overline{\beta}_1^2 \varphi^{I,(2)}_{d,2} \Bigg] \,,
         
\end{autobreak}
\nonumber\\
\begin{autobreak}
    h^I_{d,43}(\omega) =
          \frac{1}{(1-\omega)^4} \Bigg[ - \varphi^{I,(3)}_{d,4}
          + 3 L_\omega \beta_0 \overline{\beta}_1 \varphi^{I,(3)}_{d,3} \Bigg] \,,
         
\end{autobreak}
\nonumber\\
\begin{autobreak}
    h^I_{d,44}(\omega) =
        \frac{1}{(1-\omega)^4} \Bigg[ \varphi^{I,(4)}_{d,4} \Bigg] \,,
  \end{autobreak}
\end{align}
where $L_{\omega} = \log(1- \omega)$.

In summary, thanks to the remarkable simplification of mass factorised formula for the rapidity distribution for diagonal channels and the knowledge of logarithmic structure of differential soft-collinear distribution and the mass factorization kernels from fixed order results, we can systematically resum both SV and NSV terms in $z$ as well as in double Mellin space. Note that we could extend the framework that was used for the production of a single colorless state to the state of $n$-colorless particles.  This was possible simply because of the fact that IR structures of the building blocks, namely, the pure virtual contribution, soft-collinear distributions and the mass factorization kernels for $2 \rightarrow n$ is identical to those for $2 \rightarrow 1$.   In addition, the IR singularity structure of these quantities in dimensional regularisation plays an important role to understand the leading SV distributions and sub-leading NSV logarithms in the real emission processes.   In the case of SV, there exist two universal or process independent soft-collinear functions, namely for quark-anti quark ($q \bar q$) and gluon-gluon ($gg$) initiated processes.  For the NSV, we find that a part of the soft-collinear functions is partially determined by universal anomalous dimensions and the remaining part depends on the underlying hard process.  However, for a given process, the resummed results either in $z$ or 
$\{N_1,N_2\}$ can be used to predict SV and NSV terms at higher orders provided the lower order results are known.  With this in mind, we have derived a general formula for coefficient functions up to fourth order and resummed results to N$^3$LL accuracy which will be useful when the complete four loop form factor, soft-collinear functions and lower order process dependent terms become available.

\section{Conclusions and Outlook}
\label{sec:conclusion}
In today's era reducing the theoretical uncertainties remain one of the main motivations for higher order radiative corrections. It is also particularly relevant for constraining beyond SM scenarios and validation of the SM itself. Among several observables, differential cross-sections allow a wider range of comparisons with the experiment and hence several attempts were made in the past for better theoretical understanding of the same. In this article, we restrict ourselves to the discussion of differential rapidity distribution for the production of $n$-number of colorless particles in the hadronic collision within the realm of perturbative QCD. 

Through this publication, we intend to present a systematic framework for the study of  soft-plus-virtual corrections to the differential distribution with respect to the rapidity variable $y$, for the production of $n$-colorless particles in the hadron collider. The infrared structure of rapidity distribution which was earlier studied in ref.~\cite{Ravindran:2006bu} for Sudakov type processes is further extended to the case of $2\rightarrow n$ scattering. We employ the universality of the soft enhancements associated with the real emission diagrams. The main deviation from the Sudakov type formalism comes from the virtual corrections where the kinematic dependence is much more involved and hence these are now expressed in terms of scalar products of the kind $\{p_j\cdot q_k\}$. The rest of the formalism relies on the collinear factorization of the differential cross section,
the renormalization group invariance, universality of perturbative infrared structure of the 
scattering amplitudes, and the process independence of the soft-collinear distribution. Besides this, we also use an additional fact that the $N$-th Mellin moment of the differential distribution has a relation with its inclusive counterpart in the limit $N\rightarrow \infty$,  as depicted through \eqref{eq:mellin-rapidity-Xsec}. The mere use of this fact enables us to to get an all order relation between the soft-collinear distribution of inclusive cross-section and that of rapidity. Hence from the given quantity in inclusive part, we can determine it for the rapidity and  thereby avoid performing the explicit computation of the real emission processes for rapidity distribution.  The \textit{goal} of this current article is to present the general structure for the SV differential rapidity distribution up to N$^4$LO and also the resummed predictions till N$^3$LL level in QCD, which can be expressed in terms of universal anomalous dimensions along with the process dependent virtual matrix elements. The former, which comprises of process independent finite segments of soft-collinear distribution and the mass factorized kernels, remains unaltered irrespective of the number of colorless particles in the final states. Furthermore, the soft-collinear distributions for the quark and gluon initiated processes are found to be related to each other through simple quadratic Casimir scaling, known as the maximally non-Abelian property.This is explicitly verified up to N$^3$LO. However, whether the validity will remain intact beyond this order with generalized Casimir scaling, that will be an interesting thing to look into.  \textcolor{black}{Often, one finds that in certain kinematic regions, the sub-leading logarithms, namely NSV terms can not be ignored in phenomenological studies.  Our investigation on these terms for diagonal partonic channels reveals that there are similarities in the structure of IR terms with those of $2 \rightarrow 1$ process allowing us to propose resummed predictions for NSV terms within the same framework.  }

In summary, in order to obtain the fixed order as well as resummed prediction for the differential rapidity distributions of a generic $n$-colorless final states, one merely requires the form factor corresponding to the hard process under study provided the soft-collinear distribution for Sudakov type process is known.  We present the analytical results for the fixed order up to $\mathrm{N^4LO}$ and the resummed predictions up to $\mathrm{N^3LL}$ level in the appendix for the scale choice of $\mu_R^2 = \mu_F^2 = q^2$ and the same with the explicit scale dependence are provided in the Online Resource supplied with this article.

\section*{Acknowledgements}
The algebraic computations have been done with the latest version of the symbolic manipulation system {\sc Form}~\cite{Vermaseren:2000nd,Ruijl:2017dtg}. {\color{black}We would like to thank S. Catani, E. Laenen, L. Magnea, C. Duhr, B. Mistlberger, J. Michel and F.J.Tackmann for useful discussions. We also thank P. Banerjee,  A. Chakraborty,  G. Das, P.K. Dhani, M.C. Kumar,  M. Mahakhud, M.K. Mandal, P. Mathews, N. Rana, K. Samanta, S. Seth and S. Tiwari for collaborating with us on this topic at various stages.} The work of T.A. received funding from the European Research Council (ERC) under the European Union’s Horizon 2020 research and innovation programme, \textit{Novel structures in scattering amplitudes} (grant agreement No. 725110).

\appendix
\section{Soft-collinear distribution for rapidity distribution}
\label{secA:SCdist}
In this section, we present soft-collinear distribution  ${\textbf{S}}_{d,I}$, as defined in \eqref{eq:soft-collinear-operator},
in powers of $a_s(\mu_R^2)$ up to N$^2$LO.
Expanding the quantity in powers of $a_s$ as
\begin{align}
    {\textbf{S}}_{d,I}(z_1,z_2,q^2,\mu_R^2,\mu_F^2) &= \delta(\zo) \delta(\zt)
    \nonumber\\& + \sum_{i=1}^{\infty} a_s^i(\mu_R^2)\,   {\textbf{S}}^{(i)}_{d,I}(z_1,z_2,q^2,\mu_R^2,\mu_F^2) \,,
\end{align}
we present the results for $\mu_R^2=\mu_F^2=q^2$. The results up to N$^4$LO with explicit scale dependence can be found from the Online Resource supplied with this article.
\begin{align}
\begin{autobreak}

   \textbf{S}_{d,I}^{(1)} =
        {\cal D}_0 \overline{{\cal D}}_0   \bigg\{
           \frac{1}{2} A^I_1
          \bigg\}

       + \delta(\zo) \overline{{\cal D}}_1   \bigg\{
           A^I_1
          \bigg\}

       + \delta(\zo) \overline{{\cal D}}_0   \bigg\{
          - f^I_1
          \bigg\}

       + \delta(\zo) \delta(\zt)   \bigg\{ \g^{I,(1)}_{d,1} \bigg\} + (z_1 \leftrightarrow z_2) \,,

   \end{autobreak}
\nonumber \\

\begin{autobreak}

   \textbf{S}_{d,I}^{(2)} =

        {\cal D}_1 \overline{{\cal D}}_1   \bigg\{ \frac{3}{2} (A^I_1)^2
          \bigg\}

       + {\cal D}_0 \overline{{\cal D}}_2   \bigg\{
           \frac{3}{2} (A^I_1)^2
          \bigg\}

       + {\cal D}_0 \overline{{\cal D}}_1   \bigg\{ - 3 A^I_1 f^I_1 - \beta_0 A^I_1  \bigg\}

       + {\cal D}_0 \overline{{\cal D}}_0   \bigg\{
           \frac{1}{2} (f^I_1)^2
          + \frac{1}{2} A^I_2
          + A^I_1 \g^{I,(1)}_{d,1}
          - (A^I_1)^2 \zeta_2
          + \frac{1}{2} \beta_0 f^I_1
          \bigg\}

       + \delta(\zo) \overline{{\cal D}}_3   \bigg\{\frac{1}{2} (A^I_1)^2 \bigg\} + \delta(\zo) \overline{{\cal D}}_2   \bigg\{ - \frac{3}{2} A^I_1 f^I_1 
       - \frac{1}{2} \beta_0 A^I_1 \bigg\}

       + \delta(\zo) \overline{{\cal D}}_1   \bigg\{
           (f^I_1)^2
          + A^I_2
          + 2 A^I_1 \g^{I,(1)}_{d,1}
          - 2 (A^I_1)^2 \zeta_2
          + \beta_0 f^I_1
          \bigg\}

       + \delta(\zo) \overline{{\cal D}}_0   \bigg\{
          - f^I_2
          - 2 f^I_1 \g^{I,(1)}_{d,1}
          + 2 A^I_1 \zeta_2 f^I_1
          + 2 (A^I_1)^2 \zeta_3
          - 2 \beta_0 \g^{I,(1)}_{d,1}
          \bigg\}

       + \delta(\zo) \delta(\zt)   \bigg\{ \frac{1}{2} \g^{I,(2)}_{d,1}
          + (\g^{I,(1)}_{d,1})^2
          - \frac{1}{2} \zeta_2 (f^I_1)^2
          - A^I_1 \zeta_3 f^I_1
          + \frac{1}{5} (A^I_1)^2 \zeta_2^2
          + \beta_0 \g^{I,(1)}_{d,2} \bigg\} + (z_1 \leftrightarrow z_2) \,,

   \end{autobreak}
\end{align}
The symbols ${\tilde{\cal G}}^{,I,j}_{d,k}$ that appear in the aforementioned soft-collinear distributions are provided explicitly in the Online Resource with this article. 

\section{Soft-virtual partonic rapidity distribution}
\label{secA:SVDelta}
We expand the $\Delta^{\rm sv}_{d,I}$, as defined in \eqref{eq:SV-master-formula-recast-I}, in powers of $a_s(\mu_R^2)$ through
\begin{align}
\label{eqA:Sof-Coll-Expand}
    \Delta^{\rm sv}_{d,I} &= \delta(\zo) \delta(\zt) |{\cal M}^{(0)}_{I, {\rm fin}}|^2 \nonumber\\&
    +  \sum_{i=1}^{\infty} a_s^{i}(\mu_R^2) \Delta^{{\rm sv},(i)}_{d,I}\left(\{p_j\cdot q_k\}, z_1,z_2,q^2,\mu_F^2,\mu_R^2\right) .
\end{align}
Here, we present $\Delta^{\rm sv}_{d,I}$ to N$^2$LO for the specific scale choice $\mu_F^2=\mu_R^2=q^2$. 
The results up to N$^4$LO with explicit dependence on $\mu_R$ and $\mu_F$ are provided in the Online Resource supplied with this article.
\begin{align}
\begin{autobreak}

   {\Delta}^{\text{sv},(1)}_{d,I} =

        |{\cal M}^{(0)}_{I,\text{fin}}|^{2} {\cal D}_0 \overline{{\cal D}}_0   \bigg\{\frac{1}{2} A^I_1
          \bigg\}

       + \delta(\zo) |{\cal M}^{(0)}_{I,\text{fin}}|^{2} \overline{{\cal D}}_1   \bigg\{ A^I_1
          \bigg\}

       + \delta(\zo) |{\cal M}^{(0)}_{I,\text{fin}}|^{2} \overline{{\cal D}}_0   \bigg\{ - f^I_1 \bigg\}

       + \delta(\zo) \delta(\zt)   \bigg\{ {\cal M}^{(0,1)}_{I,\text{fin}}
          \bigg\}

       + \delta(\zo) \delta(\zt) |{\cal M}^{(0)}_{I,\text{fin}}|^{2}   \bigg\{\g^{I,(1)}_{d,1}\bigg\} 
       + (z_1 \leftrightarrow z_2) \,,

 \end{autobreak}
\nonumber \\

\begin{autobreak}
  
   {\Delta}^{\text{sv},(2)}_{d,I} =

       + {\cal D}_0 \overline{{\cal D}}_0   \bigg\{A^I_1 {\cal M}^{(0,1)}_{I,\text{fin}} \bigg\}

       + |{\cal M}^{(0)}_{I,\text{fin}}|^{2} {\cal D}_1 \overline{{\cal D}}_1   \bigg\{\frac{3}{2} (A^I_1)^2\bigg\}

       + |{\cal M}^{(0)}_{I,\text{fin}}|^{2} {\cal D}_0 \overline{{\cal D}}_2   \bigg\{\frac{3}{2} (A^I_1)^2
          \bigg\}

       + |{\cal M}^{(0)}_{I,\text{fin}}|^{2} {\cal D}_0 \overline{{\cal D}}_1 \bigg\{- 3 A^I_1 f^I_1- \beta_0 A^I_1 \bigg\}

       + |{\cal M}^{(0)}_{I,\text{fin}}|^{2} {\cal D}_0 \overline{{\cal D}}_0   \bigg\{\frac{1}{2} (f^I_1)^2
          + \frac{1}{2} A^I_2
          + A^I_1 \g^{I,(1)}_{d,1}
          - (A^I_1)^2 \zeta_2
          + \frac{1}{2} \beta_0 f^I_1 \bigg\}

       + \delta(\zo) \overline{{\cal D}}_1   \bigg\{2 A^I_1 {\cal M}^{(0,1)}_{I,\text{fin}} \bigg\}

       + \delta(\zo) \overline{{\cal D}}_0   \bigg\{ - 2 f^I_1 {\cal M}^{(0,1)}_{I,\text{fin}} \bigg\}

       + \delta(\zo) |{\cal M}^{(0)}_{I,\text{fin}}|^{2} \overline{{\cal D}}_3 \bigg\{\frac{1}{2} (A^I_1)^2\bigg\}

       + \delta(\zo) |{\cal M}^{(0)}_{I,\text{fin}}|^{2} \overline{{\cal D}}_2   \bigg\{
          - \frac{3}{2} A^I_1 f^I_1
          - \frac{1}{2} \beta_0 A^I_1
          \bigg\}

       + \delta(\zo) |{\cal M}^{(0)}_{I,\text{fin}}|^{2} \overline{{\cal D}}_1   \bigg\{
           (f^I_1)^2
          + A^I_2
          + 2 A^I_1 \g^{I,(1)}_{d,1}
          - 2 (A^I_1)^2 \zeta_2
          + \beta_0 f^I_1
          \bigg\}

       + \delta(\zo) |{\cal M}^{(0)}_{I,\text{fin}}|^{2} \overline{{\cal D}}_0   \bigg\{
          - f^I_2
          - 2 f^I_1 \g^{I,(1)}_{d,1}
          + 2 A^I_1 \zeta_2 f^I_1
          + 2 (A^I_1)^2 \zeta_3
          - 2 \beta_0 \g^{I,(1)}_{d,1}
          \bigg\}

       + \delta(\zo) \delta(\zt)   \bigg\{
           \frac{1}{2} {\cal M}^{(1,1)}_{I,\text{fin}}
          + {\cal M}^{(0,2)}_{I,\text{fin}}
          + 2 \g^{I,(1)}_{d,1} {\cal M}^{(0,1)}_{I,\text{fin}}
          \bigg\}

       + \delta(\zo) \delta(\zt) |{\cal M}^{(0)}_{I,\text{fin}}|^{2}   \bigg\{
           \frac{1}{2} \g^{I,(2)}_{d,1}
          + (\g^{I,(1)}_{d,1})^2
          - \frac{1}{2} \zeta_2 (f^I_1)^2
          - A^I_1 \zeta_3 f^I_1
          + \frac{1}{5} (A^I_1)^2 \zeta_2^2
          + \beta_0 \g^{I,(1)}_{d,2}
          \bigg\} + (z_1 \leftrightarrow z_2) \,,

 \end{autobreak}
\end{align}
where
\begin{eqnarray}
{\cal D}_i=\left[{\log^i(1-z_1) \over (1-z_1)}\right]_+~,
\quad \quad \quad \quad {\cal \overline D}_i=
\left[{\log^i(1-z_2) \over (1-z_2)}\right]_+ \,.
\end{eqnarray}
In the aforementioned equations, we define
\begin{align}
    {\cal M}^{(m,n)}_{I,{\rm fin}} \equiv {\rm Real} \left( \langle {\cal M}^{(m)}_I|{\cal M}^{(n)}_I \rangle_{\rm fin} \right)\,,
\end{align}
where $|{\cal M}^{(n)}_I \rangle$ is the UV renormalized pure virtual amplitude at $n$-th order in $a_s$ as introduced in \eqref{eq:M-Mfin}.
\section{Soft-collinear distribution for threshold resummation}
\label{secA:Sdelta}
The universal soft-collinear operator that is required to obtain the resummed cross-section in $z$-space, defined in \eqref{eq:sdelta}, is expanded in powers of $a_s(\mu_R^2)$ as
\begin{align}
    S_{res,\delta}^{d,I}(q^2,\mu_R^2,\mu_F^2) = 1 + \sum_{i=1}^{\infty} a_s^i(\mu_R^2) S_{res,\delta}^{d,I,(i)}(q^2,\mu_R^2,\mu_F^2)\,.
\end{align}
We present the results for $q^2=\mu_R^2=\mu_F^2$ below to fourth order in coupling constant. The result with explicit scale dependence can be obtained from the Online Resource provided with this article.
\begin{align}
\begin{autobreak}

\end{autobreak}\nonumber\\
\begin{autobreak}
    
{ S}^{d,I, (1)}_{res,\delta} =
        2 \g^{I,(1)}_{d,1} \,,
 \end{autobreak}
\nonumber \\

\begin{autobreak}
    { S}^{d,I, (2)}_{res,\delta} =
       \g^{I,(2)}_{d,1}
          + 2 (\g^{I,(1)}_{d,1})^2
          + 2 \beta_0 \g^{I,(1)}_{d,2}
         \,,
 \end{autobreak}
\nonumber \\

\begin{autobreak}
    { S}^{d,I, (3)}_{res,\delta} =
        \frac{2}{3} \g^{I,(3)}_{d,1}
          + 2 \g^{I,(1)}_{d,1} \g^{I,(2)}_{d,1}
          + \frac{4}{3} (\g^{I,(1)}_{d,1})^3
          + \frac{4}{3} \beta_1 \g^{I,(1)}_{d,2}
          + \frac{4}{3} \beta_0 \g^{I,(2)}_{d,2}
          + 4 \beta_0 \g^{I,(1)}_{d,1} \g^{I,(1)}_{d,2}
          + \frac{8}{3} \beta_0^2 \g^{I,(1)}_{d,3}
         \,,
 \end{autobreak}
\nonumber \\

\begin{autobreak}
    { S}^{d,I, (4)}_{res,\delta} =
        \frac{1}{2} \g^{I,(4)}_{d,1}
          + \frac{1}{2} (\g^{I,(2)}_{d,1})^2
          + \frac{4}{3} \g^{I,(1)}_{d,1} \g^{I,(3)}_{d,1}
          + 2 (\g^{I,(1)}_{d,1})^2 \g^{I,(2)}_{d,1}
          + \frac{2}{3} (\g^{I,(1)}_{d,1})^4
          + \beta_2 \g^{I,(1)}_{d,2}
          + \beta_1 \g^{I,(2)}_{d,2}
          + \frac{8}{3} \beta_1 \g^{I,(1)}_{d,1} \g^{I,(1)}_{d,2}
          + \beta_0 \g^{I,(3)}_{d,2}
          + 2 \beta_0 \g^{I,(1)}_{d,2} \g^{I,(2)}_{d,1}
          + \frac{8}{3} \beta_0 \g^{I,(1)}_{d,1} \g^{I,(2)}_{d,2}
          + 4 \beta_0 (\g^{I,(1)}_{d,1})^2 \g^{I,(1)}_{d,2}
          + 4 \beta_0 \beta_1 \g^{I,(1)}_{d,3}
          + 2 \beta_0^2 \g^{I,(2)}_{d,3}
          + 2 \beta_0^2 (\g^{I,(1)}_{d,2})^2
          + \frac{16}{3} \beta_0^2 \g^{I,(1)}_{d,1} \g^{I,(1)}_{d,3}
          + 4 \beta_0^3 \g^{I,(1)}_{d,4}
         .
\end{autobreak}
\nonumber \\
\end{align}

\section{Explicit results at N$^4$LO for Drell-Yan and the Higgs boson production}
\label{secD:N4LO}
In this section, we present the explicit results of $\Delta^{\rm sv}_{d,I}$, as defined in \eqref{eq:SV-master-formula-recast-I} and \eqref{eqA:Sof-Coll-Expand}, for the Drell-Yan ($I=q$), and the Higgs boson productions through gluon fusion ($I=g$) as well as bottom quark annihilation ($I=b$) at fourth order in coupling constant. Setting $\mu_F^2=\mu_R^2=q^2$, in the following, we provide only the new results, and the old results for Drell-Yan and Higgs boson productions can be found in \cite{Ravindran:2006bu,Ahmed:2014uya,Ahmed:2014era}. 
The results with explicit dependence on $\mu_R$ and $\mu_F$ are provided up to N$^4$LO in the Online Resource files supplied with this article.

\begin{align}
\begin{autobreak}

   {\Delta}^{\text{sv},(4)}_{d,g} =

     \delta(\zo)\delta(\zt)  \bigg[  n_f\frac{d_F^{abcd} d_A^{abcd}}{N_A}   \bigg\{
           \chi_{5}^{g}
          - \frac{2560}{3} \zeta_2 \zeta_5
          - \frac{512}{3} \zeta_2 \zeta_3
          + 512 \zeta_2^2
          \bigg\}

       + n_f^2\frac{d_F^{abcd} d_F^{abcd}}{N_A}   \bigg\{
          - \frac{9008}{9}
          - 960 \zeta_5
          + 1520 \zeta_3
          + 512 \zeta_3^2
          + \frac{384}{5} \zeta_2^2
          \bigg\}

       + \frac{d_A^{abcd} d_A^{abcd}}{N_A}   \bigg\{
           \chi_{7}^{g}
          + \frac{7040}{3} \zeta_2 \zeta_5
          + \frac{256}{3} \zeta_2 \zeta_3
          - 768 \zeta_2 \zeta_3^2
          - 256 \zeta_2^2
          - \frac{15872}{35} \zeta_2^4
          \bigg\}

       + n_f   \bigg\{- \frac{1}{3} \g^{g,(2)}_{3} \bigg\}

       + C_F n_f^3   \bigg\{
          - \frac{233953}{972}
          + \frac{640}{27} \zeta_5
          + \frac{4060}{27} \zeta_3
          + \frac{376}{3} \zeta_2
          - \frac{640}{9} \zeta_2 \zeta_3
          + \frac{112}{45} \zeta_2^2
          \bigg\}

       + C_F^2 n_f^2   \bigg\{\frac{11401}{54}
          + \frac{3920}{3} \zeta_5
          - \frac{5044}{3} \zeta_3
          + \frac{896}{3} \zeta_3^2
          - \frac{50}{9} \zeta_2
          + \frac{32}{3} \zeta_2 \zeta_3
          - \frac{212}{15} \zeta_2^2
          + \frac{13696}{945} \zeta_2^3
          \bigg\}

       + C_F^3 n_f   \bigg\{
           \chi_{4}^{g}
          \bigg\}

       + C_A   \bigg\{
           \frac{11}{6} \g^{g,(2)}_{3}
          \bigg\}

       + C_A n_f^3   \bigg\{ - \frac{943703}{5832}
          - \frac{566}{27} \zeta_5
          - \frac{57182}{729} \zeta_3
          + \frac{1156}{9} \zeta_2
          + \frac{3332}{81} \zeta_2 \zeta_3
          + \frac{10141}{405} \zeta_2^2
          \bigg\}

       + C_A C_F n_f^2   \bigg\{ \frac{49991435}{11664}
          - \frac{23572}{27} \zeta_5
          - \frac{357994}{243} \zeta_3
          - \frac{3214}{9} \zeta_3^2
          - \frac{6509}{18} \zeta_2
          - \frac{1640}{9} \zeta_2 \zeta_3
          - \frac{101459}{540} \zeta_2^2
          - \frac{2756}{189} \zeta_2^3
          \bigg\}

       + C_A C_F^2 n_f   \bigg\{
           \frac{153625}{162}
          + \chi_{3}^{g}
          - \frac{17656}{3} \zeta_7
          + \frac{29728}{9} \zeta_5
          + \frac{222683}{54} \zeta_3
          - \frac{3824}{3} \zeta_3^2
          + \frac{2227}{9} \zeta_2
          - 1280 \zeta_2 \zeta_5
          + 1104 \zeta_2 \zeta_3
          + \frac{4032}{5} \zeta_2^2
          - \frac{2464}{3} \zeta_2^2 \zeta_3
          - \frac{415952}{945} \zeta_2^3
          \bigg\}

       + C_A^2 n_f^2   \bigg\{\frac{3373102453}{839808}
          - \frac{5957}{27} \zeta_5
          + \frac{1583227}{1944} \zeta_3
          + \frac{4504}{27} \zeta_3^2
          - \frac{43937939}{34992} \zeta_2
          + \frac{10277}{54} \zeta_2 \zeta_3
          - \frac{2169709}{3240} \zeta_2^2
          - \frac{1567}{135} \zeta_2^3
          \bigg\}

       + C_A^2 C_F n_f   \bigg\{ - \frac{495711665}{69984}
          + \chi_{2}^{g}
          + \frac{28744}{9} \zeta_7
          + \frac{1077758}{405} \zeta_5
          + \frac{186703}{729} \zeta_3
          - \frac{178784}{81} \zeta_3^2
          - \frac{5080535}{1944} \zeta_2
          + \frac{1120}{9} \zeta_2 \zeta_5
          + \frac{194875}{81} \zeta_2 \zeta_3
          + \frac{443861}{1080} \zeta_2^2
          + \frac{76904}{45} \zeta_2^2 \zeta_3
          + \frac{24368}{105} \zeta_2^3
          \bigg\}

       + C_A^3 n_f   \bigg\{\frac{8163906247}{472392} + \chi_{1}^{g}
          - \frac{383989}{63} \zeta_7
          + \frac{7783846}{1215} \zeta_5
          - \frac{142069663}{26244} \zeta_3
          + \frac{2132341}{729} \zeta_3^2 + \frac{6135005}{8748} \zeta_2
          + \frac{77156}{135} \zeta_2 \zeta_5  - \frac{568195}{243} \zeta_2 \zeta_3
          + \frac{101328101}{29160} \zeta_2^2
          - \frac{134854}{81} \zeta_2^2 \zeta_3
          + \frac{178103}{315} \zeta_2^3\bigg\}

       + C_A^4   \bigg\{\frac{24881127343}{1889568}
          + \chi_{6}^{g}
          + \frac{1024}{45} \zeta_{5,3}
          + \frac{2029709}{42} \zeta_7
          - \frac{9370337}{486} \zeta_5
          - \frac{293538695}{13122} \zeta_3
          - \frac{304447}{45} \zeta_3 \zeta_5
          + \frac{6412558}{729} \zeta_3^2
          + \frac{339046381}{52488} \zeta_2
          + \frac{433642}{135} \zeta_2 \zeta_5
          - \frac{4939438}{729} \zeta_2 \zeta_3
          - \frac{35402}{27} \zeta_2 \zeta_3^2
          - \frac{107201743}{14580} \zeta_2^2
          + \frac{1731257}{405} \zeta_2^2 \zeta_3
          + \frac{5324869}{3780} \zeta_2^3
          - \frac{33921919}{15750} \zeta_2^4
          \bigg\}

       + \frac{1}{4}  \g^{g,(1)}_{4}
         \bigg] + 

    \delta(\zo) \overline{{\cal D}}_0 \bigg[   n_f\frac{d_F^{abcd} d_A^{abcd}}{N_A}   \bigg\{
           384
          + 2 b^q_{4,d_F^{abcd}d_F^{abcd}}
          + \frac{21760}{9} \zeta_5
          + \frac{5312}{9} \zeta_3
          - \frac{1216}{3} \zeta_3^2
          - \frac{4544}{3} \zeta_2
          - 128 \zeta_2 \zeta_3
          + \frac{320}{3} \zeta_2^2
          - \frac{9472}{315} \zeta_2^3
          \bigg\}

       + \frac{d_A^{abcd} d_A^{abcd}}{N_A}   \bigg\{- f_{4,d_F^{abcd} d_A^{abcd}}^q \bigg\}

       + C_A n_f^3   \bigg\{\frac{5216}{2187}
          + \frac{80}{81} \zeta_3
          - \frac{800}{81} \zeta_2
          + \frac{16}{9} \zeta_2^2
          \bigg\}

       + C_A C_F n_f^2   \bigg\{- \frac{155083}{486}
          + 16 \zeta_5
          + \frac{1912}{9} \zeta_3
          + \frac{1400}{9} \zeta_2
          - \frac{320}{3} \zeta_2 \zeta_3
          + \frac{32}{3} \zeta_2^2
          \bigg\}

       + C_A C_F^2 n_f   \bigg\{- \frac{27949}{108}
          + 2 b^q_{4,n_f C_F^3}
          - \frac{3872}{3} \zeta_5
          + \frac{1120}{9} \zeta_3
          - 368 \zeta_3^2
          - 320 \zeta_2
          + \frac{512}{3} \zeta_2 \zeta_3
          + \frac{668}{5} \zeta_2^2
          + \frac{117344}{315} \zeta_2^3
          \bigg\}

       + C_A^2 n_f^2   \bigg\{- \frac{1543153}{5832}
          + \frac{1936}{9} \zeta_5
          + \frac{1240}{9} \zeta_3
          + \frac{258130}{729} \zeta_2
          - \frac{3536}{27} \zeta_2 \zeta_3
          - \frac{1504}{45} \zeta_2^2
          \bigg\}

       + C_A^2 C_F n_f   \bigg\{
           \frac{2798681}{972}
          + 2 b^q_{4, n_f C_F^2 C_A}
          + \frac{10952}{9} \zeta_5
          - \frac{70426}{27} \zeta_3
          + \frac{7816}{3} \zeta_3^2
          - \frac{4925}{9} \zeta_2
          - \frac{368}{9} \zeta_2 \zeta_3
          - \frac{24400}{27} \zeta_2^2
          - \frac{1328}{7} \zeta_2^3
          \bigg\}

       + C_A^3 n_f   \bigg\{
           \frac{17665315}{5832}
          - b^q_{4, n_f C_F^2 C_A}
          - \frac{1}{2} b^q_{4,n_f C_F^3}
          - \frac{1}{24} b^q_{4,d_F^{abcd}d_F^{abcd}}
          - \frac{83128}{27} \zeta_5
          - \frac{177566}{27} \zeta_3
          + \frac{604}{9} \zeta_3^2
          - \frac{1991747}{729} \zeta_2
          + \frac{8848}{3} \zeta_2 \zeta_3
          + \frac{195944}{405} \zeta_2^2
          - \frac{2696}{945} \zeta_2^3
          \bigg\}

       + C_A^4   \bigg\{- \frac{40498399}{4374}
          + \frac{1}{24} f^q_{4, d_F^{abcd} d_A^{abcd}}
          + 9380 \zeta_7
          + \frac{83240}{9} \zeta_5
          + \frac{2047084}{81} \zeta_3
          - \frac{36652}{3} \zeta_3^2
          + \frac{4083193}{729} \zeta_2
          - 7264 \zeta_2 \zeta_5
          - \frac{312316}{27} \zeta_2 \zeta_3
          + \frac{76708}{405} \zeta_2^2
          + \frac{7568}{15} \zeta_2^2 \zeta_3
          - \frac{25036}{315} \zeta_2^3
          \bigg\} \bigg] +

     \delta(\zo) \overline{{\cal D}}_1 \bigg[  n_f\frac{d_F^{abcd} d_A^{abcd}}{N_A}   \bigg\{
          - \frac{1280}{3} \zeta_5
          - \frac{256}{3} \zeta_3
          + 256 \zeta_2
          \bigg\}

       + \frac{d_A^{abcd} d_A^{abcd}}{N_A}   \bigg\{ \frac{3520}{3} \zeta_5
          + \frac{128}{3} \zeta_3
          - 384 \zeta_3^2
          - 128 \zeta_2
          - \frac{7936}{35} \zeta_2^3
          \bigg\}

       + C_A n_f^3   \bigg\{
          - \frac{4000}{729}
          + \frac{320}{27} \zeta_2
          \bigg\}

       + C_A C_F n_f^2   \bigg\{
           \frac{57890}{81}
          - \frac{4576}{9} \zeta_3
          - \frac{976}{9} \zeta_2
          - \frac{128}{45} \zeta_2^2
          \bigg\}

       + C_A C_F^2 n_f   \bigg\{
           \frac{3004}{9}
          - 1600 \zeta_5
          + \frac{2960}{3} \zeta_3
          \bigg\}

       + C_A^2 n_f^2   \bigg\{
          \frac{544699}{729}
          - \frac{464}{9} \zeta_3
          - \frac{33344}{81} \zeta_2
          - \frac{832}{45} \zeta_2^2
          \bigg\}

       + C_A^2 C_F n_f   \bigg\{
          - \frac{410057}{81}
          + 800 \zeta_5
          + \frac{28480}{9} \zeta_3
          + \frac{2920}{9} \zeta_2
          + 256 \zeta_2 \zeta_3
          + \frac{704}{45} \zeta_2^2
          \bigg\}

       + C_A^3 n_f   \bigg\{
          - \frac{6305477}{729}
          + 2000 \zeta_5
          + \frac{136664}{27} \zeta_3
          + \frac{232048}{81} \zeta_2
          - 3040 \zeta_2 \zeta_3
          + \frac{36128}{135} \zeta_2^2
          \bigg\}

       + C_A^4   \bigg\{
           \frac{14936741}{729}
          - 6600 \zeta_5
          - \frac{798056}{27} \zeta_3
          + \frac{34816}{3} \zeta_3^2
          - \frac{413360}{81} \zeta_2
          + 15312 \zeta_2 \zeta_3
          - \frac{131072}{135} \zeta_2^2
          + \frac{1696}{105} \zeta_2^3
          \bigg\} \bigg] +

      {\cal D}_0 \overline{{\cal D}}_0 \bigg[ n_f\frac{d_F^{abcd} d_A^{abcd}}{N_A}   \bigg\{- \frac{640}{3} \zeta_5 - \frac{128}{3} \zeta_3+ 128 \zeta_2 \bigg\}

       + \frac{d_A^{abcd} d_A^{abcd}}{N_A}   \bigg\{\frac{1760}{3} \zeta_5
          + \frac{64}{3} \zeta_3
          - 192 \zeta_3^2
          - 64 \zeta_2
          - \frac{3968}{35} \zeta_2^3
          \bigg\}

       + C_A n_f^3   \bigg\{- \frac{2000}{729}
          + \frac{160}{27} \zeta_2
          \bigg\}

       + C_A C_F n_f^2   \bigg\{
           \frac{28945}{81}
          - \frac{2288}{9} \zeta_3
          - \frac{488}{9} \zeta_2
          - \frac{64}{45} \zeta_2^2
          \bigg\}

       + C_A C_F^2 n_f   \bigg\{
           \frac{1502}{9}
          - 800 \zeta_5
          + \frac{1480}{3} \zeta_3
          \bigg\}

       + C_A^2 n_f^2   \bigg\{\frac{544699}{1458}- \frac{232}{9} \zeta_3 - \frac{16672}{81} \zeta_2- \frac{416}{45} \zeta_2^2\bigg\}

       + C_A^2 C_F n_f   \bigg\{
          - \frac{410057}{162}
          + 400 \zeta_5
          + \frac{14240}{9} \zeta_3
          + \frac{1460}{9} \zeta_2
          + 128 \zeta_2 \zeta_3
          + \frac{352}{45} \zeta_2^2
          \bigg\}

       + C_A^3 n_f   \bigg\{- \frac{6305477}{1458}
          + 1000 \zeta_5
          + \frac{68332}{27} \zeta_3
          + \frac{116024}{81} \zeta_2
          - 1520 \zeta_2 \zeta_3
          + \frac{18064}{135} \zeta_2^2
          \bigg\}

       + C_A^4   \bigg\{\frac{14936741}{1458}
          - 3300 \zeta_5
          - \frac{399028}{27} \zeta_3
          + \frac{17408}{3} \zeta_3^2
          - \frac{206680}{81} \zeta_2
          + 7656 \zeta_2 \zeta_3
          - \frac{65536}{135} \zeta_2^2
          + \frac{848}{105} \zeta_2^3
          \bigg\} \bigg] + \bigg\{ z_1 \leftrightarrow z_2 \bigg\} .
\end{autobreak}
\end{align}

\begin{align}
\begin{autobreak}

 {\Delta}^{\text{sv},(4)}_{d,q} =
 
        \delta(\zo)\delta(\zt)  \bigg[ 
        n_f \frac{d_F^{abcd} d_F^{abcd}}{N_F}   \bigg\{
           \frac{3190}{3}
          - 1240 \zeta_7
          + \frac{95098}{27} \zeta_5
          - \frac{13414}{27} \zeta_3
          + \frac{680}{9} \zeta_3^2
          - \frac{21566}{9} \zeta_2
          - \frac{1568}{3} \zeta_2 \zeta_5
          - 140 \zeta_2 \zeta_3
          + \frac{30592}{45} \zeta_2^2
          - \frac{3952}{15} \zeta_2^2 \zeta_3
          + \frac{41620}{189} \zeta_2^3
          \bigg\}

       + \frac{d_F^{abcd} d_A^{abcd}}{N_F}   \bigg\{
           \chi_{10}^{q}
          + \frac{7040}{3} \zeta_2 \zeta_5
          + \frac{256}{3} \zeta_2 \zeta_3
          - 768 \zeta_2 \zeta_3^2
          - 256 \zeta_2^2
          - \frac{15872}{35} \zeta_2^4
          \bigg\}  
          
          + n_f  \bigg\{ - \frac{1}{3} \g^{q,(2)}_3\bigg\} + C_F n_f n_{fv} N_4   \bigg\{
          - \frac{392}{9}
          + \frac{440}{9} \zeta_5
          - 28 \zeta_3
          + \frac{176}{3} \zeta_3^2
          - \frac{556}{9} \zeta_2
          - \frac{56}{3} \zeta_2 \zeta_3
          + \frac{76}{9} \zeta_2^2
          + \frac{1408}{135} \zeta_2^3
          \bigg\}

       + C_F n_f^3   \bigg\{
           \frac{65633}{1944}
          - \frac{238}{27} \zeta_5
          - \frac{3416}{729} \zeta_3
          - \frac{39398}{729} \zeta_2
          + \frac{404}{81} \zeta_2 \zeta_3
          - \frac{3511}{405} \zeta_2^2
          \bigg\}

       + C_F^2 n_{fv} N_4   \bigg\{
           \frac{3334}{3}
          + \frac{1}{8} \chi_{5}^{q}
          - \frac{8542}{3} \zeta_7
          + \frac{38468}{27} \zeta_5
          + \frac{4840}{27} \zeta_3
          - \frac{4592}{9} \zeta_3^2
          + \frac{10706}{9} \zeta_2
          - 464 \zeta_2 \zeta_5
          + \frac{902}{3} \zeta_2 \zeta_3
          - \frac{4024}{45} \zeta_2^2
          - \frac{784}{3} \zeta_2^2 \zeta_3
          + \frac{11254}{135} \zeta_2^3
          \bigg\}

       + C_F^2 n_f^2   \bigg\{
          - \frac{1194071}{5832}
          + \frac{1876}{27} \zeta_5
          - \frac{425615}{729} \zeta_3
          + \frac{286}{27} \zeta_3^2
          + \frac{108874}{729} \zeta_2
          + \frac{10120}{27} \zeta_2 \zeta_3
          + \frac{5401}{60} \zeta_2^2
          + \frac{18868}{945} \zeta_2^3
          \bigg\}

       + C_F^3 n_f   \bigg\{
          - \frac{7723623865}{419904}
          + \chi_{3}^{q}
          - \frac{473078}{63} \zeta_7
          + \frac{1268018}{81} \zeta_5
          + \frac{10401064}{243} \zeta_3
          - \frac{674818}{81} \zeta_3^2
          - \frac{12237605}{648} \zeta_2
          - \frac{8464}{9} \zeta_2 \zeta_5
          + \frac{364040}{81} \zeta_2 \zeta_3
          + \frac{1994819}{1215} \zeta_2^2
          - \frac{327928}{135} \zeta_2^2 \zeta_3
          - \frac{2803408}{2835} \zeta_2^3
          \bigg\}

       + C_F^4   \bigg\{
          - \frac{6246665}{128}
          + \chi_{9}^{q}
          + \frac{5152}{3} \zeta_{5,3}
          + \frac{164740}{7} \zeta_7
          + \frac{1550732}{15} \zeta_5
          - \frac{144091}{2} \zeta_3
          - \frac{658688}{45} \zeta_3 \zeta_5
          + \frac{292424}{27} \zeta_3^2
          - \frac{1762043}{48} \zeta_2
          + \frac{2576}{5} \zeta_2 \zeta_5
          + \frac{203056}{9} \zeta_2 \zeta_3
          + \frac{59264}{27} \zeta_2 \zeta_3^2
          - \frac{90321}{10} \zeta_2^2
          + \frac{19856}{15} \zeta_2^2 \zeta_3
          + \frac{3412928}{315} \zeta_2^3
          - \frac{507968}{1575} \zeta_2^4
          \bigg\}

       + C_A   \bigg\{
           \frac{11}{6} \g_3^{q,(2)}
          \bigg\}

       + C_A C_F n_{fv} N_4   \bigg\{
          - \frac{1870}{3}
          + \frac{1}{8} \chi_{4}^{q}
          - \frac{8272}{9} \zeta_5
          - \frac{1034}{9} \zeta_3
          + \frac{3784}{3} \zeta_3^2
          - \frac{2860}{3} \zeta_2
          - 330 \zeta_2 \zeta_3
          + \frac{2156}{15} \zeta_2^2
          + \frac{107008}{315} \zeta_2^3
          \bigg\}

       + C_A C_F n_f^2   \bigg\{- \frac{667210703}{839808}
          + \frac{8152}{27} \zeta_5
          + \frac{2099687}{2916} \zeta_3
          + \frac{68}{9} \zeta_3^2
          + \frac{42932371}{34992} \zeta_2
          - \frac{13396}{27} \zeta_2 \zeta_3
          + \frac{15877}{108} \zeta_2^2
          - \frac{3517}{315} \zeta_2^3
          \bigg\}

       + C_A C_F^2 n_f   \bigg\{
           \frac{161137649299}{1889568}
          + \chi_{2}^{q}
          - 1722 \zeta_7
          - \frac{8225273}{486} \zeta_5
          - \frac{5024904581}{52488} \zeta_3
          + \frac{11407462}{729} \zeta_3^2
          + \frac{1790660321}{34992} \zeta_2
          + \frac{187004}{135} \zeta_2 \zeta_5
          - \frac{898147}{81} \zeta_2 \zeta_3
          - \frac{125343967}{29160} \zeta_2^2
          + \frac{146510}{81} \zeta_2^2 \zeta_3
          + \frac{20536}{27} \zeta_2^3
          \bigg\}

       + C_A C_F^3   \bigg\{\frac{154126124135}{839808}
          + \chi_{8}^{q}
          - \frac{23024}{15} \zeta_{5,3}
          + \frac{23357}{21} \zeta_7
          - \frac{272959751}{1620} \zeta_5
          - \frac{1562360669}{11664} \zeta_3
          + \frac{43564}{5} \zeta_3 \zeta_5
          + \frac{6415507}{162} \zeta_3^2
          + \frac{1143090157}{7776} \zeta_2
          + \frac{265036}{45} \zeta_2 \zeta_5
          - \frac{6518455}{81} \zeta_2 \zeta_3
          - \frac{28928}{9} \zeta_2 \zeta_3^2
          - \frac{52425877}{9720} \zeta_2^2
          + \frac{393476}{27} \zeta_2^2 \zeta_3
          - \frac{24147386}{2835} \zeta_2^3
          - \frac{2236966}{875} \zeta_2^4
          \bigg\}

       + C_A^2 C_F n_f   \bigg\{
          - \frac{7253202277}{104976}
          + \chi_{1}^{q}
          - \frac{211}{9} \zeta_7
          + \frac{394286}{45} \zeta_5
          + \frac{141306569}{2916} \zeta_3
          - \frac{62306}{9} \zeta_3^2
          - \frac{175686721}{4374} \zeta_2
          - \frac{2672}{3} \zeta_2 \zeta_5
          + \frac{204071}{27} \zeta_2 \zeta_3
          + \frac{361903}{216} \zeta_2^2
          - \frac{9104}{45} \zeta_2^2 \zeta_3
          - \frac{4261}{105} \zeta_2^3
          \bigg\}

       + C_A^2 C_F^2   \bigg\{
          - \frac{139605518111}{472392}
          + \chi_{7}^{q}
          - \frac{7184}{45} \zeta_{5,3}
          + \frac{112505}{9} \zeta_7
          + \frac{491406151}{4860} \zeta_5
          + \frac{31284635423}{104976} \zeta_3
          - \frac{7567}{9} \zeta_3 \zeta_5
          - \frac{45139415}{729} \zeta_3^2
          - \frac{39174091799}{209952} \zeta_2
          - \frac{1163402}{135} \zeta_2 \zeta_5
          + \frac{60145883}{729} \zeta_2 \zeta_3
          - \frac{7018}{27} \zeta_2 \zeta_3^2
          + \frac{628489141}{29160} \zeta_2^2
          - \frac{1136015}{81} \zeta_2^2 \zeta_3
          + \frac{135362}{45} \zeta_2^3
          + \frac{17432749}{15750} \zeta_2^4
          \bigg\}

       + C_A^3 C_F   \bigg\{
           \frac{29551452589}{209952}
          + \chi_{6}^{q}
          + \frac{2321}{18} \zeta_7
          - \frac{736349}{30} \zeta_5
          - \frac{175768027}{1458} \zeta_3
          + \frac{41129}{2} \zeta_3^2
          + \frac{1547277107}{17496} \zeta_2
          + \frac{19976}{3} \zeta_2 \zeta_5
          - \frac{920596}{27} \zeta_2 \zeta_3
          - 32 \zeta_2 \zeta_3^2
          - \frac{1404395}{162} \zeta_2^2
          + \frac{113432}{45} \zeta_2^2 \zeta_3
          + \frac{79827}{140} \zeta_2^3
          - \frac{40064}{105} \zeta_2^4
          \bigg\}

       + \frac{1}{4} \g_4^{q,(1)}
         \bigg] + 
         
       \delta(\zo) \overline{\cal D}_0  \bigg[ 
        n_f \frac{d_F^{abcd} d_F^{abcd}}{N_F}   \bigg\{
           384
          + 2 b^q_{4,d_F^{abcd}d_F^{abcd}}
          + \frac{21760}{9} \zeta_5
          + \frac{5312}{9} \zeta_3
          - \frac{1216}{3} \zeta_3^2
          - \frac{4544}{3} \zeta_2
          - 128 \zeta_2 \zeta_3
          + \frac{320}{3} \zeta_2^2
          - \frac{9472}{315} \zeta_2^3
          \bigg\}

       + \frac{d_F^{abcd} d_A^{abcd}}{N_F}   \bigg\{- f^q_{4, d_F^{abcd} d_A^{abcd}}
          \bigg\}

       + C_F n_f^3   \bigg\{
           \frac{5216}{2187}
          + \frac{80}{81} \zeta_3
          - \frac{800}{81} \zeta_2
          + \frac{16}{9} \zeta_2^2
          \bigg\}

       + C_F^2 n_f^2   \bigg\{
          - \frac{142769}{1458}
          + \frac{1168}{9} \zeta_5
          + \frac{8552}{27} \zeta_3
          - \frac{71900}{729} \zeta_2
          - \frac{6272}{27} \zeta_2 \zeta_3
          + \frac{1504}{27} \zeta_2^2
          \bigg\}

       + C_F^3 n_f   \bigg\{
          - \frac{80221}{108}
          + 2 b^q_{4,n_f C_F^3}
          - \frac{7712}{3} \zeta_5
          + \frac{14800}{9} \zeta_3
          + 1168 \zeta_3^2
          - \frac{8210}{27} \zeta_2
          + \frac{4576}{9} \zeta_2 \zeta_3
          + \frac{8108}{45} \zeta_2^2
          + \frac{117344}{315} \zeta_2^3
          \bigg\}

       + C_F^4   \bigg\{7680 \zeta_7
          - 6144 \zeta_5
          + 4088 \zeta_3
          - 1920 \zeta_3^2
          - 4608 \zeta_2 \zeta_5
          + 3904 \zeta_2 \zeta_3
          - \frac{6656}{5} \zeta_2^2 \zeta_3
          \bigg\}

       + C_A C_F n_f^2   \bigg\{
          - \frac{898033}{5832}
          + \frac{304}{3} \zeta_5
          + \frac{2456}{81} \zeta_3
          + \frac{75718}{243} \zeta_2
          + \frac{80}{9} \zeta_2 \zeta_3
          - \frac{3104}{45} \zeta_2^2
          \bigg\}

       + C_A C_F^2 n_f   \bigg\{ - \frac{955285}{2916}
          + 2 b^q_{4,n_f C_F^2 C_A}
          - \frac{104}{3} \zeta_5
          - \frac{25124}{27} \zeta_3
          + \frac{4072}{3} \zeta_3^2
          + \frac{1781395}{729} \zeta_2
          + \frac{12880}{9} \zeta_2 \zeta_3
          - \frac{675088}{405} \zeta_2^2
          - \frac{12976}{105} \zeta_2^3
          \bigg\}

       + C_A C_F^3   \bigg\{
          - \frac{103222}{27}
          + 8576 \zeta_5
          - \frac{57095}{9} \zeta_3
          - 5008 \zeta_3^2
          + \frac{18479}{9} \zeta_2
          - 2304 \zeta_2 \zeta_5
          - \frac{14936}{9} \zeta_2 \zeta_3
          + \frac{8968}{45} \zeta_2^2
          + \frac{4672}{5} \zeta_2^2 \zeta_3
          - \frac{704}{5} \zeta_2^3
          \bigg\}

       + C_A^2 C_F n_f   \bigg\{\frac{10761379}{5832}
          - \frac{1}{2} b^q_{4,n_f C_F^3}
          - b^q_{4,n_f C_F^2 C_A}
          - \frac{1}{24} b^q_{4,d_F^{abcd}d_F^{abcd}}
          - \frac{14776}{27} \zeta_5
          - \frac{210778}{81} \zeta_3
          - \frac{4868}{9} \zeta_3^2
          - \frac{654449}{243} \zeta_2
          + \frac{6640}{9} \zeta_2 \zeta_3
          + \frac{138808}{135} \zeta_2^2
          - \frac{65192}{945} \zeta_2^3
          \bigg\}

       + C_A^2 C_F^2   \bigg\{
           \frac{7543094}{729}
          + \frac{58624}{9} \zeta_5
          - \frac{134305}{27} \zeta_3
          - 2032 \zeta_3^2
          - \frac{7337465}{729} \zeta_2
          - 768 \zeta_2 \zeta_5
          - \frac{72640}{27} \zeta_2 \zeta_3
          + \frac{193424}{81} \zeta_2^2
          + \frac{1840}{3} \zeta_2^2 \zeta_3
          - \frac{1232}{15} \zeta_2^3
          \bigg\}

       + C_A^3 C_F   \bigg\{
          - \frac{28325071}{4374}
          + \frac{1}{24} f^q_{4,d_F^{abcd} d_A^{abcd}}
          + 1700 \zeta_7
          - \frac{24920}{9} \zeta_5
          + 9600 \zeta_3
          - \frac{2332}{3} \zeta_3^2
          + \frac{1642195}{243} \zeta_2
          + 416 \zeta_2 \zeta_5
          - \frac{30772}{9} \zeta_2 \zeta_3
          - \frac{78188}{45} \zeta_2^2
          + 288 \zeta_2^2 \zeta_3
          + \frac{45188}{315} \zeta_2^3
          \bigg\} \bigg]
          
        +\delta(\zo) \overline{\cal D}_1  \bigg[ n_f \frac{d_F^{abcd} d_F^{abcd}}{N_F}   \bigg\{- \frac{1280}{3} \zeta_5- \frac{256}{3} \zeta_3 
        + 256 \zeta_2 \bigg\}

       + \frac{d_F^{abcd} d_A^{abcd}}{N_F}   \bigg\{
           \frac{3520}{3} \zeta_5
          + \frac{128}{3} \zeta_3
          - 384 \zeta_3^2
          - 128 \zeta_2
          - \frac{7936}{35} \zeta_2^3
          \bigg\}

       + C_F n_f^3   \bigg\{ - \frac{4000}{729}
          + \frac{320}{27} \zeta_2 \bigg\} + C_F^2 n_{fv} N_4   \bigg\{32 - \frac{640}{3} \zeta_5 + \frac{112}{3} \zeta_3 
          + 80 \zeta_2 - \frac{16}{5} \zeta_2^2 \bigg\}

       + C_F^2 n_f^2   \bigg\{
          - \frac{72590}{729}
          - \frac{30688}{81} \zeta_3
          + \frac{11408}{81} \zeta_2
          + \frac{5632}{135} \zeta_2^2
          \bigg\}

       + C_F^3 n_f   \bigg\{
          - \frac{4310}{9}
          + \frac{10816}{9} \zeta_5
          + \frac{29456}{27} \zeta_3
          - \frac{5576}{27} \zeta_2
          - \frac{22400}{9} \zeta_2 \zeta_3
          + \frac{27968}{135} \zeta_2^2
          \bigg\}

       + C_F^4   \bigg\{
          - \frac{11198}{3}
          + 5312 \zeta_5
          - 1840 \zeta_3
          + \frac{18304}{3} \zeta_3^2
          - \frac{6652}{3} \zeta_2
          + 1280 \zeta_2 \zeta_3
          + \frac{592}{5} \zeta_2^2
          + \frac{97408}{315} \zeta_2^3
          \bigg\}

       + C_A C_F n_f^2   \bigg\{
           \frac{58045}{243}
          - \frac{208}{9} \zeta_3
          - \frac{7616}{27} \zeta_2
          + \frac{64}{5} \zeta_2^2
          \bigg\}

       + C_A C_F^2 n_f   \bigg\{ \frac{2300107}{729}
          + 128 \zeta_5
          + \frac{234368}{81} \zeta_3
          - \frac{18100}{9} \zeta_2
          - \frac{1408}{3} \zeta_2 \zeta_3
          - \frac{51536}{135} \zeta_2^2
          \bigg\}

       + C_A C_F^3   \bigg\{ 12062
          - \frac{110752}{9} \zeta_5
          - \frac{214592}{27} \zeta_3
          + \frac{15808}{3} \zeta_3^2
          + \frac{21554}{27} \zeta_2
          + \frac{111008}{9} \zeta_2 \zeta_3
          - \frac{211376}{135} \zeta_2^2
          - \frac{71168}{315} \zeta_2^3
          \bigg\}

       + C_A^2 C_F n_f   \bigg\{- \frac{571387}{243}
          - \frac{1360}{9} \zeta_5
          + \frac{3128}{3} \zeta_3
          + \frac{58000}{27} \zeta_2
          + 32 \zeta_2 \zeta_3
          - \frac{3872}{15} \zeta_2^2
          \bigg\}

       + C_A^2 C_F^2   \bigg\{ - \frac{11106458}{729}
          - 816 \zeta_5
          + \frac{6164}{81} \zeta_3
          + \frac{752}{3} \zeta_3^2
          + \frac{594406}{81} \zeta_2
          + 1648 \zeta_2 \zeta_3
          + \frac{58972}{135} \zeta_2^2
          + \frac{38944}{315} \zeta_2^3
          \bigg\}

       + C_A^3 C_F   \bigg\{\frac{4520317}{729}
          + \frac{15400}{9} \zeta_5
          - \frac{51032}{9} \zeta_3
          - 16 \zeta_3^2
          - \frac{140200}{27} \zeta_2
          + 528 \zeta_2 \zeta_3
          + \frac{3520}{3} \zeta_2^2
          - \frac{20032}{105} \zeta_2^3
          \bigg\} \bigg] +
          
        {\cal D}_0 \overline{{\cal D}}_0 \bigg[
        n_f \frac{d_F^{abcd} d_F^{abcd}}{N_F}   \bigg\{
          - \frac{640}{3} \zeta_5
          - \frac{128}{3} \zeta_3
          + 128 \zeta_2
          \bigg\}

       + \frac{d_F^{abcd} d_A^{abcd}}{N_F}   \bigg\{
           \frac{1760}{3} \zeta_5
          + \frac{64}{3} \zeta_3
          - 192 \zeta_3^2
          - 64 \zeta_2
          - \frac{3968}{35} \zeta_2^3
          \bigg\}

       + C_F n_f^3   \bigg\{  - \frac{2000}{729}
          + \frac{160}{27} \zeta_2
        \bigg\}

       + C_F^2 n_{fv} N_4   \bigg\{
           16
          - \frac{320}{3} \zeta_5
          + \frac{56}{3} \zeta_3
          + 40 \zeta_2
          - \frac{8}{5} \zeta_2^2
          \bigg\}

       + C_F^2 n_f^2   \bigg\{
          - \frac{36295}{729}
          - \frac{15344}{81} \zeta_3
          + \frac{5704}{81} \zeta_2
          + \frac{2816}{135} \zeta_2^2
          \bigg\}

       + C_F^3 n_f   \bigg\{ - \frac{2155}{9}
          + \frac{5408}{9} \zeta_5
          + \frac{14728}{27} \zeta_3
          - \frac{2788}{27} \zeta_2
          - \frac{11200}{9} \zeta_2 \zeta_3
          + \frac{13984}{135} \zeta_2^2\bigg\}

       + C_F^4   \bigg\{
          - \frac{5599}{3}
          + 2656 \zeta_5
          - 920 \zeta_3
          + \frac{9152}{3} \zeta_3^2
          - \frac{3326}{3} \zeta_2
          + 640 \zeta_2 \zeta_3
          + \frac{296}{5} \zeta_2^2
          + \frac{48704}{315} \zeta_2^3
          \bigg\}

       + C_A C_F n_f^2   \bigg\{
           \frac{58045}{486}
          - \frac{104}{9} \zeta_3
          - \frac{3808}{27} \zeta_2
          + \frac{32}{5} \zeta_2^2
          \bigg\}

       + C_A C_F^2 n_f   \bigg\{
           \frac{2300107}{1458}
          + 64 \zeta_5
          + \frac{117184}{81} \zeta_3
          - \frac{9050}{9} \zeta_2
          - \frac{704}{3} \zeta_2 \zeta_3
          - \frac{25768}{135} \zeta_2^2
          \bigg\}

       + C_A C_F^3   \bigg\{
           6031
          - \frac{55376}{9} \zeta_5
          - \frac{107296}{27} \zeta_3
          + \frac{7904}{3} \zeta_3^2
          + \frac{10777}{27} \zeta_2
          + \frac{55504}{9} \zeta_2 \zeta_3
          - \frac{105688}{135} \zeta_2^2
          - \frac{35584}{315} \zeta_2^3
          \bigg\}

       + C_A^2 C_F n_f   \bigg\{
          - \frac{571387}{486}
          - \frac{680}{9} \zeta_5
          + \frac{1564}{3} \zeta_3
          + \frac{29000}{27} \zeta_2
          + 16 \zeta_2 \zeta_3
          - \frac{1936}{15} \zeta_2^2
          \bigg\}

       + C_A^2 C_F^2   \bigg\{
          - \frac{5553229}{729}
          - 408 \zeta_5
          + \frac{3082}{81} \zeta_3
          + \frac{376}{3} \zeta_3^2
          + \frac{297203}{81} \zeta_2
          + 824 \zeta_2 \zeta_3
          + \frac{29486}{135} \zeta_2^2
          + \frac{19472}{315} \zeta_2^3
          \bigg\}

       + C_A^3 C_F   \bigg\{\frac{4520317}{1458}+ \frac{7700}{9} \zeta_5
          - \frac{25516}{9} \zeta_3 - 8 \zeta_3^2  - \frac{70100}{27} \zeta_2
          + 264 \zeta_2 \zeta_3  + \frac{1760}{3} \zeta_2^2
          - \frac{10016}{105} \zeta_2^3  \bigg\} \bigg] + \bigg\{ z_1 \leftrightarrow z_2 \bigg\} .

 \end{autobreak}
\end{align}

\begin{align}
\begin{autobreak}
 {\Delta}^{\text{sv},(4)}_{d,b} =

      \delta(\zo)\delta(\zt)  
       \bigg[ n_f \frac{d_F^{abcd} d_F^{abcd}}{N_F}   \bigg\{
          - \frac{2560}{3} \zeta_2 \zeta_5
          - \frac{512}{3} \zeta_2 \zeta_3
          + 512 \zeta_2^2
          \bigg\}

       + \frac{d_F^{abcd} d_A^{abcd}}{N_F}   \bigg\{ \frac{7040}{3} \zeta_2 \zeta_5 + \frac{256}{3} \zeta_2 \zeta_3 - 768 \zeta_2 \zeta_3^2
          - 256 \zeta_2^2 - \frac{15872}{35} \zeta_2^4 \bigg\}

       + n_f   \bigg\{- \frac{1}{3} \g_3^{b,(2)} \bigg\}

       + C_F n_f^3   \bigg\{
          - \frac{580}{729}
          - \frac{238}{27} \zeta_5
          - \frac{3686}{729} \zeta_3
          - \frac{3758}{729} \zeta_2
          + \frac{404}{81} \zeta_2 \zeta_3
          - \frac{335}{81} \zeta_2^2
          \bigg\}

       + C_F^2 n_f^2   \bigg\{
          - \frac{1341097}{11664}
          - \frac{3488}{27} \zeta_5
          - \frac{245390}{729} \zeta_3
          + \frac{286}{27} \zeta_3^2
          + \frac{241525}{729} \zeta_2
          + \frac{12856}{27} \zeta_2 \zeta_3
          - \frac{3065}{36} \zeta_2^2
          + \frac{18868}{945} \zeta_2^3
          \bigg\}

       + C_F^3 n_f   \bigg\{\frac{42292165}{52488}
          - \frac{473078}{63} \zeta_7
          + \frac{5271502}{405} \zeta_5
          + \frac{3163859}{486} \zeta_3
          - \frac{271510}{81} \zeta_3^2
          - \frac{352162}{81} \zeta_2
          - \frac{8464}{9} \zeta_2 \zeta_5
          + \frac{317420}{81} \zeta_2 \zeta_3
          - \frac{778774}{1215} \zeta_2^2
          - \frac{327928}{135} \zeta_2^2 \zeta_3
          + \frac{179864}{405} \zeta_2^3
          \bigg\}

       + C_F^4   \bigg\{
           \frac{31207}{2}
          + \frac{5152}{3} \zeta_{5,3}
          + 17448 \zeta_7
          + \frac{579184}{15} \zeta_5
          - \frac{236447}{3} \zeta_3
          - \frac{658688}{45} \zeta_3 \zeta_5
          + \frac{363374}{27} \zeta_3^2
          - \frac{56399}{6} \zeta_2
          + \frac{33008}{5} \zeta_2 \zeta_5
          + \frac{57568}{9} \zeta_2 \zeta_3
          + \frac{59264}{27} \zeta_2 \zeta_3^2
          - \frac{50856}{5} \zeta_2^2
          + \frac{14248}{5} \zeta_2^2 \zeta_3
          + \frac{256712}{45} \zeta_2^3
          - \frac{507968}{1575} \zeta_2^4
          \bigg\}

       + C_A   \bigg\{
           \frac{11}{6} \g_3^{b,(2)}
          \bigg\}

       + C_A C_F n_f^2   \bigg\{\frac{198202909}{839808}
          + \frac{7396}{27} \zeta_5
          + \frac{1477931}{2916} \zeta_3
          + \frac{68}{9} \zeta_3^2
          + \frac{3988867}{34992} \zeta_2
          - \frac{12460}{27} \zeta_2 \zeta_3
          + \frac{34049}{540} \zeta_2^2
          - \frac{3517}{315} \zeta_2^3
          \bigg\}

       + C_A C_F^2 n_f   \bigg\{
           \frac{893866105}{472392}
          - 1722 \zeta_7
          - \frac{19235759}{1215} \zeta_5
          - \frac{767605141}{26244} \zeta_3
          + \frac{8355031}{729} \zeta_3^2
          + \frac{71947313}{8748} \zeta_2
          + \frac{187004}{135} \zeta_2 \zeta_5
          - \frac{861269}{81} \zeta_2 \zeta_3
          + \frac{254215}{5832} \zeta_2^2
          + \frac{146510}{81} \zeta_2^2 \zeta_3
          + \frac{874568}{945} \zeta_2^3
          \bigg\}

       + C_A C_F^3   \bigg\{ - \frac{61739114}{6561}
          - \frac{23024}{15} \zeta_{5,3}
          + \frac{432101}{21} \zeta_7
          - \frac{36539051}{405} \zeta_5
          + \frac{44371336}{729} \zeta_3
          + \frac{43564}{5} \zeta_3 \zeta_5
          + \frac{699449}{81} \zeta_3^2
          + \frac{7935479}{243} \zeta_2
          - \frac{48124}{9} \zeta_2 \zeta_5
          - \frac{3288814}{81} \zeta_2 \zeta_3
          - \frac{28928}{9} \zeta_2 \zeta_3^2
          + \frac{16395079}{1215} \zeta_2^2
          + \frac{1575124}{135} \zeta_2^2 \zeta_3
          - \frac{20261636}{2835} \zeta_2^3
          - \frac{2236966}{875} \zeta_2^4
          \bigg\}

       + C_A^2 C_F n_f   \bigg\{ \frac{11380577}{6561}
          - \frac{211}{9} \zeta_7
          + \frac{345056}{45} \zeta_5
          + \frac{47301911}{2916} \zeta_3
          - \frac{55286}{9} \zeta_3^2
          - \frac{17328527}{2187} \zeta_2
          - \frac{2672}{3} \zeta_2 \zeta_5
          + \frac{186359}{27} \zeta_2 \zeta_3
          + \frac{268843}{216} \zeta_2^2
          - \frac{9104}{45} \zeta_2^2 \zeta_3
          - \frac{1237}{15} \zeta_2^3
          \bigg\}

       + C_A^2 C_F^2   \bigg\{
           \frac{4432795339}{472392}
          - \frac{7184}{45} \zeta_{5,3}
          + \frac{95771}{9} \zeta_7
          + \frac{14489696}{243} \zeta_5
          + \frac{261351839}{6561} \zeta_3
          - \frac{7567}{9} \zeta_3 \zeta_5
          - \frac{49177675}{1458} \zeta_3^2
          - \frac{462645803}{13122} \zeta_2
          - \frac{653282}{135} \zeta_2 \zeta_5
          + \frac{40089554}{729} \zeta_2 \zeta_3
          - \frac{7018}{27} \zeta_2 \zeta_3^2
          - \frac{17427029}{7290} \zeta_2^2
          - \frac{1030139}{81} \zeta_2^2 \zeta_3
          + \frac{117904}{315} \zeta_2^3
          + \frac{17432749}{15750} \zeta_2^4
          \bigg\}

       + C_A^3 C_F   \bigg\{- \frac{634801943}{52488}
          + \frac{2321}{18} \zeta_7
          - \frac{570359}{30} \zeta_5
          - \frac{92733521}{2916} \zeta_3
          + \frac{32549}{2} \zeta_3^2
          + \frac{164602591}{8748} \zeta_2
          + \frac{19976}{3} \zeta_2 \zeta_5
          - \frac{725170}{27} \zeta_2 \zeta_3
          - 32 \zeta_2 \zeta_3^2
          - \frac{8169839}{1620} \zeta_2^2
          + \frac{113432}{45} \zeta_2^2 \zeta_3
          + \frac{112079}{140} \zeta_2^3
          - \frac{40064}{105} \zeta_2^4 \bigg\} + \frac{1}{4} \g_4^{b,(1)} + n_f \bigg\{ \frac{1}{4} \chi^b_1  \bigg\} 
          + n_f^0 \bigg\{ \frac{1}{4} \chi^b_2 \bigg\}\bigg]  + \delta(\zo) \overline{\cal D}_0 \bigg[ n_f \frac{d_F^{abcd} d_F^{abcd}}{N_F}   
          \bigg\{384 + 2  b^q_{4,d_F^{abcd}d_F^{abcd}}
          + \frac{21760}{9} \zeta_5 + \frac{5312}{9} \zeta_3
          - \frac{1216}{3} \zeta_3^2- \frac{4544}{3} \zeta_2
          - 128 \zeta_2 \zeta_3 + \frac{320}{3} \zeta_2^2
          - \frac{9472}{315} \zeta_2^3  \bigg\}

       + \frac{d_F^{abcd} d_A^{abcd}}{N_F}   \bigg\{ - f^q_{4, d_F^{abcd} d_A^{abcd}} \bigg\}

       + C_F n_f^3   \bigg\{
           \frac{5216}{2187}
          + \frac{80}{81} \zeta_3
          - \frac{800}{81} \zeta_2
          + \frac{16}{9} \zeta_2^2
          \bigg\}

       + C_F^2 n_f^2   \bigg\{
          - \frac{309953}{1458}
          + \frac{1168}{9} \zeta_5
          + \frac{8168}{27} \zeta_3
          + \frac{43552}{729} \zeta_2
          - \frac{6272}{27} \zeta_2 \zeta_3
          + \frac{928}{27} \zeta_2^2
          \bigg\}

       + C_F^3 n_f   \bigg\{ - \frac{48157}{108}
          + 2 b^q_{4,n_f C_F^3}
          - \frac{7712}{3} \zeta_5
          - \frac{2368}{9} \zeta_3
          + 1168 \zeta_3^2
          - \frac{292}{27} \zeta_2
          + \frac{6880}{9} \zeta_2 \zeta_3
          + \frac{2012}{45} \zeta_2^2
          + \frac{117344}{315} \zeta_2^3
          \bigg\}

       + C_F^4   \bigg\{ 7680 \zeta_7
          - 1536 \zeta_5
          + 512 \zeta_3
          - 1920 \zeta_3^2
          - 4608 \zeta_2 \zeta_5
          + 1536 \zeta_2 \zeta_3
          - \frac{6656}{5} \zeta_2^2 \zeta_3  \bigg\}

       + C_A C_F n_f^2   \bigg\{ - \frac{898033}{5832}
          + \frac{304}{3} \zeta_5
          + \frac{2456}{81} \zeta_3
          + \frac{75718}{243} \zeta_2
          + \frac{80}{9} \zeta_2 \zeta_3
          - \frac{3104}{45} \zeta_2^2
          \bigg\}

       + C_A C_F^2 n_f   \bigg\{\frac{5590667}{2916}
          + 2 b^q_{4,n_f C_F^2 C_A}
          - \frac{104}{3} \zeta_5
          - \frac{63782}{27} \zeta_3
          + \frac{4072}{3} \zeta_3^2
          + \frac{146491}{729} \zeta_2
          + \frac{15760}{9} \zeta_2 \zeta_3
          - \frac{516976}{405} \zeta_2^2
          - \frac{12976}{105} \zeta_2^3
          \bigg\}

       + C_A C_F^3   \bigg\{
          - \frac{12928}{27}
          + 8576 \zeta_5
          + 928 \zeta_3
          - 6160 \zeta_3^2
          - \frac{6592}{27} \zeta_2
          - 2304 \zeta_2 \zeta_5
          - \frac{34736}{9} \zeta_2 \zeta_3
          + \frac{23488}{45} \zeta_2^2
          + \frac{4672}{5} \zeta_2^2 \zeta_3
          - \frac{704}{5} \zeta_2^3
          \bigg\}

       + C_A^2 C_F n_f   \bigg\{
           \frac{10761379}{5832}
          - \frac{1}{2} b^q_{4,n_f C_F^3}
          - b^q_{4, n_f C_F^2 C_A}
          - \frac{1}{24} b^q_{4,d_F^{abcd}d_F^{abcd}}
          - \frac{14776}{27} \zeta_5
          - \frac{210778}{81} \zeta_3
          - \frac{4868}{9} \zeta_3^2
          - \frac{654449}{243} \zeta_2
          + \frac{6640}{9} \zeta_2 \zeta_3
          + \frac{138808}{135} \zeta_2^2
          - \frac{65192}{945} \zeta_2^3
          \bigg\}

       + C_A^2 C_F^2   \bigg\{
           \frac{785732}{729}
          + \frac{37888}{9} \zeta_5
          + \frac{176600}{27} \zeta_3
          - 3040 \zeta_3^2
          - \frac{1902716}{729} \zeta_2
          - 768 \zeta_2 \zeta_5
          - \frac{136144}{27} \zeta_2 \zeta_3
          + \frac{529936}{405} \zeta_2^2
          + \frac{1840}{3} \zeta_2^2 \zeta_3
          - \frac{1232}{15} \zeta_2^3
          \bigg\}

       + C_A^3 C_F   \bigg\{
          - \frac{28325071}{4374}
          + \frac{1}{24} f^q_{4,d_F^{abcd} d_A^{abcd}}
          + 1700 \zeta_7
          - \frac{24920}{9} \zeta_5
          + 9600 \zeta_3
          - \frac{2332}{3} \zeta_3^2
          + \frac{1642195}{243} \zeta_2
          + 416 \zeta_2 \zeta_5
          - \frac{30772}{9} \zeta_2 \zeta_3
          - \frac{78188}{45} \zeta_2^2
          + 288 \zeta_2^2 \zeta_3
          + \frac{45188}{315} \zeta_2^3
          \bigg\} \bigg]

           + \delta(\zo) \overline{\cal D}_1 \bigg[ n_f \frac{d_F^{abcd} d_F^{abcd}}{N_F}   \bigg\{- \frac{1280}{3} \zeta_5
          - \frac{256}{3} \zeta_3 + 256 \zeta_2 \bigg\}

       + \frac{d_F^{abcd} d_A^{abcd}}{N_F}   \bigg\{
          \frac{3520}{3} \zeta_5
          + \frac{128}{3} \zeta_3
          - 384 \zeta_3^2
          - 128 \zeta_2
          - \frac{7936}{35} \zeta_2^3
          \bigg\}

       + C_F n_f^3   \bigg\{ - \frac{4000}{729} + \frac{320}{27} \zeta_2
          \bigg\}

       + C_F^2 n_f^2   \bigg\{\frac{123010}{729}
          - \frac{30112}{81} \zeta_3
          - \frac{4720}{81} \zeta_2
          + \frac{5632}{135} \zeta_2^2
          \bigg\}

       + C_F^3 n_f   \bigg\{
           108
          + \frac{10816}{9} \zeta_5
          + \frac{60560}{27} \zeta_3
          - \frac{14624}{27} \zeta_2
          - \frac{22400}{9} \zeta_2 \zeta_3
          + \frac{58208}{135} \zeta_2^2
          \bigg\}

       + C_F^4   \bigg\{\frac{4312}{3}
          + 3392 \zeta_5
          - 4752 \zeta_3
          + \frac{18304}{3} \zeta_3^2
          - \frac{2968}{3} \zeta_2
          + 704 \zeta_2 \zeta_3
          - \frac{1056}{5} \zeta_2^2
          + \frac{97408}{315} \zeta_2^3
          \bigg\}

       + C_A C_F n_f^2   \bigg\{
           \frac{58045}{243}
          - \frac{208}{9} \zeta_3
          - \frac{7616}{27} \zeta_2
          + \frac{64}{5} \zeta_2^2
          \bigg\}

       + C_A C_F^2 n_f   \bigg\{
          - \frac{1243577}{729}
          + 128 \zeta_5
          + \frac{281600}{81} \zeta_3
          + \frac{2416}{3} \zeta_2
          - \frac{1408}{3} \zeta_2 \zeta_3
          - \frac{64064}{135} \zeta_2^2
          \bigg\}

       + C_A C_F^3   \bigg\{
          - \frac{7496}{9}
          - \frac{102112}{9} \zeta_5
          - \frac{180032}{27} \zeta_3
          + \frac{15808}{3} \zeta_3^2
          + \frac{102728}{27} \zeta_2
          + \frac{112736}{9} \zeta_2 \zeta_3
          - \frac{340112}{135} \zeta_2^2
          - \frac{71168}{315} \zeta_2^3
          \bigg\}

       + C_A^2 C_F n_f   \bigg\{- \frac{571387}{243}
          - \frac{1360}{9} \zeta_5
          + \frac{3128}{3} \zeta_3
          + \frac{58000}{27} \zeta_2
          + 32 \zeta_2 \zeta_3
          - \frac{3872}{15} \zeta_2^2
          \bigg\}

       + C_A^2 C_F^2   \bigg\{ \frac{2420680}{729}
          - 336 \zeta_5
          - \frac{638128}{81} \zeta_3
          + \frac{752}{3} \zeta_3^2
          - \frac{178496}{81} \zeta_2
          + 1696 \zeta_2 \zeta_3
          + \frac{151312}{135} \zeta_2^2
          + \frac{38944}{315} \zeta_2^3
          \bigg\}

       + C_A^3 C_F   \bigg\{
           \frac{4520317}{729}
          + \frac{15400}{9} \zeta_5
          - \frac{51032}{9} \zeta_3
          - 16 \zeta_3^2
          - \frac{140200}{27} \zeta_2
          + 528 \zeta_2 \zeta_3
          + \frac{3520}{3} \zeta_2^2
          - \frac{20032}{105} \zeta_2^3
          \bigg\} \bigg]

       + {\cal D}_0 \overline{\cal D}_0 \bigg[ n_f \frac{d_F^{abcd} d_F^{abcd}}{N_F}   \bigg\{
          - \frac{640}{3} \zeta_5
          - \frac{128}{3} \zeta_3
          + 128 \zeta_2
          \bigg\}

       + \frac{d_F^{abcd} d_A^{abcd}}{N_F}   \bigg\{
           \frac{1760}{3} \zeta_5
          + \frac{64}{3} \zeta_3
          - 192 \zeta_3^2
          - 64 \zeta_2
          - \frac{3968}{35} \zeta_2^3
          \bigg\}

       + C_F n_f^3   \bigg\{
          - \frac{2000}{729}
          + \frac{160}{27} \zeta_2
          \bigg\}

       + C_F^2 n_f^2   \bigg\{
           \frac{61505}{729}
          - \frac{15056}{81} \zeta_3
          - \frac{2360}{81} \zeta_2
          + \frac{2816}{135} \zeta_2^2
          \bigg\}

       + C_F^3 n_f   \bigg\{
           54
          + \frac{5408}{9} \zeta_5
          + \frac{30280}{27} \zeta_3
          - \frac{7312}{27} \zeta_2
          - \frac{11200}{9} \zeta_2 \zeta_3
          + \frac{29104}{135} \zeta_2^2
          \bigg\}

       + C_F^4   \bigg\{ \frac{2156}{3}
          + 1696 \zeta_5
          - 2376 \zeta_3
          + \frac{9152}{3} \zeta_3^2
          - \frac{1484}{3} \zeta_2
          + 352 \zeta_2 \zeta_3
          - \frac{528}{5} \zeta_2^2
          + \frac{48704}{315} \zeta_2^3
          \bigg\}

       + C_A C_F n_f^2   \bigg\{
           \frac{58045}{486}
          - \frac{104}{9} \zeta_3
          - \frac{3808}{27} \zeta_2
          + \frac{32}{5} \zeta_2^2
          \bigg\}

       + C_A C_F^2 n_f   \bigg\{- \frac{1243577}{1458}
          + 64 \zeta_5
          + \frac{140800}{81} \zeta_3
          + \frac{1208}{3} \zeta_2
          - \frac{704}{3} \zeta_2 \zeta_3
          - \frac{32032}{135} \zeta_2^2
          \bigg\}

       + C_A C_F^3   \bigg\{- \frac{3748}{9}
          - \frac{51056}{9} \zeta_5
          - \frac{90016}{27} \zeta_3
          + \frac{7904}{3} \zeta_3^2
          + \frac{51364}{27} \zeta_2
          + \frac{56368}{9} \zeta_2 \zeta_3
          - \frac{170056}{135} \zeta_2^2
          - \frac{35584}{315} \zeta_2^3
          \bigg\}

       + C_A^2 C_F n_f   \bigg\{
          - \frac{571387}{486}
          - \frac{680}{9} \zeta_5
          + \frac{1564}{3} \zeta_3
          + \frac{29000}{27} \zeta_2
          + 16 \zeta_2 \zeta_3
          - \frac{1936}{15} \zeta_2^2
          \bigg\}

       + C_A^2 C_F^2   \bigg\{
          \frac{1210340}{729}
          - 168 \zeta_5
          - \frac{319064}{81} \zeta_3
          + \frac{376}{3} \zeta_3^2
          - \frac{89248}{81} \zeta_2
          + 848 \zeta_2 \zeta_3
          + \frac{75656}{135} \zeta_2^2
          + \frac{19472}{315} \zeta_2^3
          \bigg\}

       + C_A^3 C_F   \bigg\{
           \frac{4520317}{1458}
          + \frac{7700}{9} \zeta_5
          - \frac{25516}{9} \zeta_3
          - 8 \zeta_3^2
          - \frac{70100}{27} \zeta_2
          + 264 \zeta_2 \zeta_3
          + \frac{1760}{3} \zeta_2^2
          - \frac{10016}{105} \zeta_2^3
          \bigg\} \bigg] + \bigg\{ z_1 \leftrightarrow z_2 \bigg\} .
\end{autobreak}
\end{align}

As a  consequence of \eqref{eq:mellin-rapidity-Xsec}, the finite coefficients $\g^{I,(j)}_{k}$ which contribute to the soft part of the partonic cross-section are related to the coefficients  $\g^{I,(j)}_{d,k}$ in rapidity distribution and the relations are provided in terms of the anomalous dimensions in the Online Resource files (For further details, see Eq.~(44) and Eq.~(34) in  \cite{Ravindran:2005vv} and \cite{Ravindran:2006bu}, respectively.). The symbols $\chi^q_j$ and $\chi^g_j$ denote the unknown coefficients of the color factors in four loop form factors of the Drell-Yan and of the Higgs boson production through gluon fusion, respectively. For the case of Higgs boson production in bottom quark annihilation, only the $n_f^3$ and $n_f^2$ contributions to the four loop form factor are available in the literature~\cite{Lee:2017mip}. As a result, the unknown coefficients corresponding to $O(n_f)$ and $O(n_f^0)$ color factors are denoted by $\chi^b_1$ and $\chi^b_2$, respectively. Also the symbols $f^q_{4,d_F^{abcd} d_A^{abcd}}$ and $b^q_{4,j}$, where $j = \big\{d_F^{abcd} d_F^{abcd}, n_f C_F^3, \nonumber\\
n_f C_F^2 C_A, d_F^{abcd} d_A^{abcd}, C_F^2 C_A^2, C_F^3 C_A, C_F^4 \big\}$ are the unknown coefficients of the color factors in four loop soft and collinear anomalous dimensions.
In the aforementioned equations, $n_{fv}$ is proportional to the charge weighted sum of the quark flavours and $N_4 = (n_c^2-4)/n_c$ \cite{Gehrmann:2010ue}.
Following \cite{Henn:2019swt}, we have
\begin{align}
  \frac{d_{A}^{abcd}d_{A}^{abcd}}{N_{A}} &= \frac{n_c^2 (n_c^2+36)}{24} \,,\quad
 \frac{d_{F}^{abcd}d_{A}^{abcd}}{N_{A}} =\frac{n_c (n_c^2+6)}{48} \,,\nonumber\\& \frac{d_{F}^{abcd}d_{F}^{abcd}}{N_{A}} = \frac{n_c^4-6 n_c^2+18}{96 n_c^2} \,,\nn\\&
 C_{A} = n_c \,,\quad C_{F} =\frac{n_c^2-1}{2 n_c} \,,\quad N_{A} = n_c^2-1 \,,\nn\\&N_{F} = n_c . 
 \end{align}

\section{Explicit results at N$^3$LL for Drell-Yan and the Higgs boson production}
\label{secE:N3LL}
Here, we provide the full explicit result of the resummation constant  ${ g}^I_{d,4}$, given in \eqref{eq:GNbexp}, at N$^{3}$LL for the Drell-Yan and the Higgs boson productions.
 \begin{align}
\begin{autobreak}

{ g}^I_{d,4} = 

          \frac{1}{(1-\omega)^2} \bigg[ C_R \bigg[ \frac{\beta_1^3}{\beta_0^5}   \bigg\{
           \frac{4}{3}  \omega^3
          + 2  L_\omega \omega^2
          - \frac{2}{3}  L_\omega^3
          \bigg\}

       + \frac{\beta_1 \beta_2}{\beta_0^4}   \bigg\{
          - 2  \omega
          + 3  \omega^2
          - \frac{8}{3}  \omega^3
          - 2  L_\omega
          + 4  L_\omega \omega
          - 4  L_\omega \omega^2
          \bigg\}

       + \frac{\beta_1^2}{\beta_0^4} n_f  \bigg\{
          - \frac{20}{9}  \omega
          + \frac{10}{9}  \omega^2
          + \frac{40}{27}  \omega^3
          - \frac{20}{9}  L_\omega
          - \frac{20}{9}  L_\omega^2
          \bigg\}

       + \frac{\beta_1^2}{\beta_0^4} C_A   \bigg\{ \frac{134}{9}  \omega
          - \frac{67}{9}  \omega^2
          - \frac{268}{27}  \omega^3
          + \frac{134}{9}  L_\omega
          + \frac{134}{9} L_\omega^2
          - 4 \zeta_2 \omega
          + 2  \zeta_2 \omega^2
          + \frac{8}{3}  \zeta_2 \omega^3
          - 4  \zeta_2 L_\omega
          - 4  \zeta_2 L_\omega^2 \bigg\}

       + \frac{\beta_3}{\beta_0^3} \bigg\{ 2 \omega
          - 3  \omega^2
          + \frac{4}{3}  \omega^3
          + 2  L_\omega
          - 4  L_\omega \omega
          + 2  L_\omega \omega^2
          \bigg\}

       + \frac{\beta_2}{\beta_0^3} n_f   \bigg\{- \frac{40}{27}  \omega^3 \bigg\}

       + \frac{\beta_1}{\beta_0^3} n_f^2    \bigg\{
           \frac{8}{27}  \omega
          + \frac{4}{27}  \omega^2
          - \frac{16}{81}  \omega^3
          + \frac{8}{27}  L_\omega
          \bigg\}

       + \frac{\beta_1}{\beta_0^3} C_F n_f   \bigg\{
           \frac{55}{3}  \omega
          + \frac{55}{6}  \omega^2
          - \frac{110}{9} \omega^3
          + \frac{55}{3}  L_\omega
          - 16  \zeta_3 \omega
          - 8  \zeta_3 \omega^2
          + \frac{32}{3}  \zeta_3 \omega^3
          - 16  \zeta_3 L_\omega
          \bigg\}

       + \frac{\beta_2}{\beta_0^3} C_A    \bigg\{
           \frac{268}{27}  \omega^3
          - \frac{8}{3}  \zeta_2 \omega^3
          \bigg\}

       + \frac{\beta_1}{\beta_0^3} C_A n_f  \bigg\{
           \frac{418}{27}  \omega
          + \frac{209}{27}  \omega^2
          - \frac{836}{81}  \omega^3
          + \frac{418}{27}  L_\omega
          + \frac{56}{3}  \zeta_3 \omega
          + \frac{28}{3}  \zeta_3 \omega^2
          - \frac{112}{9}  \zeta_3 \omega^3
          + \frac{56}{3}  \zeta_3 L_\omega
          - \frac{80}{9}  \zeta_2 \omega
          - \frac{40}{9}  \zeta_2 \omega^2
          + \frac{160}{27}  \zeta_2 \omega^3
          - \frac{80}{9}  \zeta_2 L_\omega
          \bigg\}

       + \frac{\beta_1}{\beta_0^3} C_A^2   \bigg\{
          - \frac{245}{3}  \omega
          - \frac{245}{6}  \omega^2
          + \frac{490}{9}  \omega^3
          - \frac{245}{3}  L_\omega
          - 44  \zeta_4 \omega
          - 22  \zeta_4 \omega^2
          + \frac{88}{3}  \zeta_4 \omega^3
          - 44  \zeta_4 L_\omega
          - \frac{44}{3}  \zeta_3 \omega
          - \frac{22}{3}  \zeta_3 \omega^2
          + \frac{88}{9}  \zeta_3 \omega^3
          - \frac{44}{3}  \zeta_3 L_\omega
          + \frac{536}{9}  \zeta_2 \omega
          + \frac{268}{9}  \zeta_2 \omega^2
          - \frac{1072}{27}  \zeta_2 \omega^3
          + \frac{536}{9}  \zeta_2 L_\omega
          \bigg\}

       + \frac{\beta_1}{\beta_0^2} n_f   \bigg\{\frac{112}{27}  \omega
          - \frac{56}{27}  \omega^2
          + \frac{112}{27}  L_\omega
          - \frac{4}{3}  \zeta_2 \omega
          + \frac{2}{3}  \zeta_2 \omega^2
          - \frac{4}{3}  \zeta_2 L_\omega \bigg\}

       + \frac{1}{\beta_0^2} n_f^3    \bigg\{
          - \frac{16}{81} \omega^2
          + \frac{32}{243} \omega^3
          + \frac{32}{27} \zeta_3 \omega^2
          - \frac{64}{81} \zeta_3 \omega^3
          \bigg\}

       + \frac{1}{\beta_0^2} C_F n_f^2   \bigg\{
           \frac{1196}{81} \omega^2
          - \frac{2392}{243} \omega^3
          + 16 \zeta_4 \omega^2
          - \frac{32}{3} \zeta_4 \omega^3
          - \frac{320}{9} \zeta_3 \omega^2
          + \frac{640}{27} \zeta_3 \omega^3
          \bigg\}

       + \frac{1}{\beta_0^2} C_F^2 n_f    \bigg\{
           \frac{286}{9} \omega^2
          - \frac{572}{27} \omega^3
          - 160 \zeta_5 \omega^2
          + \frac{320}{3} \zeta_5 \omega^3
          + \frac{296}{3} \zeta_3 \omega^2
          - \frac{592}{9} \zeta_3 \omega^3
          \bigg\}

       + \frac{1}{\beta_0^2} C_A   \bigg\{
          - \frac{808}{27} \beta_1 \omega
          + \frac{404}{27} \beta_1 \omega^2
          - \frac{808}{27} \beta_1 L_\omega
          + 28 \beta_1 \zeta_3 \omega
          - 14 \beta_1 \zeta_3 \omega^2
          + 28 \beta_1 \zeta_3 L_\omega
          + \frac{22}{3} \beta_1 \zeta_2 \omega
          - \frac{11}{3} \beta_1 \zeta_2 \omega^2
          + \frac{22}{3} \beta_1 \zeta_2 L_\omega
          \bigg\}

       + \frac{1}{\beta_0^2} C_A n_f^2    \bigg\{
           \frac{923}{162} \omega^2
          - \frac{923}{243} \omega^3
          - \frac{56}{3} \zeta_4 \omega^2
          + \frac{112}{9} \zeta_4 \omega^3
          + \frac{1120}{27} \zeta_3 \omega^2
          - \frac{2240}{81} \zeta_3 \omega^3
          - \frac{304}{81} \zeta_2 \omega^2
          + \frac{608}{243} \zeta_2 \omega^3
          \bigg\}

       + \frac{1}{\beta_0^2} C_A C_F n_f    \bigg\{
          - \frac{17033}{81} \omega^2
          + \frac{34066}{243} \omega^3
          + 80 \zeta_5 \omega^2
          - \frac{160}{3} \zeta_5 \omega^3
          - 88 \zeta_4 \omega^2
          + \frac{176}{3} \zeta_4 \omega^3
          + \frac{1856}{9} \zeta_3 \omega^2
          - \frac{3712}{27} \zeta_3 \omega^3
          + \frac{220}{3} \zeta_2 \omega^2
          - \frac{440}{9} \zeta_2 \omega^3
          - 64 \zeta_2 \zeta_3 \omega^2
          + \frac{128}{3} \zeta_2 \zeta_3 \omega^3
          \bigg\}

       + \frac{1}{\beta_0^2} C_A^2 n_f   \bigg\{
          - \frac{24137}{162} \omega^2
          + \frac{24137}{243} \omega^3
          + \frac{1048}{9} \zeta_5 \omega^2
          - \frac{2096}{27} \zeta_5 \omega^3
          - \frac{88}{3} \zeta_4 \omega^2
          + \frac{176}{9} \zeta_4 \omega^3
          - \frac{11552}{27} \zeta_3 \omega^2
          + \frac{23104}{81} \zeta_3 \omega^3
          + \frac{10160}{81} \zeta_2 \omega^2
          - \frac{20320}{243} \zeta_2 \omega^3
          + \frac{224}{3} \zeta_2 \zeta_3 \omega^2
          - \frac{448}{9} \zeta_2 \zeta_3 \omega^3
          \bigg\}

       + \frac{1}{\beta_0^2} C_A^3   \bigg\{
           \frac{42139}{81} \omega^2
          - \frac{84278}{243} \omega^3
          - \frac{1804}{9} \zeta_5 \omega^2
          + \frac{3608}{27} \zeta_5 \omega^3
          + 902 \zeta_4 \omega^2
          - \frac{1804}{3} \zeta_4 \omega^3
          + \frac{10472}{27} \zeta_3 \omega^2
          - \frac{20944}{81} \zeta_3 \omega^3
          - 8 \zeta_3^2 \omega^2
          + \frac{16}{3} \zeta_3^2 \omega^3
          - \frac{44200}{81} \zeta_2 \omega^2
          + \frac{88400}{243} \zeta_2 \omega^3
          - \frac{5008}{21} \zeta_2 \zeta_4 \omega^2
          + \frac{10016}{63} \zeta_2 \zeta_4 \omega^3
          - \frac{176}{3} \zeta_2 \zeta_3 \omega^2
          + \frac{352}{9} \zeta_2 \zeta_3 \omega^3
          \bigg\}

       + \frac{\beta_1}{\beta_0}   \bigg\{
          - 2  \zeta_2 L_\omega
          \bigg\}

       + \frac{1}{\beta_0} n_f^2   \bigg\{
          - \frac{2080}{729} \omega
          + \frac{1040}{729} \omega^2
          + \frac{112}{27} \zeta_3 \omega
          - \frac{56}{27} \zeta_3 \omega^2
          - \frac{40}{27} \zeta_2 \omega
          + \frac{20}{27} \zeta_2 \omega^2
          \bigg\}

       + \frac{1}{\beta_0} C_F n_f    \bigg\{
          - \frac{1711}{27} \omega
          + \frac{1711}{54} \omega^2
          + 16 \zeta_4 \omega
          - 8 \zeta_4 \omega^2
          + \frac{304}{9} \zeta_3 \omega
          - \frac{152}{9} \zeta_3 \omega^2
          + 4 \zeta_2 \omega
          - 2 \zeta_2 \omega^2
          \bigg\}

       + \frac{1}{\beta_0} C_A n_f    \bigg\{
          - \frac{11842}{729} \omega
          + \frac{5921}{729} \omega^2
          - 48 \zeta_4 \omega
          + 24 \zeta_4 \omega^2
          + \frac{728}{27} \zeta_3 \omega
          - \frac{364}{27} \zeta_3 \omega^2
          + \frac{2828}{81} \zeta_2 \omega
          - \frac{1414}{81} \zeta_2 \omega^2
          \bigg\}

       + \frac{1}{\beta_0} C_A^2   \bigg\{
           \frac{136781}{729} \omega
          - \frac{136781}{1458} \omega^2
          + 192 \zeta_5 \omega
          - 96 \zeta_5 \omega^2
          + 176 \zeta_4 \omega
          - 88 \zeta_4 \omega^2
          - \frac{1316}{3} \zeta_3 \omega
          + \frac{658}{3} \zeta_3 \omega^2
          - \frac{12650}{81} \zeta_2 \omega
          + \frac{6325}{81} \zeta_2 \omega^2
          + \frac{176}{3} \zeta_2 \zeta_3 \omega
          - \frac{88}{3} \zeta_2 \zeta_3 \omega^2
          \bigg\}

       + n_f   \bigg\{
          - \frac{656}{81} \omega
          + \frac{328}{81} \omega^2
          + \frac{64}{3} \zeta_3 \omega
          - \frac{32}{3} \zeta_3 \omega^2
          - \frac{20}{9} \zeta_2 \omega
          + \frac{10}{9} \zeta_2 \omega^2\bigg\}

       + C_A   \bigg\{
           \frac{4856}{81} \omega
          - \frac{2428}{81} \omega^2
          - 60 \zeta_4 \omega
          + 30 \zeta_4 \omega^2
          - \frac{352}{3} \zeta_3 \omega
          + \frac{176}{3} \zeta_3 \omega^2
          + \frac{134}{9} \zeta_2 \omega
          - \frac{67}{9} \zeta_2 \omega^2
          \bigg\}

       + \beta_0   \bigg\{
           \frac{92}{3} \zeta_3 \omega
          - \frac{46}{3} \zeta_3 \omega^2
          \bigg\} \bigg]

           + \frac{1}{\beta_0^2}   {n_f d_R^{abcd} d_F^{abcd} \over N_R} \bigg\{
          - \frac{640}{3} \zeta_5 \omega^2
          + \frac{1280}{9} \zeta_5 \omega^3
          - \frac{128}{3} \zeta_3 \omega^2
          + \frac{256}{9} \zeta_3 \omega^3
          + 128 \zeta_2 \omega^2  - \frac{256}{3} \zeta_2 \omega^3 \bigg\} 
          
          + \frac{1}{\beta_0^2}  {d_R^{abcd} d_A^{abcd} \over N_R} \bigg\{ \frac{1760}{3} \zeta_5 \omega^2
          - \frac{3520}{9} \zeta_5 \omega^3
          + \frac{64}{3} \zeta_3 \omega^2
          - \frac{128}{9} \zeta_3 \omega^3
          - 192 \zeta_3^2 \omega^2
          + 128 \zeta_3^2 \omega^3
          - 64 \zeta_2 \omega^2
          + \frac{128}{3} \zeta_2 \omega^3
          - \frac{1984}{7} \zeta_2 \zeta_4 \omega^2
          + \frac{3968}{21} \zeta_2 \zeta_4 \omega^3 \bigg\} \bigg].
\end{autobreak}
\end{align}
In the aforementioned equation, $L_\omega = \ln(1-\omega)$, $R = A$ for gluons($I=g$) and $R = F$ for quarks($I=q,b$). 
 The general results of the resummation constants in terms of universal quantities are presented to N$^3$LL accuracy in the Online Resource file supplied with this article.

\section{NSV partonic rapidity distribution}
\label{secF:NSVDelta}
We expand the $\Delta^{\rm nsv}_{d,I}$, as defined in \eqref{eq:SVNSV-master-formula-recast}, in powers of $a_s(\mu_R^2)$ through
\begin{align}
\label{eqA:Sof-Coll-Expand}
    \Delta^{\rm nsv}_{d,I}\left(\{p_j\cdot q_k\},z_1,z_2,q^2,\mu_F^2\right) &=  \sum_{i=1}^{\infty} a_s^{i}(\mu_R^2) \Delta^{{\rm nsv},(i)}_{d,I} .
\end{align}
Here, we present $\Delta^{\rm nsv}_{d,I}$ to N$^2$LO for the specific scale choice $\mu_F^2=\mu_R^2=q^2$. 

\begin{align}
\begin{autobreak}

\end{autobreak}\nonumber\\
\begin{autobreak}
    
   \Delta^{\rm{nsv},(1)}_{d,I} =
        
          {\cal D}_0 L_{z_2}   \bigg\{
           C_1^I |{\cal M}^{(0)}_{I,\text{fin}}|^{2}
          \bigg\}
          + {\cal D}_0    \bigg\{
          |{\cal M}^{(0)}_{I,\text{fin}}|^{2} D_1^I
          \bigg\}

       + \delta(\overline{z}_1) L^2_{z_2}   \bigg\{
           C_1^I |{\cal M}^{(0)}_{I,\text{fin}}|^{2}
          \bigg\}

       + \delta(\overline{z}_1) L_{z_2}   \bigg\{
           |{\cal M}^{(0)}_{I,\text{fin}}|^{2} D_1^I
          + \varphi^{I,(1)}_{d,1} |{\cal M}^{(0)}_{I,\text{fin}}|^{2}
          \bigg\}

       + \delta(\overline{z}_1)   \bigg\{ \varphi^{I,(0)}_{d,1} |{\cal M}^{(0)}_{I,\text{fin}}|^{2} \bigg\}  + (z_1 \leftrightarrow z_2) \,,

\end{autobreak}\nonumber\\
\begin{autobreak}
   \Delta^{\rm{nsv},(2)}_{d,I} =
           
         L_{z_2} {\cal D}_2   \bigg\{
           \frac{3}{2} A^I_1 C_1^I |{\cal M}^{(0)}_{I,\text{fin}}|^{2}
          \bigg\}

       + {\cal D}_1 L^2_{z_2}   \bigg\{
           3 A^I_1 C_1^I |{\cal M}^{(0)}_{I,\text{fin}}|^{2}
          \bigg\}

       + {\cal D}_1 L_{z_2}   \bigg\{ - 2 f^I_1 C_1^I |{\cal M}^{(0)}_{I,\text{fin}}|^{2}
          + 3 A^I_1 |{\cal M}^{(0)}_{I,\text{fin}}|^{2} D_1^I
          + A^I_1 \varphi^{I,(1)}_{d,1} |{\cal M}^{(0)}_{I,\text{fin}}|^{2}
          - \beta_0 C_1^I |{\cal M}^{(0)}_{I,\text{fin}}|^{2}
          \bigg\}

       +  {\cal D}_2   \bigg\{
           \frac{3}{2} A^I_1 |{\cal M}^{(0)}_{I,\text{fin}}|^{2} D_1^I
          \bigg\}

       +  {\cal D}_1   \bigg\{
          - 2 f^I_1 |{\cal M}^{(0)}_{I,\text{fin}}|^{2} D_1^I
          + A^I_1 \varphi^{I,(0)}_{d,1} |{\cal M}^{(0)}_{I,\text{fin}}|^{2}
          - 2 A^I_1 \zeta_2 C_1^I |{\cal M}^{(0)}_{I,\text{fin}}|^{2}
          + (A^I_1)^2 |{\cal M}^{(0)}_{I,\text{fin}}|^{2}
          - \beta_0 |{\cal M}^{(0)}_{I,\text{fin}}|^{2} D_1^I
          \bigg\}

       + {\cal D}_0 L^3_{z_2}   \bigg\{
           \frac{3}{2} A^I_1 C_1^I |{\cal M}^{(0)}_{I,\text{fin}}|^{2}
          \bigg\}

       + {\cal D}_0 L^2_{z_2}   \bigg\{
          - 2 f^I_1 C_1^I |{\cal M}^{(0)}_{I,\text{fin}}|^{2}
          + \frac{3}{2} A^I_1 |{\cal M}^{(0)}_{I,\text{fin}}|^{2} D_1^I
          + A^I_1 \varphi^{I,(1)}_{d,1} |{\cal M}^{(0)}_{I,\text{fin}}|^{2}
          - \beta_0 C_1^I |{\cal M}^{(0)}_{I,\text{fin}}|^{2}
          \bigg\}
 + {\cal D}_0 L_{z_2}   \bigg\{
           C_2^I |{\cal M}^{(0)}_{I,\text{fin}}|^{2}
          + 2 C_1^I {\cal M}^{(0,1)}_{I,\text{fin}}
          + 2 C_1^I \g^{I,(1)}_{d,1} |{\cal M}^{(0)}_{I,\text{fin}}|^{2}
          - 2 f^I_1 |{\cal M}^{(0)}_{I,\text{fin}}|^{2} D_1^I
          - f^I_1 \varphi^{I,(1)}_{d,1} |{\cal M}^{(0)}_{I,\text{fin}}|^{2}
          + A^I_1 \varphi^{I,(0)}_{d,1} |{\cal M}^{(0)}_{I,\text{fin}}|^{2}
          - 4 A^I_1 \zeta_2 C_1^I |{\cal M}^{(0)}_{I,\text{fin}}|^{2}
          + (A^I_1)^2 |{\cal M}^{(0)}_{I,\text{fin}}|^{2}
          - \beta_0 |{\cal M}^{(0)}_{I,\text{fin}}|^{2} D_1^I
          - \beta_0 \varphi^{I,(1)}_{d,1} |{\cal M}^{(0)}_{I,\text{fin}}|^{2}
          \bigg\}

       + {\cal D}_0  \bigg\{
           |{\cal M}^{(0)}_{I,\text{fin}}|^{2} D_2^I
          + 2 {\cal M}^{(0,1)}_{I,\text{fin}} D_1^I
          + 2 \g^{I,(1)}_{d,1} |{\cal M}^{(0)}_{I,\text{fin}}|^{2} D_1^I
          - f^I_1 \varphi^{I,(0)}_{d,1} |{\cal M}^{(0)}_{I,\text{fin}}|^{2}
          + \zeta_2 f^I_1 C_1^I |{\cal M}^{(0)}_{I,\text{fin}}|^{2}
          - A^I_1 f^I_1 |{\cal M}^{(0)}_{I,\text{fin}}|^{2}
          + 3 A^I_1 \zeta_3 C_1^I |{\cal M}^{(0)}_{I,\text{fin}}|^{2}
          - 2 A^I_1 \zeta_2 |{\cal M}^{(0)}_{I,\text{fin}}|^{2} D_1^I
          - A^I_1 \zeta_2 \varphi^{I,(1)}_{d,1} |{\cal M}^{(0)}_{I,\text{fin}}|^{2}
          - \beta_0 \varphi^{I,(0)}_{d,1} |{\cal M}^{(0)}_{I,\text{fin}}|^{2}
          \bigg\}

       + \delta(\overline{z}_1) L^4_{z_2}   \bigg\{
           \frac{1}{2} A^I_1 C_1^I |{\cal M}^{(0)}_{I,\text{fin}}|^{2}
          \bigg\}

       + \delta(\overline{z}_1) L^3_{z_2}   \bigg\{
          - f^I_1 C_1^I |{\cal M}^{(0)}_{I,\text{fin}}|^{2}
          + \frac{1}{2} A^I_1 |{\cal M}^{(0)}_{I,\text{fin}}|^{2} D_1^I
          + \frac{1}{2} A^I_1 \varphi^{I,(1)}_{d,1} |{\cal M}^{(0)}_{I,\text{fin}}|^{2}
          - \frac{1}{2} \beta_0 C_1^I |{\cal M}^{(0)}_{I,\text{fin}}|^{2}
          \bigg\}

       + \delta(\overline{z}_1) L^2_{z_2}   \bigg\{
           \varphi^{I,(2)}_{d,2} |{\cal M}^{(0)}_{I,\text{fin}}|^{2}
          + C_2^I |{\cal M}^{(0)}_{I,\text{fin}}|^{2}
          + 2 C_1^I {\cal M}^{(0,1)}_{I,\text{fin}}
          + 2 C_1^I \g^{I,(1)}_{d,1} |{\cal M}^{(0)}_{I,\text{fin}}|^{2}
          - f^I_1 |{\cal M}^{(0)}_{I,\text{fin}}|^{2} D_1^I
          - f^I_1 \varphi^{I,(1)}_{d,1} |{\cal M}^{(0)}_{I,\text{fin}}|^{2}
          + \frac{1}{2} A^I_1 \varphi^{I,(0)}_{d,1} |{\cal M}^{(0)}_{I,\text{fin}}|^{2}
          - 3 A^I_1 \zeta_2 C_1^I |{\cal M}^{(0)}_{I,\text{fin}}|^{2}
          + \frac{1}{2} (A^I_1)^2 |{\cal M}^{(0)}_{I,\text{fin}}|^{2}
          - \frac{1}{2} \beta_0 |{\cal M}^{(0)}_{I,\text{fin}}|^{2} D_1^I
          - \beta_0 \varphi^{I,(1)}_{d,1} |{\cal M}^{(0)}_{I,\text{fin}}|^{2}
          \bigg\}
       + \delta(\overline{z}_1) L_{z_2}   \bigg\{
           |{\cal M}^{(0)}_{I,\text{fin}}|^{2} D_2^I
          + 2 {\cal M}^{(0,1)}_{I,\text{fin}} D_1^I
          + \varphi^{I,(1)}_{d,2} |{\cal M}^{(0)}_{I,\text{fin}}|^{2}
          + 2 \varphi^{I,(1)}_{d,1} {\cal M}^{(0,1)}_{I,\text{fin}}
          + 2 \g^{I,(1)}_{d,1} |{\cal M}^{(0)}_{I,\text{fin}}|^{2} D_1^I
          + 2 \g^{I,(1)}_{d,1} \varphi^{I,(1)}_{d,1} |{\cal M}^{(0)}_{I,\text{fin}}|^{2}
          - f^I_1 \varphi^{I,(0)}_{d,1} |{\cal M}^{(0)}_{I,\text{fin}}|^{2}
          + 3 \zeta_2 f^I_1 C_1^I |{\cal M}^{(0)}_{I,\text{fin}}|^{2}
          - A^I_1 f^I_1 |{\cal M}^{(0)}_{I,\text{fin}}|^{2}
          + 5 A^I_1 \zeta_3 C_1^I |{\cal M}^{(0)}_{I,\text{fin}}|^{2}
          - 2 A^I_1 \zeta_2 |{\cal M}^{(0)}_{I,\text{fin}}|^{2} D_1^I
          - A^I_1 \zeta_2 \varphi^{I,(1)}_{d,1} |{\cal M}^{(0)}_{I,\text{fin}}|^{2}
          - \beta_0 \varphi^{I,(0)}_{d,1} |{\cal M}^{(0)}_{I,\text{fin}}|^{2}
          \bigg\}

       + \delta(\overline{z}_1)   \bigg\{
           \varphi^{I,(0)}_{d,2} |{\cal M}^{(0)}_{I,\text{fin}}|^{2}
          + 2 \varphi^{I,(0)}_{d,1} {\cal M}^{(0,1)}_{I,\text{fin}}
          + 2 \g^{I,(1)}_{d,1} \varphi^{I,(0)}_{d,1} |{\cal M}^{(0)}_{I,\text{fin}}|^{2}
          + \frac{1}{2} (f^I_1)^2 |{\cal M}^{(0)}_{I,\text{fin}}|^{2}
          - 2 \zeta_3 f^I_1 C_1^I |{\cal M}^{(0)}_{I,\text{fin}}|^{2}
          + 2 \zeta_2 f^I_1 |{\cal M}^{(0)}_{I,\text{fin}}|^{2} D_1^I
          + \zeta_2 f^I_1 \varphi^{I,(1)}_{d,1} |{\cal M}^{(0)}_{I,\text{fin}}|^{2}
          + 2 A^I_1 \zeta_3 |{\cal M}^{(0)}_{I,\text{fin}}|^{2} D_1^I
          + A^I_1 \zeta_3 \varphi^{I,(1)}_{d,1} |{\cal M}^{(0)}_{I,\text{fin}}|^{2}
          + \frac{4}{5} A^I_1 \zeta_2^2 C_1^I |{\cal M}^{(0)}_{I,\text{fin}}|^{2}
          - (A^I_1)^2 \zeta_2 |{\cal M}^{(0)}_{I,\text{fin}}|^{2}
          \bigg\}  + (z_1 \leftrightarrow z_2) \,,

\end{autobreak}
\end{align}
 where $L_{z_2} = \log(1-z_2)$. The results up to N$^4$LO with explicit dependence on $\mu_R^2$ and $\mu_F^2$ are provided in the Online Resource supplied with this article.   
      
\section{Next-to-soft-collinear distribution for rapidity distribution}
\label{secG:SCdistNSV}
In this section, we present soft-collinear distribution  ${\tilde{\textbf{S}}}_{d,I}$, as defined in \eqref{eq:nsoft-collinear-operator},
in powers of $a_s(\mu_R^2)$ up to N$^2$LO.
Expanding the quantity in powers of $a_s$ as
\begin{align}
    \tilde{{\textbf{S}}}_{d,I}(z_1,z_2,q^2,\mu_R^2,\mu_F^2) =  \sum_{i=1}^{\infty} a_s^i(\mu_R^2)\,   \tilde{{\textbf{S}}}^{(i)}_{d,I}(z_1,z_2,q^2,\mu_R^2,\mu_F^2) \,,
\end{align}
we present the results for $\mu_R^2=\mu_F^2=q^2$. 
\begin{align}
\begin{autobreak}
     \tilde{\textbf{S}}_{d,I}^{(1)} =
            {\cal D}_0 L_{z_2}   \bigg\{ C_1^I \bigg\}

       + {\cal D}_0    \bigg\{
           D_1^I
          \bigg\}

       + \delta(\overline{z}_1) L^2_{z_2}   \bigg\{
           C_1^I
          \bigg\}

       + \delta(\overline{z}_1) L_{z_2}   \bigg\{
           D_1^I
          + \varphi^{I,(1)}_{d,1}
          \bigg\}

       + \delta(\overline{z}_1)    \bigg\{ \varphi^{I,(0)}_{d,1}  \bigg\}
        + (z_1 \leftrightarrow z_2) \,,

\end{autobreak}
\nonumber\\
\begin{autobreak}
    \tilde{\textbf{S}}_{d,I}^{(2)} =

         L_{z_2} {\cal D}_2   \bigg\{
           \frac{3}{2} A^I_1 C_1^I
          \bigg\}

       + {\cal D}_1 L^2_{z_2}   \bigg\{
          3 A^I_1 C_1^I
          \bigg\}

       + {\cal D}_1 L_{z_2}   \bigg\{
          - 2 f^I_1 C_1^I
          + 3 A^I_1 D_1^I
          + A^I_1 \varphi^{I,(1)}_{d,1}
          - \beta_0 C_1^I
          \bigg\}

       +  {\cal D}_2   \bigg\{
          \frac{3}{2} A^I_1 D_1^I
          \bigg\}

       +  {\cal D}_1   \bigg\{
          - 2 f^I_1 D_1^I
          + A^I_1 \varphi^{I,(0)}_{d,1}
          - 2 A^I_1 \zeta_2 C_1^I
          + (A^I_1)^2
          - \beta_0 D_1^I
          \bigg\}

       + {\cal D}_0 L^3_{z_2}   \bigg\{
          \frac{3}{2} A^I_1 C_1^I
          \bigg\}

       + {\cal D}_0 L^2_{z_2}   \bigg\{
          - 2 f^I_1 C_1^I
          + \frac{3}{2} A^I_1 D_1^I
          + A^I_1 \varphi^{I,(1)}_{d,1}
          - \beta_0 C_1^I
          \bigg\}

       + {\cal D}_0 L_{z_2}   \bigg\{  C_2^I
          + 2 C_1^I \g^{I,(1)}_{d,1}
          - 2 f^I_1 D_1^I
          - f^I_1 \varphi^{I,(1)}_{d,1}
          + A^I_1 \varphi^{I,(0)}_{d,1}
          - 4 A^I_1 \zeta_2 C_1^I
          + (A^I_1)^2
          - \beta_0 D_1^I
          - \beta_0 \varphi^{I,(1)}_{d,1}  \bigg\}

           + {\cal D}_0  \bigg\{ D_2^I
          + 2 \g^{I,(1)}_{d,1} D_1^I
          - f^I_1 \varphi^{I,(0)}_{d,1}
          + \zeta_2 f^I_1 C_1^I
          - A^I_1 f^I_1
          + 3 A^I_1 \zeta_3 C_1^I
          - 2 A^I_1 \zeta_2 D_1^I
          - A^I_1 \zeta_2 \varphi^{I,(1)}_{d,1}
          - \beta_0 \varphi^{I,(0)}_{d,1}   \bigg\}

       + \delta(\overline{z}_1) L^4_{z_2}   \bigg\{
           \frac{1}{2} A^I_1 C_1^I
          \bigg\}

       + \delta(\overline{z}_1) L^3_{z_2}   \bigg\{
          - f^I_1 C_1^I
          + \frac{1}{2} A^I_1 D_1^I
          + \frac{1}{2} A^I_1 \varphi^{I,(1)}_{d,1}
          - \frac{1}{2} \beta_0 C_1^I
          \bigg\}

       + \delta(\overline{z}_1) L^2_{z_2}   \bigg\{
           \varphi^{I,(2)}_{d,2}
          + C_2^I
          + 2 C_1^I \g^{I,(1)}_{d,1}
          - f^I_1 D_1^I
          - f^I_1 \varphi^{I,(1)}_{d,1}
          + \frac{1}{2} A^I_1 \varphi^{I,(0)}_{d,1}
          - 3 A^I_1 \zeta_2 C_1^I
          + \frac{1}{2} (A^I_1)^2
          - \frac{1}{2} \beta_0 D_1^I
          - \beta_0 \varphi^{I,(1)}_{d,1}
          \bigg\}

       + \delta(\overline{z}_1) L_{z_2}   \bigg\{
           D_2^I
          + \varphi^{I,(1)}_{d,2}
          + 2 \g^{I,(1)}_{d,1} D_1^I
          + 2 \g^{I,(1)}_{d,1} \varphi^{I,(1)}_{d,1}
          - f^I_1 \varphi^{I,(0)}_{d,1}
          + 3 \zeta_2 f^I_1 C_1^I
          - A^I_1 f^I_1
          + 5 A^I_1 \zeta_3 C_1^I
          - 2 A^I_1 \zeta_2 D_1^I
          - A^I_1 \zeta_2 \varphi^{I,(1)}_{d,1}
          - \beta_0 \varphi^{I,(0)}_{d,1}
          \bigg\}
          + \delta(\overline{z}_1)   \bigg\{  \varphi^{I,(0)}_{d,2}
          + 2 \g^{I,(1)}_{d,1} \varphi^{I,(0)}_{d,1}
          + \frac{1}{2} (f^I_1)^2
          - 2 \zeta_3 f^I_1 C_1^I
          + 2 \zeta_2 f^I_1 D_1^I
          + \zeta_2 f^I_1 \varphi^{I,(1)}_{d,1}
          + 2 A^I_1 \zeta_3 D_1^I
          + A^I_1 \zeta_3 \varphi^{I,(1)}_{d,1}
          + \frac{4}{5} A^I_1 \zeta_2^2 C_1^I
          - (A^I_1)^2 \zeta_2  \bigg\}  + (z_1 \leftrightarrow z_2) \,.
\end{autobreak}
\end{align}
The results up to N$^4$LO with explicit scale dependence are provided in the Online Resource supplied with this article.


\begin{thebibliography}{9}
\bibitem{Anastasiou:2003yy}
C.~Anastasiou, L.~J. Dixon, K.~Melnikov and F.~Petriello, \emph{{Dilepton
  rapidity distribution in the Drell-Yan process at NNLO in QCD}},
  \href{https://doi.org/10.1103/PhysRevLett.91.182002}{\emph{Phys. Rev. Lett.}
  {\bfseries 91} (2003) 182002}
  [\href{https://arxiv.org/abs/hep-ph/0306192}{{\ttfamily hep-ph/0306192}}].

\bibitem{Anastasiou:2003ds}
C.~Anastasiou, L.~J. Dixon, K.~Melnikov and F.~Petriello, \emph{{High precision
  QCD at hadron colliders: Electroweak gauge boson rapidity distributions at
  NNLO}}, \href{https://doi.org/10.1103/PhysRevD.69.094008}{\emph{Phys. Rev. D}
  {\bfseries 69} (2004) 094008}
  [\href{https://arxiv.org/abs/hep-ph/0312266}{{\ttfamily hep-ph/0312266}}].

\bibitem{Anastasiou:2004xq}
C.~Anastasiou, K.~Melnikov and F.~Petriello, \emph{{Higgs boson production at
  hadron colliders: Differential cross sections through next-to-next-to-leading
  order}}, \href{https://doi.org/10.1103/PhysRevLett.93.262002}{\emph{Phys.
  Rev. Lett.} {\bfseries 93} (2004) 262002}
  [\href{https://arxiv.org/abs/hep-ph/0409088}{{\ttfamily hep-ph/0409088}}].

\bibitem{Dulat:2018bfe}
F.~Dulat, B.~Mistlberger and A.~Pelloni, \emph{{Precision predictions at
  N$^3$LO for the Higgs boson rapidity distribution at the LHC}},
  \href{https://doi.org/10.1103/PhysRevD.99.034004}{\emph{Phys. Rev.}
  {\bfseries D99} (2019) 034004}
  [\href{https://arxiv.org/abs/1810.09462}{{\ttfamily 1810.09462}}].

\bibitem{Cieri:2018oms}
L.~Cieri, X.~Chen, T.~Gehrmann, E.~N. Glover and A.~Huss, \emph{{Higgs boson
  production at the LHC using the $q_T$ subtraction formalism at N$^3$LO QCD}},
  \href{https://doi.org/10.1007/JHEP02(2019)096}{\emph{JHEP} {\bfseries 02}
  (2019) 096} [\href{https://arxiv.org/abs/1807.11501}{{\ttfamily
  1807.11501}}].

\bibitem{Buehler:2012cu}
S.~Bühler, F.~Herzog, A.~Lazopoulos and R.~Müller, \emph{{The fully
  differential hadronic production of a Higgs boson via bottom quark fusion at
  NNLO}}, \href{https://doi.org/10.1007/JHEP07(2012)115}{\emph{JHEP} {\bfseries
  07} (2012) 115} [\href{https://arxiv.org/abs/1204.4415}{{\ttfamily
  1204.4415}}].

\bibitem{Ravindran:2006bu}
V.~Ravindran, J.~Smith and W.~L. van Neerven, \emph{{QCD threshold corrections
  to di-lepton and Higgs rapidity distributions beyond $N^{2}$ LO}},
  \href{https://doi.org/10.1016/j.nuclphysb.2007.01.005}{\emph{Nucl. Phys.}
  {\bfseries B767} (2007) 100}
  [\href{https://arxiv.org/abs/hep-ph/0608308}{{\ttfamily hep-ph/0608308}}].

\bibitem{Ravindran:2007sv}
V.~Ravindran and J.~Smith, \emph{{Threshold corrections to rapidity
  distributions of $Z$ and $W^\pm$ bosons beyond $N^{2}$ LO at hadron
  colliders}}, \href{https://doi.org/10.1103/PhysRevD.76.114004}{\emph{Phys.
  Rev.} {\bfseries D76} (2007) 114004}
  [\href{https://arxiv.org/abs/0708.1689}{{\ttfamily 0708.1689}}].

\bibitem{Mukherjee:2006uu}
A.~Mukherjee and W.~Vogelsang, \emph{{Threshold resummation for W-boson
  production at RHIC}},
  \href{https://doi.org/10.1103/PhysRevD.73.074005}{\emph{Phys. Rev. D}
  {\bfseries 73} (2006) 074005}
  [\href{https://arxiv.org/abs/hep-ph/0601162}{{\ttfamily hep-ph/0601162}}].

\bibitem{Laenen:1992ey}
E.~Laenen and G.~F. Sterman, \emph{{Resummation for Drell-Yan differential
  distributions}},  in \emph{{7th Meeting of the APS Division of Particles
  Fields}}, pp.~987--989, 11, 1992.

\bibitem{Bolzoni:2006ky}
P.~Bolzoni, \emph{{Threshold resummation of Drell-Yan rapidity distributions}},
  \href{https://doi.org/10.1016/j.physletb.2006.10.064}{\emph{Phys. Lett. B}
  {\bfseries 643} (2006) 325}
  [\href{https://arxiv.org/abs/hep-ph/0609073}{{\ttfamily hep-ph/0609073}}].

\bibitem{Bonvini:2010tp}
M.~Bonvini, S.~Forte and G.~Ridolfi, \emph{{Soft gluon resummation of Drell-Yan
  rapidity distributions: Theory and phenomenology}},
  \href{https://doi.org/10.1016/j.nuclphysb.2011.01.023}{\emph{Nucl. Phys. B}
  {\bfseries 847} (2011) 93} [\href{https://arxiv.org/abs/1009.5691}{{\ttfamily
  1009.5691}}].

\bibitem{Catani:1989ne}
S.~Catani and L.~Trentadue, \emph{{Resummation of the QCD Perturbative Series
  for Hard Processes}},
  \href{https://doi.org/10.1016/0550-3213(89)90273-3}{\emph{Nucl. Phys. B}
  {\bfseries 327} (1989) 323}.

\bibitem{Banerjee:2017cfc}
P.~Banerjee, G.~Das, P.~K. Dhani and V.~Ravindran, \emph{{Threshold resummation
  of the rapidity distribution for Higgs production at NNLO+NNLL}},
  \href{https://doi.org/10.1103/PhysRevD.97.054024}{\emph{Phys. Rev. D}
  {\bfseries 97} (2018) 054024}
  [\href{https://arxiv.org/abs/1708.05706}{{\ttfamily 1708.05706}}].

\bibitem{Lustermans:2019cau}
G.~Lustermans, J.~K. Michel and F.~J. Tackmann, \emph{{Generalized Threshold
  Factorization with Full Collinear Dynamics}},
  \href{https://arxiv.org/abs/1908.00985}{{\ttfamily 1908.00985}}.


\bibitem{Ahmed:2014uya}
T.~Ahmed, M.~K. Mandal, N.~Rana and V.~Ravindran, \emph{{Rapidity Distributions
  in Drell-Yan and Higgs Productions at Threshold to Third Order in QCD}},
  \href{https://doi.org/10.1103/PhysRevLett.113.212003}{\emph{Phys. Rev. Lett.}
  {\bfseries 113} (2014) 212003}
  [\href{https://arxiv.org/abs/1404.6504}{{\ttfamily 1404.6504}}].

\bibitem{Ahmed:2014cla}
T.~Ahmed, M.~Mahakhud, N.~Rana and V.~Ravindran, \emph{{Drell-Yan Production at
  Threshold to Third Order in QCD}},
  \href{https://doi.org/10.1103/PhysRevLett.113.112002}{\emph{Phys. Rev. Lett.}
  {\bfseries 113} (2014) 112002}
  [\href{https://arxiv.org/abs/1404.0366}{{\ttfamily 1404.0366}}].

\bibitem{Anastasiou:2014vaa}
C.~Anastasiou, C.~Duhr, F.~Dulat, E.~Furlan, T.~Gehrmann, F.~Herzog et~al.,
  \emph{{Higgs boson gluon–fusion production at threshold in N$^3$LO QCD}},
  \href{https://doi.org/10.1016/j.physletb.2014.08.067}{\emph{Phys. Lett.}
  {\bfseries B737} (2014) 325}
  [\href{https://arxiv.org/abs/1403.4616}{{\ttfamily 1403.4616}}].

\bibitem{Ahmed:2014era}
T.~Ahmed, M.~K. Mandal, N.~Rana and V.~Ravindran, \emph{{Higgs Rapidity
  Distribution in $b {\bar b}$ Annihilation at Threshold in N$^{3}$LO QCD}},
  \href{https://doi.org/10.1007/JHEP02(2015)131}{\emph{JHEP} {\bfseries 02}
  (2015) 131} [\href{https://arxiv.org/abs/1411.5301}{{\ttfamily 1411.5301}}].

\bibitem{Banerjee:2018vvb}
P.~Banerjee, G.~Das, P.~K. Dhani and V.~Ravindran, \emph{{Threshold resummation
  of the rapidity distribution for Drell-Yan production at NNLO+NNLL}},
  \href{https://doi.org/10.1103/PhysRevD.98.054018}{\emph{Phys. Rev. D}
  {\bfseries 98} (2018) 054018}
  [\href{https://arxiv.org/abs/1805.01186}{{\ttfamily 1805.01186}}].

\bibitem{Laenen:2008ux}
E.~Laenen, L.~Magnea and G.~Stavenga, \emph{{On next-to-eikonal corrections to
  threshold resummation for the Drell-Yan and DIS cross sections}},
  \href{https://doi.org/10.1016/j.physletb.2008.09.037}{\emph{Phys. Lett. B}
  {\bfseries 669} (2008) 173}
  [\href{https://arxiv.org/abs/0807.4412}{{\ttfamily 0807.4412}}].

\bibitem{Laenen:2010uz}
E.~Laenen, L.~Magnea, G.~Stavenga and C.~D. White, \emph{{Next-to-Eikonal
  Corrections to Soft Gluon Radiation: A Diagrammatic Approach}},
  \href{https://doi.org/10.1007/JHEP01(2011)141}{\emph{JHEP} {\bfseries 01}
  (2011) 141} [\href{https://arxiv.org/abs/1010.1860}{{\ttfamily 1010.1860}}].

\bibitem{Bonocore:2014wua}
D.~Bonocore, E.~Laenen, L.~Magnea, L.~Vernazza and C.~D. White, \emph{{The
  method of regions and next-to-soft corrections in Drell--Yan production}},
  \href{https://doi.org/10.1016/j.physletb.2015.02.008}{\emph{Phys. Lett. B}
  {\bfseries 742} (2015) 375}
  [\href{https://arxiv.org/abs/1410.6406}{{\ttfamily 1410.6406}}].

\bibitem{Bonocore:2015esa}
D.~Bonocore, E.~Laenen, L.~Magnea, S.~Melville, L.~Vernazza and C.~White,
  \emph{{A factorization approach to next-to-leading-power threshold
  logarithms}}, \href{https://doi.org/10.1007/JHEP06(2015)008}{\emph{JHEP}
  {\bfseries 06} (2015) 008}
  [\href{https://arxiv.org/abs/1503.05156}{{\ttfamily 1503.05156}}].

\bibitem{Bonocore:2016awd}
D.~Bonocore, E.~Laenen, L.~Magnea, L.~Vernazza and C.~White, \emph{{Non-abelian
  factorisation for next-to-leading-power threshold logarithms}},
  \href{https://doi.org/10.1007/JHEP12(2016)121}{\emph{JHEP} {\bfseries 12}
  (2016) 121} [\href{https://arxiv.org/abs/1610.06842}{{\ttfamily
  1610.06842}}].

\bibitem{DelDuca:2017twk}
V.~Del~Duca, E.~Laenen, L.~Magnea, L.~Vernazza and C.~White,
  \emph{{Universality of next-to-leading power threshold effects for colourless
  final states in hadronic collisions}},
  \href{https://doi.org/10.1007/JHEP11(2017)057}{\emph{JHEP} {\bfseries 11}
  (2017) 057} [\href{https://arxiv.org/abs/1706.04018}{{\ttfamily
  1706.04018}}].

\bibitem{Bahjat-Abbas:2019fqa}
N.~Bahjat-Abbas, D.~Bonocore, J.~Sinninghe~Damst\'e, E.~Laenen, L.~Magnea,
  L.~Vernazza et~al., \emph{{Diagrammatic resummation of leading-logarithmic
  threshold effects at next-to-leading power}},
  \href{https://doi.org/10.1007/JHEP11(2019)002}{\emph{JHEP} {\bfseries 11}
  (2019) 002} [\href{https://arxiv.org/abs/1905.13710}{{\ttfamily
  1905.13710}}].

\bibitem{Soar:2009yh}
G.~Soar, S.~Moch, J.~Vermaseren and A.~Vogt, \emph{{On Higgs-exchange DIS,
  physical evolution kernels and fourth-order splitting functions at large x}},
  \href{https://doi.org/10.1016/j.nuclphysb.2010.02.003}{\emph{Nucl. Phys. B}
  {\bfseries 832} (2010) 152}
  [\href{https://arxiv.org/abs/0912.0369}{{\ttfamily 0912.0369}}].

\bibitem{Moch:2009hr}
S.~Moch and A.~Vogt, \emph{{On non-singlet physical evolution kernels and
  large-x coefficient functions in perturbative QCD}},
  \href{https://doi.org/10.1088/1126-6708/2009/11/099}{\emph{JHEP} {\bfseries
  11} (2009) 099} [\href{https://arxiv.org/abs/0909.2124}{{\ttfamily
  0909.2124}}].

\bibitem{deFlorian:2014vta}
D.~de~Florian, J.~Mazzitelli, S.~Moch and A.~Vogt, \emph{{Approximate N$^{3}$LO
  Higgs-boson production cross section using physical-kernel constraints}},
  \href{https://doi.org/10.1007/JHEP10(2014)176}{\emph{JHEP} {\bfseries 10}
  (2014) 176} [\href{https://arxiv.org/abs/1408.6277}{{\ttfamily 1408.6277}}].

\bibitem{Beneke:2018gvs}
M.~Beneke, A.~Broggio, M.~Garny, S.~Jaskiewicz, R.~Szafron, L.~Vernazza et~al.,
  \emph{{Leading-logarithmic threshold resummation of the Drell-Yan process at
  next-to-leading power}},
  \href{https://doi.org/10.1007/JHEP03(2019)043}{\emph{JHEP} {\bfseries 03}
  (2019) 043} [\href{https://arxiv.org/abs/1809.10631}{{\ttfamily
  1809.10631}}].

\bibitem{Beneke:2019mua}
M.~Beneke, M.~Garny, S.~Jaskiewicz, R.~Szafron, L.~Vernazza and J.~Wang,
  \emph{{Leading-logarithmic threshold resummation of Higgs production in gluon
  fusion at next-to-leading power}},
  \href{https://doi.org/10.1007/JHEP01(2020)094}{\emph{JHEP} {\bfseries 01}
  (2020) 094} [\href{https://arxiv.org/abs/1910.12685}{{\ttfamily
  1910.12685}}].

\bibitem{Beneke:2019oqx}
M.~Beneke, A.~Broggio, S.~Jaskiewicz and L.~Vernazza, \emph{{Threshold
  factorization of the Drell-Yan process at next-to-leading power}},
  \href{https://arxiv.org/abs/1912.01585}{{\ttfamily 1912.01585}}.

\bibitem{Ajjath:2020ulr}
A.~H. Ajjath, P.~Mukherjee and V.~Ravindran, \emph{{On next to soft corrections
  to Drell-Yan and Higgs Boson productions}},
  \href{https://arxiv.org/abs/2006.06726}{{\ttfamily 2006.06726}}.

\bibitem{Ajjath:2020lwb}
A.~H. Ajjath, P.~Mukherjee, V.~Ravindran, A.~Sankar and S.~Tiwari, \emph{{On
  next to soft corrections for Drell-Yan and Higgs boson rapidity distributions
  beyond N$^3$LO}},
  \href{https://doi.org/10.1103/PhysRevD.103.L111502}{\emph{Phys. Rev. D}
  {\bfseries 103} (2021) L111502}
  [\href{https://arxiv.org/abs/2010.00079}{{\ttfamily 2010.00079}}].

\bibitem{Mueller:1979ih}
A.~H. Mueller, \emph{{On the Asymptotic Behavior of the Sudakov Form-factor}},
  \href{https://doi.org/10.1103/PhysRevD.20.2037}{\emph{Phys. Rev.} {\bfseries
  D20} (1979) 2037}.

\bibitem{Sen:1981sd}
A.~Sen, \emph{{Asymptotic Behavior of the Sudakov Form-Factor in QCD}},
  \href{https://doi.org/10.1103/PhysRevD.24.3281}{\emph{Phys. Rev.} {\bfseries
  D24} (1981) 3281}.

\bibitem{Magnea:1990zb}
L.~Magnea and G.~F. Sterman, \emph{{Analytic continuation of the Sudakov
  form-factor in QCD}},
  \href{https://doi.org/10.1103/PhysRevD.42.4222}{\emph{Phys. Rev. D}
  {\bfseries 42} (1990) 4222}.

\bibitem{Catani:1998bh}
S.~Catani, \emph{{The Singular behavior of QCD amplitudes at two loop order}},
  \href{https://doi.org/10.1016/S0370-2693(98)00332-3}{\emph{Phys. Lett.}
  {\bfseries B427} (1998) 161}
  [\href{https://arxiv.org/abs/hep-ph/9802439}{{\ttfamily hep-ph/9802439}}].

\bibitem{Sterman:2002qn}
G.~F. Sterman and M.~E. Tejeda-Yeomans, \emph{{Multiloop amplitudes and
  resummation}},
  \href{https://doi.org/10.1016/S0370-2693(02)03100-3}{\emph{Phys. Lett.}
  {\bfseries B552} (2003) 48}
  [\href{https://arxiv.org/abs/hep-ph/0210130}{{\ttfamily hep-ph/0210130}}].

\bibitem{Ravindran:2004mb}
V.~Ravindran, J.~Smith and W.~L. van Neerven, \emph{{Two-loop corrections to
  Higgs boson production}},
  \href{https://doi.org/10.1016/j.nuclphysb.2004.10.039}{\emph{Nucl. Phys.}
  {\bfseries B704} (2005) 332}
  [\href{https://arxiv.org/abs/hep-ph/0408315}{{\ttfamily hep-ph/0408315}}].

\bibitem{Moch:2005tm}
S.~Moch, J.~A.~M. Vermaseren and A.~Vogt, \emph{{Three-loop results for quark
  and gluon form-factors}},
  \href{https://doi.org/10.1016/j.physletb.2005.08.067}{\emph{Phys. Lett.}
  {\bfseries B625} (2005) 245}
  [\href{https://arxiv.org/abs/hep-ph/0508055}{{\ttfamily hep-ph/0508055}}].

\bibitem{Becher:2009cu}
T.~Becher and M.~Neubert, \emph{{Infrared singularities of scattering
  amplitudes in perturbative QCD}},
  \href{https://doi.org/10.1103/PhysRevLett.102.162001,
  10.1103/PhysRevLett.111.199905}{\emph{Phys. Rev. Lett.} {\bfseries 102}
  (2009) 162001} [\href{https://arxiv.org/abs/0901.0722}{{\ttfamily
  0901.0722}}].

\bibitem{Gardi:2009qi}
E.~Gardi and L.~Magnea, \emph{{Factorization constraints for soft anomalous
  dimensions in QCD scattering amplitudes}},
  \href{https://doi.org/10.1088/1126-6708/2009/03/079}{\emph{JHEP} {\bfseries
  03} (2009) 079} [\href{https://arxiv.org/abs/0901.1091}{{\ttfamily
  0901.1091}}].

\bibitem{Dixon:2008gr}
L.~J. Dixon, L.~Magnea and G.~F. Sterman, \emph{{Universal structure of
  subleading infrared poles in gauge theory amplitudes}},
  \href{https://doi.org/10.1088/1126-6708/2008/08/022}{\emph{JHEP} {\bfseries
  08} (2008) 022} [\href{https://arxiv.org/abs/0805.3515}{{\ttfamily
  0805.3515}}].

\bibitem{Ahmed:2020nci}
T.~Ahmed, A.~A. H., G.~Das, P.~Mukherjee, V.~Ravindran and S.~Tiwari,
  \emph{{Soft-virtual correction and threshold resummation for $n$-colorless
  particles to fourth order in QCD: Part I}},
  \href{https://arxiv.org/abs/2010.02979}{{\ttfamily 2010.02979}}.

\bibitem{Collins:1980ih}
J.~C. Collins, \emph{{Algorithm to Compute Corrections to the Sudakov
  Form-factor}}, \href{https://doi.org/10.1103/PhysRevD.22.1478}{\emph{Phys.
  Rev.} {\bfseries D22} (1980) 1478}.

\bibitem{Ravindran:2005vv}
V.~Ravindran, \emph{{On Sudakov and soft resummations in QCD}},
  \href{https://doi.org/10.1016/j.nuclphysb.2006.04.008}{\emph{Nucl. Phys.}
  {\bfseries B746} (2006) 58}
  [\href{https://arxiv.org/abs/hep-ph/0512249}{{\ttfamily hep-ph/0512249}}].

\bibitem{Ravindran:2006cg}
V.~Ravindran, \emph{{Higher-order threshold effects to inclusive processes in
  QCD}}, \href{https://doi.org/10.1016/j.nuclphysb.2006.06.025}{\emph{Nucl.
  Phys.} {\bfseries B752} (2006) 173}
  [\href{https://arxiv.org/abs/hep-ph/0603041}{{\ttfamily hep-ph/0603041}}].

\bibitem{Ahmed:2017gyt}
T.~Ahmed, J.~M. Henn and M.~Steinhauser, \emph{{High energy behaviour of form
  factors}}, \href{https://doi.org/10.1007/JHEP06(2017)125}{\emph{JHEP}
  {\bfseries 06} (2017) 125}
  [\href{https://arxiv.org/abs/1704.07846}{{\ttfamily 1704.07846}}].

\bibitem{Moch:2018wjh}
S.~Moch, B.~Ruijl, T.~Ueda, J.~M. Vermaseren and A.~Vogt, \emph{{On quartic
  colour factors in splitting functions and the gluon cusp anomalous
  dimension}},
  \href{https://doi.org/10.1016/j.physletb.2018.06.017}{\emph{Phys. Lett. B}
  {\bfseries 782} (2018) 627}
  [\href{https://arxiv.org/abs/1805.09638}{{\ttfamily 1805.09638}}].

\bibitem{Korchemsky:1987wg}
G.~Korchemsky and A.~Radyushkin, \emph{{Renormalization of the Wilson Loops
  Beyond the Leading Order}},
  \href{https://doi.org/10.1016/0550-3213(87)90277-X}{\emph{Nucl. Phys. B}
  {\bfseries 283} (1987) 342}.

\bibitem{Moch:2004pa}
S.~Moch, J.~A.~M. Vermaseren and A.~Vogt, \emph{{The Three loop splitting
  functions in QCD: The Nonsinglet case}},
  \href{https://doi.org/10.1016/j.nuclphysb.2004.03.030}{\emph{Nucl. Phys.}
  {\bfseries B688} (2004) 101}
  [\href{https://arxiv.org/abs/hep-ph/0403192}{{\ttfamily hep-ph/0403192}}].

\bibitem{Vogt:2004mw}
A.~Vogt, S.~Moch and J.~A.~M. Vermaseren, \emph{{The Three-loop splitting
  functions in QCD: The Singlet case}},
  \href{https://doi.org/10.1016/j.nuclphysb.2004.04.024}{\emph{Nucl. Phys.}
  {\bfseries B691} (2004) 129}
  [\href{https://arxiv.org/abs/hep-ph/0404111}{{\ttfamily hep-ph/0404111}}].

\bibitem{Henn:2019swt}
J.~M. Henn, G.~P. Korchemsky and B.~Mistlberger, \emph{{The full four-loop cusp
  anomalous dimension in $\mathcal{N}=4$ super Yang-Mills and QCD}},
  \href{https://doi.org/10.1007/JHEP04(2020)018}{\emph{JHEP} {\bfseries 04}
  (2020) 018} [\href{https://arxiv.org/abs/1911.10174}{{\ttfamily
  1911.10174}}].

\bibitem{vonManteuffel:2020vjv}
A.~von Manteuffel, E.~Panzer and R.~M. Schabinger, \emph{{Analytic four-loop
  anomalous dimensions in massless QCD from form factors}},
  \href{https://doi.org/10.1103/PhysRevLett.124.162001}{\emph{Phys. Rev. Lett.}
  {\bfseries 124} (2020) 162001}
  [\href{https://arxiv.org/abs/2002.04617}{{\ttfamily 2002.04617}}].

\bibitem{Becher:2006mr}
T.~Becher, M.~Neubert and B.~D. Pecjak, \emph{{Factorization and Momentum-Space
  Resummation in Deep-Inelastic Scattering}},
  \href{https://doi.org/10.1088/1126-6708/2007/01/076}{\emph{JHEP} {\bfseries
  01} (2007) 076} [\href{https://arxiv.org/abs/hep-ph/0607228}{{\ttfamily
  hep-ph/0607228}}].

\bibitem{Becher:2009qa}
T.~Becher and M.~Neubert, \emph{{On the Structure of Infrared Singularities of
  Gauge-Theory Amplitudes}},
  \href{https://doi.org/10.1088/1126-6708/2009/06/081,
  10.1007/JHEP11(2013)024}{\emph{JHEP} {\bfseries 06} (2009) 081}
  [\href{https://arxiv.org/abs/0903.1126}{{\ttfamily 0903.1126}}].

\bibitem{Davies:2016jie}
J.~Davies, A.~Vogt, B.~Ruijl, T.~Ueda and J.~Vermaseren, \emph{{Large-$n_f$
  contributions to the four-loop splitting functions in QCD}},
  \href{https://doi.org/10.1016/j.nuclphysb.2016.12.012}{\emph{Nucl. Phys. B}
  {\bfseries 915} (2017) 335}
  [\href{https://arxiv.org/abs/1610.07477}{{\ttfamily 1610.07477}}].

\bibitem{Das:2019btv}
G.~Das, S.-O. Moch and A.~Vogt, \emph{{Soft corrections to inclusive
  deep-inelastic scattering at four loops and beyond}},
  \href{https://arxiv.org/abs/1912.12920}{{\ttfamily 1912.12920}}.

\bibitem{Das:2020adl}
G.~Das, S.~Moch and A.~Vogt, \emph{{Approximate four-loop QCD corrections to
  the Higgs-boson production cross section}},
  \href{https://arxiv.org/abs/2004.00563}{{\ttfamily 2004.00563}}.

\bibitem{Henn:2016men}
J.~M. Henn, A.~V. Smirnov, V.~A. Smirnov and M.~Steinhauser, \emph{{A planar
  four-loop form factor and cusp anomalous dimension in QCD}},
  \href{https://doi.org/10.1007/JHEP05(2016)066}{\emph{JHEP} {\bfseries 05}
  (2016) 066} [\href{https://arxiv.org/abs/1604.03126}{{\ttfamily
  1604.03126}}].

\bibitem{Lee:2016ixa}
J.~Henn, A.~V. Smirnov, V.~A. Smirnov, M.~Steinhauser and R.~N. Lee,
  \emph{{Four-loop photon quark form factor and cusp anomalous dimension in the
  large-$N_c$ limit of QCD}},
  \href{https://doi.org/10.1007/JHEP03(2017)139}{\emph{JHEP} {\bfseries 03}
  (2017) 139} [\href{https://arxiv.org/abs/1612.04389}{{\ttfamily
  1612.04389}}].

\bibitem{vonManteuffel:2016xki}
A.~von Manteuffel and R.~M. Schabinger, \emph{{Quark and gluon form factors to
  four-loop order in QCD: the $N_f^3$ contributions}},
  \href{https://doi.org/10.1103/PhysRevD.95.034030}{\emph{Phys. Rev. D}
  {\bfseries 95} (2017) 034030}
  [\href{https://arxiv.org/abs/1611.00795}{{\ttfamily 1611.00795}}].

\bibitem{Lee:2017mip}
R.~N. Lee, A.~V. Smirnov, V.~A. Smirnov and M.~Steinhauser, \emph{{The $n_f^2$
  contributions to fermionic four-loop form factors}},
  \href{https://doi.org/10.1103/PhysRevD.96.014008}{\emph{Phys. Rev. D}
  {\bfseries 96} (2017) 014008}
  [\href{https://arxiv.org/abs/1705.06862}{{\ttfamily 1705.06862}}].

\bibitem{vonManteuffel:2019wbj}
A.~von Manteuffel and R.~M. Schabinger, \emph{{Quark and gluon form factors in
  four loop QCD: The $N_f^2$ and $N_{q\gamma} N_f$ contributions}},
  \href{https://doi.org/10.1103/PhysRevD.99.094014}{\emph{Phys. Rev. D}
  {\bfseries 99} (2019) 094014}
  [\href{https://arxiv.org/abs/1902.08208}{{\ttfamily 1902.08208}}].

\bibitem{Gehrmann:2010ue}
T.~Gehrmann, E.~W.~N. Glover, T.~Huber, N.~Ikizlerli and C.~Studerus,
  \emph{{Calculation of the quark and gluon form factors to three loops in
  QCD}}, \href{https://doi.org/10.1007/JHEP06(2010)094}{\emph{JHEP} {\bfseries
  06} (2010) 094} [\href{https://arxiv.org/abs/1004.3653}{{\ttfamily
  1004.3653}}].

\bibitem{Gehrmann:2010tu}
T.~Gehrmann, E.~Glover, T.~Huber, N.~Ikizlerli and C.~Studerus, \emph{{The
  quark and gluon form factors to three loops in QCD through to
  O(eps\textasciicircum{}2)}},
  \href{https://doi.org/10.1007/JHEP11(2010)102}{\emph{JHEP} {\bfseries 11}
  (2010) 102} [\href{https://arxiv.org/abs/1010.4478}{{\ttfamily 1010.4478}}].

\bibitem{Gehrmann:2014vha}
T.~Gehrmann and D.~Kara, \emph{{The $Hb\bar{b}$ form factor to three loops in
  QCD}}, \href{https://doi.org/10.1007/JHEP09(2014)174}{\emph{JHEP} {\bfseries
  09} (2014) 174} [\href{https://arxiv.org/abs/1407.8114}{{\ttfamily
  1407.8114}}].

\bibitem{Chetyrkin:2017bjc}
K.~G. Chetyrkin, G.~Falcioni, F.~Herzog and J.~A.~M. Vermaseren,
  \emph{{Five-loop renormalisation of QCD in covariant gauges}},
  \href{https://doi.org/10.1007/JHEP10(2017)179}{\emph{JHEP} {\bfseries 10}
  (2017) 179} [\href{https://arxiv.org/abs/1709.08541}{{\ttfamily
  1709.08541}}].

\bibitem{Luthe:2017ttg}
T.~Luthe, A.~Maier, P.~Marquard and Y.~Schroder, \emph{{The five-loop Beta
  function for a general gauge group and anomalous dimensions beyond Feynman
  gauge}}, \href{https://doi.org/10.1007/JHEP10(2017)166}{\emph{JHEP}
  {\bfseries 10} (2017) 166}
  [\href{https://arxiv.org/abs/1709.07718}{{\ttfamily 1709.07718}}].

\bibitem{Herzog:2017ohr}
F.~Herzog, B.~Ruijl, T.~Ueda, J.~A.~M. Vermaseren and A.~Vogt, \emph{{The
  five-loop beta function of Yang-Mills theory with fermions}},
  \href{https://doi.org/10.1007/JHEP02(2017)090}{\emph{JHEP} {\bfseries 02}
  (2017) 090} [\href{https://arxiv.org/abs/1701.01404}{{\ttfamily
  1701.01404}}].

\bibitem{Baikov:2016tgj}
P.~A. Baikov, K.~G. Chetyrkin and J.~H. K\"uhn, \emph{{Five-Loop Running of the
  QCD coupling constant}},
  \href{https://doi.org/10.1103/PhysRevLett.118.082002}{\emph{Phys. Rev. Lett.}
  {\bfseries 118} (2017) 082002}
  [\href{https://arxiv.org/abs/1606.08659}{{\ttfamily 1606.08659}}].

\bibitem{Vermaseren:2000nd}
J.~A.~M. Vermaseren, \emph{{New features of FORM}},
  \href{https://arxiv.org/abs/math-ph/0010025}{{\ttfamily math-ph/0010025}}.

\bibitem{Ruijl:2017dtg}
B.~Ruijl, T.~Ueda and J.~Vermaseren, \emph{{FORM version 4.2}},
  \href{https://arxiv.org/abs/1707.06453}{{\ttfamily 1707.06453}}.
  
  
\bibitem{Catani:2014uta}
S.~Catani, L.~Cieri, D.~de~Florian, G.~Ferrera and M.~Grazzini,
  \emph{{Threshold resummation at N$^3$LL accuracy and soft-virtual cross
  sections at N$^3$LO}},
  \href{https://doi.org/10.1016/j.nuclphysb.2014.09.012}{\emph{Nucl. Phys.}
  {\bfseries B888} (2014) 75}
  [\href{https://arxiv.org/abs/1405.4827}{{\ttfamily 1405.4827}}].

\bibitem{Ajjath:2020sjk}
A.~H. Ajjath, P.~Mukherjee, V.~Ravindran, A.~Sankar and S.~Tiwari, \emph{{On
  next to soft threshold corrections to DIS and SIA processes}},
  \href{https://doi.org/10.1007/JHEP04(2021)131}{\emph{JHEP} {\bfseries 04}
  (2021) 131}, [\href{https://arxiv.org/abs/2007.12214}{{\ttfamily
  2007.12214}}].
\end{thebibliography}
\end{document}